\documentclass{article}

    \PassOptionsToPackage{numbers, compress}{natbib}

\usepackage[main, final]{neurips_2026}

\usepackage[utf8]{inputenc} 
\usepackage[T1]{fontenc}    
\usepackage{hyperref}       
\usepackage{url}            
\usepackage{booktabs}       
\usepackage{amsfonts}       
\usepackage{nicefrac}       
\usepackage{microtype}      
\usepackage[table]{xcolor}        
\usepackage{placeins}
\usepackage{graphicx}
\usepackage{subcaption}

\usepackage{amsmath,amsfonts,bm}

\def\eqref#1{equation~\ref{#1}}

\def\1{\bm{1}}

\DeclareMathAlphabet{\mathsfit}{\encodingdefault}{\sfdefault}{m}{sl}
\SetMathAlphabet{\mathsfit}{bold}{\encodingdefault}{\sfdefault}{bx}{n}

\def\gA{{\mathcal{A}}}

\def\gN{{\mathcal{N}}}
\def\gO{{\mathcal{O}}}

\def\gS{{\mathcal{S}}}

\def\gY{{\mathcal{Y}}}

\newcommand{\E}{\mathbb{E}}

\newcommand{\softmax}{\mathrm{softmax}}

\usepackage{paralist}

\usepackage{algorithm}
\usepackage{algorithmic}
\usepackage{enumitem}

\usepackage{amsmath}
\usepackage{amssymb}
\usepackage{mathtools}
\usepackage{amsthm}
\newcommand{\cm}[1]{{\fontfamily{cmtt}\selectfont #1}}

\usepackage{adjustbox}
\usepackage{makecell}
\usepackage{colortbl}

\definecolor{seenbg}{RGB}{243,248,255}    
\definecolor{unseenbg}{RGB}{255,245,245}  
\definecolor{bluetext}{RGB}{30,30,150}    
\definecolor{redtext}{RGB}{220,30,80}  
\definecolor{graytext}{RGB}{50,50,50}
\definecolor{blacktext}{RGB}{0,0,0}

\newcommand{\me}[2]{#1{\scriptsize$\pm$#2}}
\newcommand{\best}[2]{\color{blacktext}\textbf{#1}{\scriptsize$\pm$#2}}
\newcommand{\pmsem}[1]{{\scriptsize$\pm$#1}}

\theoremstyle{plain}
\newtheorem{theorem}{Theorem}[section]

\newtheorem{lemma}[theorem]{Lemma}

\theoremstyle{definition}
\newtheorem{definition}[theorem]{Definition}

\theoremstyle{remark}

\usepackage{multirow}
\usepackage{wrapfig}

\title{Addressing Exogenous Variability in \\Cooperative Multi-Agent Reinforcement Learning}

\author{%
  \textbf{Seongmin Kim}\textsuperscript{1} \quad
  \textbf{Woohyeon Byeon}\textsuperscript{1} \quad
  \textbf{Jiwon Jeon}\textsuperscript{1}
  \\
  \textbf{Seungyul Han}\textsuperscript{2}\thanks{%
    Correspondence to: Seungyul Han (\texttt{syhan@unist.ac.kr}) or
    Youngchul Sung (\texttt{ycsung@kaist.ac.kr}).%
  }
  \quad
  \textbf{Youngchul Sung}\textsuperscript{1}\footnotemark[1]
  \\[1ex]
  \textsuperscript{1}KAIST \quad \textsuperscript{2}UNIST
  \\
  \texttt{\{seongmin.kim, woohyeon.byeon, jiwon.jeon, ycsung\}@kaist.ac.kr}
  \\
  \texttt{syhan@unist.ac.kr}
}
\begin{document}

\maketitle

\begin{abstract}
  Cooperative multi-agent reinforcement learning (MARL) often fails under exogenous variability such as shifts in opponents or environment regimes that cannot be controlled by the team but alter the transition dynamics. We formalize this challenge as Exogenous Dec-POMDP (ED-POMDP), in which an episode-level exogenous regime, selected independently of the team, indexes the dynamics, and the global state is partitioned into components that are directly responsive to the current team action and components that are not directly action-responsive at the current step. This formulation exposes mediated influence: while team actions do not directly update the latter at the current step, they can shape their later evolution through induced changes in the former. Based on this, we propose \textbf{LEICA}, a CTDE-compatible algorithm that learns history-conditioned context representations for both components and shapes policy updates using agent-wise, influence-weighted counterfactual intrinsic rewards.
  Across SMAX benchmarks with opponent strategy shifts, LEICA consistently improves both training performance and generalization to unseen opponents' strategies over existing baselines, supporting the usefulness of influence-weighted shaping under exogenous train-test regime shifts.
\end{abstract}

\section{Introduction}
\label{sec:intro}

Multi-Agent Reinforcement Learning (MARL) has achieved success across a wide range of domains including games \citep{silver2016mastering, vinyals2019grandmaster}, autonomous robotics, and large-scale decision-making. 
In particular, cooperative benchmarks such as the StarCraft Multi-Agent Challenge (SMAC) \citep{samvelyan2019starcraft} designed to model battles against opponent teams, have spurred the development of various  MARL algorithms including value decomposition methods \citep{sunehag2017value, rashid2018qmix} and actor-critic methods \citep{yu2022surprising,zhong2024heterogeneous}. 

Despite such progress, current cooperative MARL methods remain brittle under task variations, such as changes in opponent strategies, scenario parameters, or latent environment regimes. Policies trained on a narrow distribution often overfit to specific training regimes and collapse at test time \citep{deng2024smac}. Crucially, these variations are typically exogenous: they are not controlled by the agent team, but they alter the environment dynamics \citep{mahajan2022generalization}. While previous MARL methods  treat these variations as ordinary task changes, we explicitly leverage their exogenous nature to improve both training performance and generalization.

Our idea is to separate the state components that the current team action updates directly from those that are updated mainly by the opponent or the environment, and to learn which directly induced changes influence the later evolution of the latter under the current regime. We formalize this view as {\em Exogenous Dec-POMDPs (ED-POMDPs)}: the term \emph{exogenous} refers to the episode-level regime, which is selected independently of the team, while the global state is partitioned into \emph{directly action-responsive} components and components that are \emph{not directly action-responsive at the current step}. The latter are not independent of the team over a trajectory; team actions can affect them later through the evolving state, for example when ally movement changes how a fixed but state-reactive opponent moves and attacks. Based on ED-POMDP, we propose Learning Exogenous Influence for Coordination and Adaptation (LEICA), a CTDE-compatible method whose influence-weighted intrinsic rewards concentrate policy updates on directly induced state changes that can shape the later response of the opponent or environment. 

We evaluated LEICA on SMAX \citep{rutherford2024jaxmarl} with train/test shifts in opponent strategies and initial conditions. 
Across multiple settings, LEICA improves performance over existing CTDE, context-inference, and robust MARL baselines that do not exploit this structure, and additional experiments on MABrax and MPE-Tag provide complementary evidence beyond SMAX.

\paragraph{Contributions}
\begin{itemize}[leftmargin=*, nosep]
    \item We formalize ED-POMDP for cooperative MARL under exogenous variability, which separates the exogenous regime from an approximate one-step action-responsiveness partition of the state and highlights influence mediated by directly action-responsive transitions.
    \item We propose \textbf{LEICA}, a CTDE method that shapes policy updates with influence-weighted, policy-relative changes of directly action-responsive state. Its multi-agent contribution is an agent-wise marginal transition predictor and counterfactual credit signal that attribute team-level effects to individual policies without enumerating joint actions, while execution remains decentralized.
    \item We present a SMAX evaluation protocol with train/test shifts in opponent strategies and initial conditions as our main evaluation, where LEICA improves performance over existing MARL baselines, complemented by MABrax and MPE-Tag experiments.
\end{itemize}
\section{Background}
A cooperative multi-agent environment is formulated as a Decentralized Partially-Observable Markov Decision Process (Dec-POMDP) \citep{bernstein2002complexity, oliehoek2016concise}, defined by the tuple $\langle \gN, \gS, \{\gA^i\}, P, R, \{\gO^i\}, O, \gamma \rangle$, where $\mathcal{N}$ is the set of agents, $\gS$ the state space, $\gA^i$ the action space of agent $i$, $P$ transition dynamics, $R$ the reward function, $\gO^i$ the observation space of agent $i$, and $\gamma$ the discount factor.  At each timestep $t$, the environment is in global state $S_t \in \gS$. Each agent $i \in \gN$ receives a local observation $o_t^i \sim O(\cdot |S_t, i)$ and selects an action $a_t^i \in \gA^i$. The joint action $A_t = (a_t^i)_{i \in \gN} \in \boldsymbol{\gA} = \prod_{i \in \gN}\gA^i$ induces a transition $S_{t+1} \sim P(\cdot \mid S_t, A_t)$ and produces a team reward $r_t = R(S_t, A_t)$.

Each agent $i$ has an action-observation history $\tau_t^i = (o_0^i, a_0^i, \ldots, o_t^i) \in T^i$, and follows a stochastic policy $\pi^i: T^i \to \Delta(\gA^i)$, where  $T^i$ denotes the set of histories and $\Delta(\gA^i)$ denotes the set of probability distributions on $\gA^i$. 
The objective is to maximize the expected discounted return $J(\boldsymbol{\pi}) = \E_{\boldsymbol{\pi}} [ \sum_{t=0}^\infty \gamma^t r_t ]$. We adopt centralized training with decentralized execution (CTDE): global state information is accessible in training time, whereas in test time agent policies rely on local observations.

\section{Related Works}

\paragraph{MARL under CTDE}
CTDE is the standard paradigm for cooperative MARL with partial observability. Major actor-critic variants include MADDPG and MAA2C \citep{lowe2017multi, papoudakis2020benchmarking}, on-policy MAPPO and IPPO \citep{yu2022surprising, de2020independent}, and theory-based trust-region methods HATRPO and HAPPO with heterogeneous-agent extensions \citep{kuba2021trust, zhong2024heterogeneous}. For value-based methods,  VDN and QMIX implement additive and monotonic mixing based on Individual-Global-Maximum principle \citep{sunehag2017value, rashid2018qmix}, and many subsequent studies investigate the value decomposition structure of MARL \citep{wang2020qplex, son2019qtran, rashid2020weighted, yang2020qatten, zheng2021episodic, son2022disentangling}. 

\paragraph{Influence-based coordination}
Influence-based methods use how much an agent affects other agents or the environment as an intrinsic signal for coordination, exploration, or opponent-aware behavior. Prior work has studied influence through communication, where messages shape other agents' behavior \citep{strouse2018learning, foerster2016learning}, and has proposed metrics for causal influence among agents \citep{lowe2019pitfalls,kim2023variational}. Other methods define influence from agents' actions and mainly use it to promote exploration \citep{wang2019influence, ma2022elign, liu2023lazy}. Opponent modeling infers other agents' policies or types \citep{he2016opponent}, and opponent-learning-aware methods explicitly model or anticipate another learner's policy updates \citep{foerster2018lola, zhao2022proximal, aghajohari2024loqa, duque2024advantage}. LEICA differs from these lines in that it does not use a separate communication channel or model concurrently learning opponents; it measures the influence of directly induced state changes on the later response of fixed but state-reactive opponents and uses it to support both adaptation to regime variation and team coordination.

\paragraph{Context-based adaptation and generalization in RL}
Several neighboring areas study generalization under distributional or strategic variation. Robust MARL seeks policies that remain effective under train-to-test shifts, for example through minimax optimization against altered opponents \citep{li2019robust} or robust value factorization under environmental uncertainty \citep{qu2026distributionally}. Meta-RL adapts to task variation by encoding interaction history into recurrent states or latent task variables, as in RL$^2$, PEARL, and VariBAD \citep{duan2016rl2,rakelly2019pearl,zintgraf2019varibad}; a few extensions to multi-agent settings are available in decentralized and game-theoretic formulations \citep{kayaalp2022dif,mao2023multi}. LEICA shares the idea that history can reveal the current regime, and our MAPPO+ctx baseline follows this spirit by adding a VariBAD-style context encoder. The key difference is that LEICA does not rely on context inference alone but further exploits state decomposition for better generalization. 
Zero-shot coordination  focuses on compatibility with novel cooperative partners \citep{hu2020other,strouse2021collaborating}, and ad hoc teamwork studies cooperation with previously unseen teammates, often without prior joint training \citep{stone2010ad,rahman2021towards}. LEICA instead assumes a jointly trained team and studies shifts in opponent or environment regimes; these settings are complementary and not subsumed by our formulation.

\paragraph{Structured state representations}
Factored MDPs exploit conditional independence among state variables for planning \citep{boutilier2000stochastic}, and Block MDPs and exogenous-information MDPs recover latent controllable states or discard exogenous noise for sample efficiency \citep{efroni2022sample,sinclair2023hindsight}. Rather than removing the not directly action-responsive component for single-agent efficiency, LEICA estimates how directly induced changes steer its later evolution and attributes this influence to individual agents.

\section{Motivation and Problem Setup}
\label{sec:motivation_setup}
We first show that cooperative MARL policies can overfit to training regimes and collapse under exogenous shifts, and that multi-regime training with context inference is still insufficient; we then formally define the problem setting.

\begin{figure*}[ht]
  \centering
  \begin{subfigure}[t]{0.27\linewidth}
    \centering
    \includegraphics[width=\linewidth]{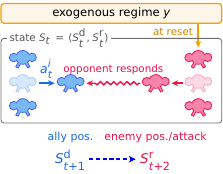}
    \caption{\textbf{Exogenous variability.}}
    \label{fig:motivation:a}
  \end{subfigure}
  \begin{subfigure}[t]{0.34\linewidth}
    \centering
    \includegraphics[width=\linewidth]{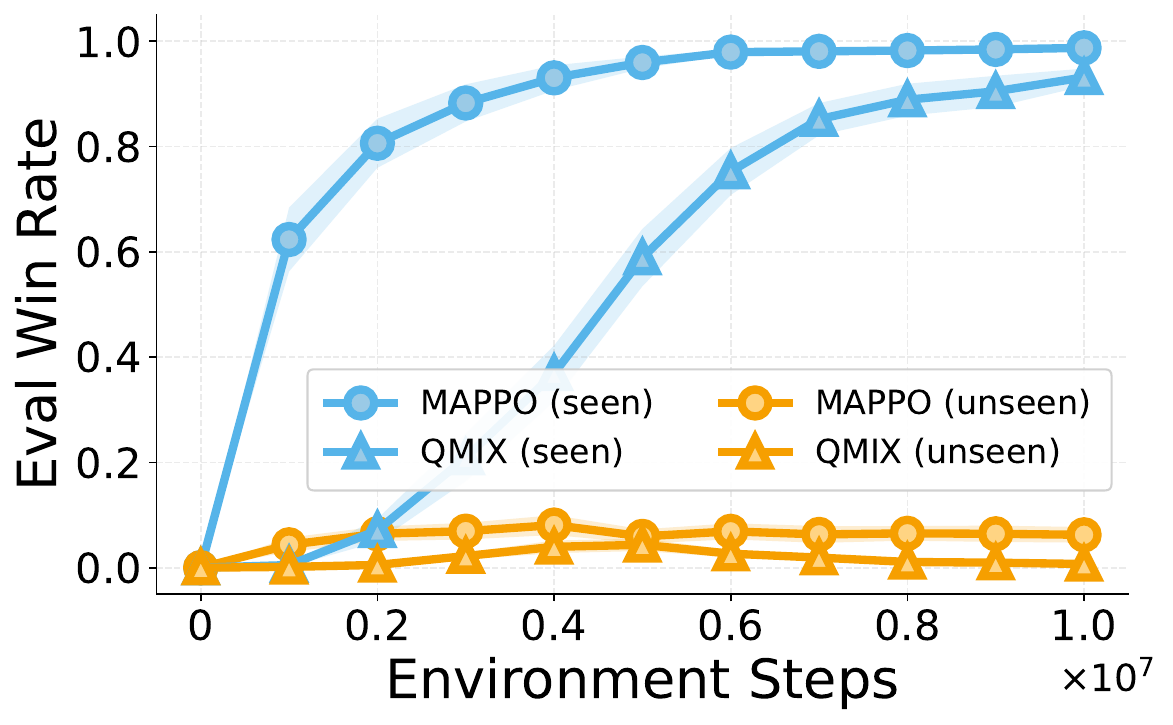}%
    \caption{\textbf{Opponent variation.}}
    \label{fig:motivation:b}
  \end{subfigure}
  \begin{subfigure}[t]{0.34\linewidth}
    \centering
    \includegraphics[width=\linewidth]{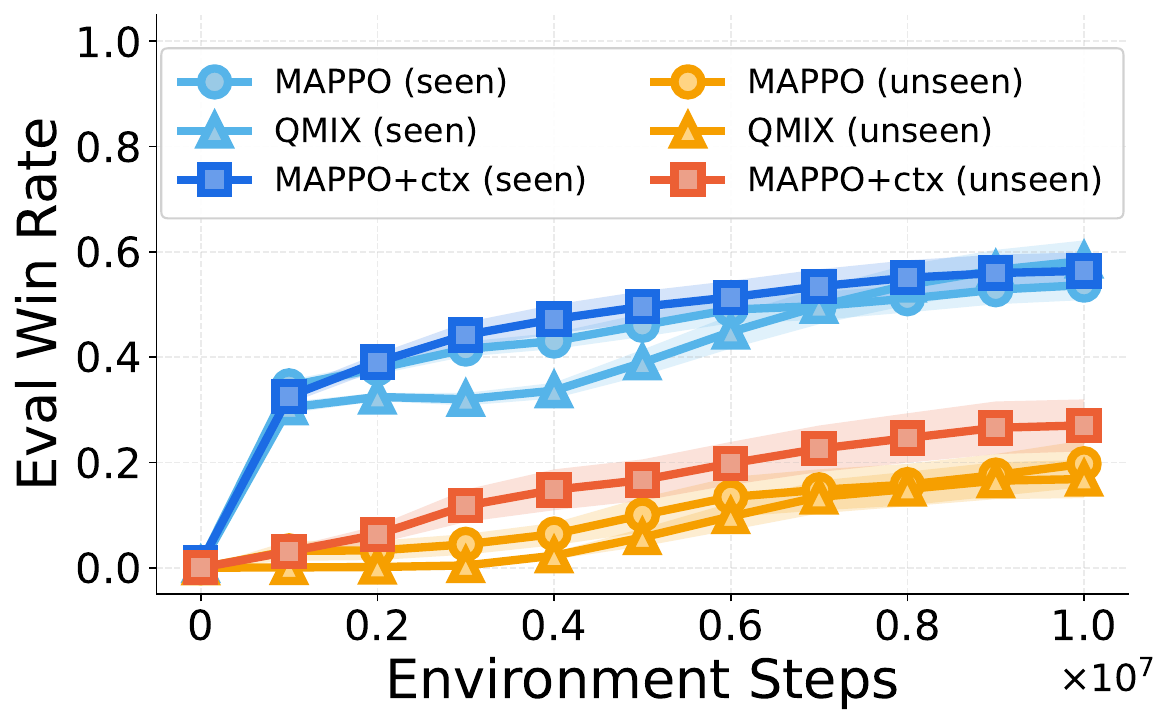}
    \caption{\textbf{Context inference is not enough.}}
    \label{fig:motivation:c}
  \end{subfigure}
  \caption{\textbf{Motivation.}
  (a) An episode-level exogenous regime (e.g., opponent strategy and initial distance) changes how a fixed but state-reactive opponent responds. Ally actions directly change ally state (e.g., positions), which then affects the opponent's later movement and attacks.
  (b) Policies trained against the default SMAX opponent achieve high performance but collapse under unseen opponent strategies.
  (c) Multi-regime training with VariBAD-style context inference \citep{zintgraf2019varibad} (MAPPO+ctx) still suffers under exogenous variation.}
  \label{fig:motivation}
\end{figure*}

\subsection{Why Exogenous Variation Breaks Coordination}
\label{sec:motivation_failure}

In cooperative MARL, policy updates are driven by the interactions experienced during training.
This process reinforces coordination patterns  tailored to the training opponents and environment dynamics. While such policies may exhibit high performance in training, they fail when encountering shifts in factors outside the team's control at test time.
We refer to such externally driven shifts as \emph{exogenous variability} in multi-agent systems.

Figure~\ref{fig:motivation:a} illustrates this failure situation, and Figure~\ref{fig:motivation:b} shows an empirical example in SMAX \citep{rutherford2024jaxmarl}.
MAPPO and QMIX achieve near-perfect win rates against the seen opponent, but their performance collapses under modified opponent strategies.
Thus, high training-regime performance does not imply robust coordination under exogenous shifts.

A natural response to this collapse is to train on multiple regimes and infer the current regime from history.
However, this context inference helps recognize which regime the team is in, but does not specify which directly induced state changes are useful for adaptation.
Figure~\ref{fig:motivation:c} shows this gap: MAPPO and QMIX reach about 60\% win rate on seen regimes but drop to roughly 20\% on unseen regimes, and MAPPO with VariBAD-style context inference \citep{zintgraf2019varibad} (MAPPO+ctx) only slightly narrows the gap.
Therefore, the key challenge lies not only in identifying the exogenous regime, but also in learning how the team should intervene through directly action-responsive state variables.
What is missing is an influence signal from directly induced state changes to the later response of the opponent or environment, which our method learns and uses to reward team actions that steer this response.

\subsection{Problem Setup}
\label{sec:problem_setup}

We formalize exogenous variability as episode-to-episode changes in the environment that the team cannot control, but that alter the interaction dynamics.
Formally, we consider a family of cooperative Dec-POMDPs $\{\mathrm{M}_{y}\}_{y\in\gY}$ indexed by a variation index $y$ that specifies the exogenous regime.
Across different $y$, the state, action, and observation spaces $(\mathcal{S},\{\mathcal{A}^i\}_{i=1}^N,\{\mathcal{O}^i\}_{i=1}^N)$ and the discount factor $\gamma$ are fixed, while the transition kernel, reward function, and initial-state distribution may vary.
We write these components as $P^{y}(S_{t+1}\!\mid S_t,A_t)$, $R^{y}(S_t,A_t)$, and $\rho_0^{y}(S_0)$.

The goal is to learn a joint policy that performs well under training regimes $y \in \mathcal{Y}_{train}$ while remaining effective under possibly shifted regimes $y \in \mathcal{Y}_{test}$ at evaluation.
Let $Y_t^y\sim P_Y(\cdot\mid S_t,y)$ denote a regime-conditioned disturbance or opponent response. Under this conditional-exogeneity model, the regime-specific transition kernel is the marginal
\[
P^y(s'\mid s,a)
=\int P(s'\mid s,a,u)\,P_Y(du\mid s,y).
\]
If $Y_t^y$ denotes the episode-level regime itself, $P_Y$ is a point mass at $y$. If it denotes a reactive opponent response, its realization may vary with the state even though the opponent policy parameters remain fixed within the episode. Our main setting uses such fixed-but-reactive opponents; concurrently learning opponents, best-response dynamics, and equilibrium adaptation are outside the main scope.

\paragraph{Training and evaluation}
\label{sec:protocol_setup}
During training, $y$ is sampled from $\mathcal{Y}_{train}$, and during evaluation from $\mathcal{Y}_{test}$.
With $J(\boldsymbol{\pi}; y)$ the expected discounted team return under joint policy $\boldsymbol{\pi}$, the training objective is $\max_{\boldsymbol{\pi}} \mathbb{E}_{y \sim \mathcal{Y}_{train}} [ J(\boldsymbol{\pi}; y) ]$, while generalization under exogenous shift is assessed on $\mathcal{Y}_{test}$, which may contain unseen regimes.

\section{Exogenous Dec-POMDP}
\label{sec:edpomdp}

We formalize exogenous variability with an \emph{Exogenous Dec-POMDP (ED-POMDP)}.
Throughout, \emph{exogenous} refers to the external selection of the episode-level regime $y$. A state-dependent response $Y_t^y$ need not be independent of the team's actions over a trajectory.
In ED-POMDP, the global state is written as $S_t=(S_t^{\mathrm{d}},S_t^{\mathrm{r}})$, where $S_t^{\mathrm{d}}$ denotes \emph{directly action-responsive} components, whose next-step values are directly updated by the current team action, and $S_t^{\mathrm{r}}$ denotes components that are \emph{not directly action-responsive at the current step}; they are primarily updated by the regime-dependent opponent or environment, and may still be influenced by earlier team actions through the evolving state.
Here, the partition can be given by domain knowledge or estimated as shown in Section~\ref{sec:method}.

\begin{definition}[Approximate one-step action-responsiveness]
\label{def:team_exo}
Let $\Pi_{\mathrm{r}}:\mathcal{S}\to\mathcal{S}^{\mathrm{r}}$ be a partition map and define $S_t^{\mathrm{r}}=\Pi_{\mathrm{r}}(S_t)$ and $S_t^{\mathrm{d}}=\Pi_{\mathrm{d}}(S_t)$,
where $\Pi_{\mathrm{d}}$ returns the complementary coordinates.
The partition satisfies the approximate one-step action-responsiveness condition with constants $\varepsilon_{\mathrm{r}},\varepsilon_{\mathrm{d}}\ge0$ if
\begin{equation}
I(S^{\mathrm{r}}_{t+1}; A_t \mid S_t, Y_t^y) \le \varepsilon_{\mathrm{r}}
\quad\text{and}\quad
I(S^{\mathrm{d}}_{t+1}; Y_t^y \mid S_t, A_t) \le \varepsilon_{\mathrm{d}} \quad \forall t,
\end{equation}
where $I(\cdot;\cdot|\cdot)$ denotes conditional mutual information.
\end{definition}
These conditions are conditional and one-step: given the current state and regime, the current team action carries little additional information about the next $S^{\mathrm{r}}$, and given the current state and action, the regime carries little additional information about the next $S^{\mathrm{d}}$. They do not imply that $S^{\mathrm{r}}$ is independent of the team's actions over a trajectory.

\begin{lemma}[Marginal invariance and conditional factorization]
\label{thm:factorization}
Let $\varepsilon_{\mathrm{r}} = \varepsilon_{\mathrm{d}} = 0$. (a) The one-step marginals satisfy
\begin{equation}
P(S^{\mathrm{r}}_{t+1} \mid S_t, A_t, Y_t^y)=P_{\mathrm{r}}(S^{\mathrm{r}}_{t+1} \mid S_t, Y_t^y),
\quad
P(S^{\mathrm{d}}_{t+1} \mid S_t, A_t, Y_t^y)=P_{\mathrm{d}}(S^{\mathrm{d}}_{t+1} \mid S_t, A_t).
\end{equation}
(b) If, in addition, $S^{\mathrm{d}}_{t+1} \perp S^{\mathrm{r}}_{t+1} \mid (S_t, A_t, Y_t^y)$, the transition factorizes as
\begin{equation}
P(S_{t+1} \mid S_t, A_t, Y_t^y) =
P_{\mathrm{d}}(S^{\mathrm{d}}_{t+1} \mid S_t, A_t)\,
P_{\mathrm{r}}(S^{\mathrm{r}}_{t+1} \mid S_t, Y_t^y).
\end{equation}
\end{lemma}
Part (a) follows directly from Definition~\ref{def:team_exo}; part (b) requires the additional conditional-independence assumption (proof in Appendix~\ref{app:method}).
LEICA does not require or enforce the exact factorization in (b): its predictors and influence signal are learned directly from rollouts and rely only on approximate one-step action-responsiveness.
In SMAX, the domain-informed split is $\varepsilon$-approximate, since contacts and sub-step interactions can couple the components.

\begin{wrapfigure}[12]{r}{0.45\textwidth}
    \vspace{-1em}
    \centering
    \includegraphics[width=1\linewidth]{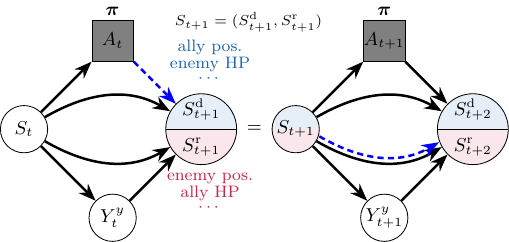}
    \caption{\textbf{Illustration of delayed mediated influence.} $A_t$ can influence a later response through $S^{\mathrm{d}}_{t+1}$.}
    \label{fig:edpomdp_cg}
\end{wrapfigure}

Under Definition~\ref{def:team_exo} with $\varepsilon_{\mathrm{r}}=0$, the conditional marginal distribution of $S_{t+1}^{\mathrm{r}}$ is invariant to the current action $A_t$ on the support of the data distribution.
As shown in Figure~\ref{fig:edpomdp_cg}, however, this invariance does not rule out delayed dependence on the action.
The action can first update the directly action-responsive state ($A_t \!\to\! S^{\mathrm{d}}_{t+1}$), which may subsequently influence $S_{t+2}^{\mathrm{r}}$.
Thus, $A_t \rightarrow S_{t+1}^{\mathrm{d}} \rightarrow S_{t+2}^{\mathrm{r}}$ represents one possible mediated influence chain.
In SMAX, for example, ally movement changes ally positions, which can subsequently alter the movement, targeting, or attacks of the fixed but state-reactive opponent.

Consequently, the actionable adaptation problem is to learn which directly induced changes can steer the subsequent evolution of $S^{\mathrm{r}}$ via this chain, and to concentrate policy updates on directly action-responsive coordinates with high mediated influence.
LEICA implements this principle with a scalable agent-wise influence signal in the next section.

\section{Proposed Method}
\label{sec:method}

\begin{figure}[h]
    
    \centering
    \includegraphics[width=0.65\linewidth]{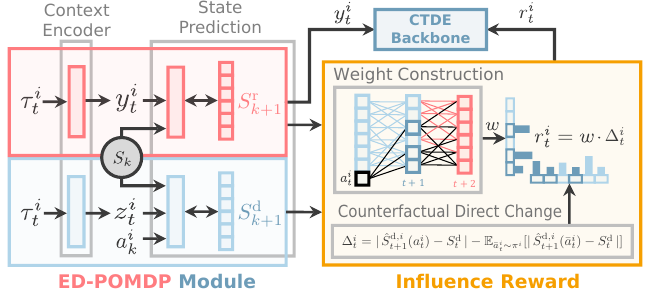}
    \caption{\textbf{Overview of LEICA.}}
    \label{fig:framework}

\end{figure}

We present \emph{Learning Exogenous Influence for Coordination and Adaptation (LEICA)}, a CTDE-compatible algorithm for ED-POMDPs.
As illustrated in Figure~\ref{fig:framework}, LEICA consists of two components:
(1) an ED-POMDP module that infers history-conditioned contexts for predicting the directly action-responsive and the not directly action-responsive state components, and
(2) an influence reward that highlights directly induced changes predictive of the later response of the opponent or environment.
\subsection{Context Inference for Direct and Response Dynamics}
\label{sec:rep}

\textbf{State partition} ~In an ED-POMDP, the global state decomposes as $S_t = (S_t^{\mathrm{d}}, S_t^{\mathrm{r}})$.
Here, a \emph{state coordinate} is one scalar feature of the centralized training state.
With domain knowledge, coordinates whose next-step update is directly affected by the current team action are assigned to $S^{\mathrm{d}}$ and those primarily updated by the opponent or environment to $S^{\mathrm{r}}$.
Otherwise, LEICA constructs the split automatically: an auxiliary one-step transition predictor is trained, and \emph{counterfactual action perturbations} replace one agent's action with another valid action while holding the state and other agents' actions fixed.
A coordinate's \emph{action-sensitivity score} is its normalized predicted next-step change under these perturbations, and two-cluster splitting assigns high- and low-sensitivity coordinates to $S^{\mathrm{d}}$ and $S^{\mathrm{r}}$.
The partition is estimated during the first 10\% of training and then fixed, assuming that the action-responsiveness structure stays sufficiently stable.

\textbf{Latent context inference} ~Given the resulting partition, each agent $i$ infers two  latent contexts from its local trajectory history $\tau_t^i$: a direct context $z_t^i$ for action-conditioned transitions of $S^{\mathrm{d}}$, and a regime context $y_t^i$ for action-free transitions of $S^{\mathrm{r}}$.
Then, the direct and response dynamics for agent $i$ are modeled as
\begin{equation}
\label{eq:41preditors}
\hat S^{\mathrm{d},i}_{k+1|t} = p_{\psi_z^i}(S_k, a_k^i, z_t^i),
\quad
\hat S^{\mathrm{r},i}_{k+1|t} = p_{\psi_y^i}(S_k, y_t^i),~~~~
k \in \mathcal{K}_t
\end{equation}
where $p_{\psi_z^i}$ and $ p_{\psi_y^i}$ are implemented with function approximators, and $\mathcal{K}_t$ denotes the set of steps $\{k\}$ associated with the context time $t$.
The direct predictor $p_{\psi_z^i}$ conditions on the individual action $a_k^i$, while the response predictor $p_{\psi_y^i}$ does not take actions, reflecting the one-step condition in Definition~\ref{def:team_exo}.
During centralized training, both predictors receive the centralized state, which provides the current physical configuration, and a context inferred from agent $i$'s local history, which supplies regime-relevant information; no predictor infers the global transition from local history alone.
The direct predictor is an agent-wise marginal model under the empirical on-policy behavior of other agents rather than a joint-action model, which avoids enumerating joint actions.
The implementation details and training objectives for the context extraction and the dynamics models are provided in Appendix~\ref{app:method}.

\subsection{Influence Scores and Weight Construction}
\label{sec:shaping}

We use the learned predictors $p_{\psi_z^i}$ and $p_{\psi_y^i}$  to construct rewards that favor directly induced changes that influence the later response (adaptation) and are sensitive to team value (coordination).
This follows the delayed influence link in Figure~\ref{fig:edpomdp_cg}.

\paragraph{Mediated influence scores}
At timestep $t+1$, agent $i$'s response predictor outputs
$\hat S^{\mathrm{r},i}_{t+2}=p_{\psi_y^i}(S_{t+1},y_t^i)
=p_{\psi_y^i}(S^{\mathrm{d}}_{t+1},S^{\mathrm{r}}_{t+1},y_t^i)$.
All agent-specific predictors share the same centralized next-state target $S^{\mathrm{r}}_{t+2}$ but may infer different contexts $y_t^i$.
We compute the Jacobian with respect to the directly action-responsive state:
\begin{equation}
\label{eq:jacobian}
    J^i_t
    =
    \frac{\partial \hat S^{\mathrm{r},i}_{t+2}}{\partial S^{\mathrm{d}}_{t+1}}
    =
    \frac{\partial\,p_{\psi_y^i}(S^{\mathrm{d}}_{t+1},S^{\mathrm{r}}_{t+1},y_t^i)}
    {\partial S^{\mathrm{d}}_{t+1}}
    \in\mathbb{R}^{d_{\mathrm{r}}\times d_{\mathrm{d}}}.
\end{equation}
Eq.~(\ref{eq:jacobian}) is a \emph{local mediated-influence estimator} of the path $S^{\mathrm{d}}_{t+1}\!\rightarrow\!S^{\mathrm{r}}_{t+2}$ under context $y_t^i$; recomputed along trajectories, it can also highlight intermediate changes in longer chains when informative intermediate responses exist.
For each coordinate $j$ of $S^{\mathrm{d}}$, we define its influence score by aggregating the agent-wise Jacobian column norms
\begin{equation}
\label{eq:cx}
c^{\mathrm{r}}_j
=
\E_{t\in\mathcal{B},\,i\in\mathcal{N}}
\Big[ \big\| J^i_t[:,j] \big\|_2 \Big],
\quad j=1,\ldots,d_{\mathrm{d}},
\end{equation}
where $\mathcal{B}$ is a training batch and $d_{\mathrm{d}}$ is the dimension of $S^{\mathrm{d}}$.
Eq.~(\ref{eq:cx}) aggregates sensitivities rather than averaging predicted states, so it uses multiple local histories without requiring their contexts to be identical.

\paragraph{Value sensitivity}
Note that influence alone can emphasize coordinates that only affect opponent or environment responses but are not useful for improving task performance.
We therefore compute a \emph{value-sensitivity score}, the absolute centralized-critic gradient, for each coordinate $j$ of $S^{\mathrm{d}}$:
\begin{equation}
\label{eq:cv}
c^v_j
=
\E_{t\in\mathcal{B}}
\Big[
\big|
(\nabla_{S^{\mathrm{d}}_{t+1}} V_\phi(S_{t+1}))_j
\big|
\Big],
\quad j=1,\ldots,d_{\mathrm{d}} .
\end{equation}

\paragraph{Weight construction}
We then standardize $c^{\mathrm{r}}$ in (\ref{eq:cx}) 
and $c^v$ in (\ref{eq:cv}) across coordinates, yielding $\bar c^{\mathrm{r}},\bar c^v\in\mathbb{R}^{d_{\mathrm{d}}}$. We then form a combined score vector
\begin{equation}
\label{eq:combine}
c ~=~ \alpha\,\bar c^{\mathrm{r}} ~+~ (1-\alpha)\,\bar c^v,
\end{equation}
where $\alpha\in[0,1]$ controls the trade-off between adaptation-relevant influence and value-sensitive coordination.
We decay $\alpha$ over training updates, prioritizing mediated influence early when the critic is less reliable and shifting toward value sensitivity later.
Finally, $c$ is converted into a nonnegative coordinate influence weight vector $w\in\mathbb{R}_{\ge 0}^{d_{\mathrm{d}}}$, which carries the essence of the score vector $c$.
For details, please refer to Appendix~\ref{app:exp:weight_impl}.

\subsection{Influence-weighted Intrinsic Reward}
\label{sec:reward}

We now quantify how agent $i$'s  action $a^i_t$ changes the directly action-responsive coordinates relative to the expectation under the current policy $\pi^i$. Using the learned direct predictor in Eq.~(\ref{eq:41preditors}), we  compute
\begin{equation}
\label{eq:cf}
\Delta^{i}_{t}
=
\big| \hat S^{\mathrm{d},i}_{t+1}(a^i_t) - S^{\mathrm{d}}_t \big|
-
\E_{\bar{a}_t^i\sim\pi^i}
\Big[
\big| \hat S^{\mathrm{d},i}_{t+1}(\bar a^i_t) - S^{\mathrm{d}}_t \big|
\Big]
\in \mathbb{R}^{d_{\mathrm{d}}},
\end{equation}
where $|\cdot|$ is element-wise absolute value, and $\hat S^{\mathrm{d},i}_{t+1}(a)=p_{\psi_z^i}(S_t,a,z_t^i)$ is a model prediction, not an unrealized simulator transition.
For each recorded transition, we keep $S_t$ and $z_t^i$ fixed and compare the realized action $a_t^i$ with valid alternatives under the agent's current recurrent policy. In SMAX, the expectation is computed exactly over valid discrete actions, with invalid actions masked. Since $p_{\psi_z^i}$ does not take $A_t^{-i}$ as input, other agents' behavior is marginalized through the on-policy training distribution rather than conditioned on their realized actions.
We then  combine $\Delta_t^i$ and the influence weight vector $w$ in Sec. \ref{sec:shaping}, to construct the  intrinsic reward, given as the weighted direct change:
\begin{equation}
\label{eq:rint}
r^{\mathrm{int},i}_t ~=~ \langle w, \Delta_t^i \rangle.
\end{equation}
Let $r^{\mathrm{ext}}_t=R^{y}(S_t,A_t)$ denote the shared extrinsic team reward from the environment. Our shaped reward is given by
\begin{equation}
\label{eq:finalreward}
\tilde r^{\,i}_t ~=~ r^{\mathrm{ext}}_t + \lambda\, r^{\mathrm{int},i}_t,
\end{equation}
where $\lambda$ controls the intrinsic reward scale.

Actors are trained with the MAPPO clipped surrogate objective \citep{yu2022surprising} using individual shaped rewards $\tilde r^i_t$, while the centralized critic regresses returns from the team-aggregated shaped reward
$\tilde r_t = \frac{1}{N}\sum_{i=1}^N \tilde r^i_t$.
Actor-critic gradients are not backpropagated into the predictors; the context encoders and predictors are updated only by their predictive losses.
Implementation details and the full algorithm are provided in Appendix~\ref{app:method}.
\section{Experiments}
\label{sec:exp}
\noindent
We compare LEICA with strong CTDE, intrinsic-reward, context-inference, and robust MARL baselines, inspect the learned influence weights, and ablate the main components.
SMAX serves as our main evaluation; experiments on continuous physical perturbations and a mixed setting provide complementary evidence on whether the mediated-influence principle extends beyond SMAX-style opponent shifts.

\subsection{Setup}
We evaluate LEICA on SMAX \citep{rutherford2024jaxmarl}, a JAX-based implementation of SMAC \citep{samvelyan2019starcraft}. Its GPU-native implementation makes it straightforward to alter the policies of opponent units, which is essential for studying exogenous variability.

We use five cooperative scenarios: \cm{5m\_vs\_6m}, \cm{10m\_vs\_11m}, \cm{3s5z\_vs\_3s6z}, \cm{6h\_vs\_9z}, and \cm{3s\_vs\_6z}.
The first three are standard super-hard SMAX tasks, while the last two are \emph{extreme-hard} variants obtained by increasing enemy sizes.

\paragraph{Exogenous configurations}
We define a discrete opponent strategy space with five strategic axes, yielding $63$ opponent strategies, and independently vary the initial distance between teams.
This gives $189$ exogenous configurations in total.
We group them into three scales: \cm{small} (train $3$ / test $3$), \cm{medium} (train $10$ / test $5$), and \cm{large} (train $30$ / test $10$).
Training uses only configurations in the training set, while evaluation reports performance separately on seen training configurations and unseen held-out configurations.
Details are provided in Appendix~\ref{app:environment}.

\paragraph{Training and evaluation protocol}
All methods are trained for $10$M environment steps.
The evaluation unit is an independently trained run, scored by its final win rate on the seen or unseen configurations; we report mean $\pm$ SEM over five tasks $\times$ eight seeds. For Table~\ref{tab:main_results}, we also use unpaired stratified permutation tests with Holm correction (Appendix~\ref{app:stats}); other conclusions rest on mean $\pm$ SEM.
Policies are checkpointed every $0.5$M steps and evaluated on multiple episodes per checkpoint. While hyperparameters and compute details are reported in Appendix~\ref{app:implementation}, we note that LEICA’s additional Jacobian and predictor computations are training-time only; at execution, agents use only local histories and the learned context encoder, yielding less than 5\% runtime overhead. 

\subsection{Baselines}
\label{sec:baselines}

We compare against standard CTDE baselines MAPPO~\citep{yu2022surprising}, QMIX~\citep{rashid2018qmix}, and SHAQ~\citep{wang2022shaq}.
We include LAIES~\citep{liu2023lazy}, a partition-informed intrinsic-reward method, and MAPPO+ctx, a VariBAD-style context-inference baseline \citep{zintgraf2019varibad}, to test whether context recognition alone is sufficient.
We also include robust MARL baselines M3DDPG~\citep{li2019robust} and M3PPO, our PPO-based variant, to evaluate whether generic max-min training improves generalization under exogenous variability.
For M3DDPG and M3PPO, we instantiate the uncertainty set through action perturbations and tune the adversarial rate under the same protocol as other hyperparameters.

We report the results of the default variant of \textbf{LEICA}, which employs the automatic partitioning module. We additionally provide the results of \textbf{LEICA (manual)}, which instead uses a domain-informed action-responsiveness partition when such knowledge is available. Both partition procedures are detailed in Appendix~\ref{app:auto-partition} and Appendix~\ref{app:exp:di_partition}.

\subsection{Performance Comparison}
\label{sec:main_results}

\begin{table*}[t]
\centering
\small
\setlength{\tabcolsep}{6pt}
\renewcommand{\arraystretch}{1.2}

\caption{\textbf{Scale-wise SMAX results under opponent strategy variation.}
Win rate (\%, mean $\pm$ SEM) over five tasks and eight training seeds per task and scale, evaluated on training (\cm{Seen}) and held-out (\cm{Unseen}) configurations. Bold indicates the highest mean in each column.
Full per-task results are provided in Appendix~\ref{app:experiments}.}
\label{tab:main_results}
\resizebox{0.99\textwidth}{!}{
\begin{tabular}{l|*{2}{>{\color{graytext}}c} |*{2}{>{\color{graytext}}c}|*{2}{>{\color{graytext}}c}|*{2}{>{\color{graytext}}c}}
\toprule
\multirow{2}{*}{\cm{Method}}
& \multicolumn{2}{c|}{\cm{Small}}
& \multicolumn{2}{c|}{\cm{Medium}}
& \multicolumn{2}{c|}{\cm{Large}}
& \multicolumn{2}{c}{\cm{Overall}} \\
\cmidrule(lr){2-3}
\cmidrule(lr){4-5}
\cmidrule(lr){6-7}
\cmidrule(lr){8-9}
& \cm{Seen} & \cm{Unseen}
& \cm{Seen} & \cm{Unseen}
& \cm{Seen} & \cm{Unseen}
& \cm{Seen} & \cm{Unseen} \\
\midrule
MAPPO        & \me{75.7}{3.8} & \me{66.5}{3.7} & \me{53.7}{2.9} & \me{19.7}{4.4} & \me{53.5}{1.7} & \me{17.2}{2.6} & \me{61.0}{1.9} & \me{34.5}{3.0} \\
MAPPO+ctx    & \me{83.3}{2.2} & \me{75.2}{1.9} & \me{56.4}{3.5} & \me{27.0}{4.9} & \me{53.6}{1.7} & \me{18.2}{2.5} & \me{64.4}{1.9} & \me{40.1}{3.0} \\
QMIX         & \me{63.7}{2.8} & \me{61.5}{2.9} & \me{58.3}{3.9} & \me{16.8}{3.6} & \me{52.2}{3.4} & \me{23.4}{4.9} & \me{58.1}{2.0} & \me{33.9}{2.9} \\
LAIES        & \me{44.7}{4.7} & \me{40.9}{4.7} & \me{59.8}{2.4} & \me{9.8}{1.1} & \me{57.2}{1.5} & \me{14.0}{1.8} & \me{53.9}{1.9} & \me{21.5}{2.1} \\
SHAQ         & \me{2.5}{0.5} & \me{1.8}{0.4} & \me{22.0}{1.0} & \me{15.0}{0.9} & \me{33.3}{1.2} & \me{24.6}{1.1} & \me{19.3}{1.3} & \me{13.8}{1.0} \\
M3DDPG       & \me{7.0}{2.3} & \me{6.7}{2.2} & \me{15.6}{1.4} & \me{0.3}{0.2} & \me{15.1}{1.8} & \me{1.0}{0.3} & \me{12.6}{1.1} & \me{2.7}{0.8} \\
M3PPO        & \me{42.1}{5.4} & \me{34.5}{5.6} & \me{31.7}{2.5} & \me{1.8}{1.0} & \me{46.5}{1.0} & \me{9.5}{0.5} & \me{40.1}{2.1} & \me{15.3}{2.3} \\
\midrule
\textbf{LEICA}   & \best{94.9}{1.1} & \best{89.4}{1.4} & \best{76.1}{3.7} & \me{55.1}{5.7} & \me{63.5}{2.7} & \me{33.9}{4.6} & \me{78.2}{2.0} & \me{59.5}{3.2}\\

\textbf{LEICA (manual)}   & \me{92.2}{1.3} & \me{84.0}{1.9} & \me{76.0}{3.1} & \best{56.5}{4.6} & \best{69.7}{3.0} & \best{45.9}{5.1} & \best{79.3}{1.7} & \best{62.1}{2.8}\\
\bottomrule
\end{tabular}
}
\end{table*}

\begin{figure}[t]
    \centering
    \includegraphics[width=0.99\linewidth]{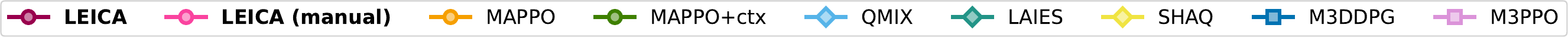}
    \vspace{0.05in}
    
    \raisebox{0.1in}{\includegraphics[width=0.02\linewidth]{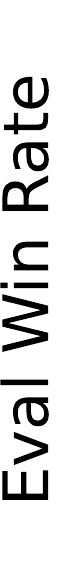}}
    \includegraphics[width=0.27\linewidth]{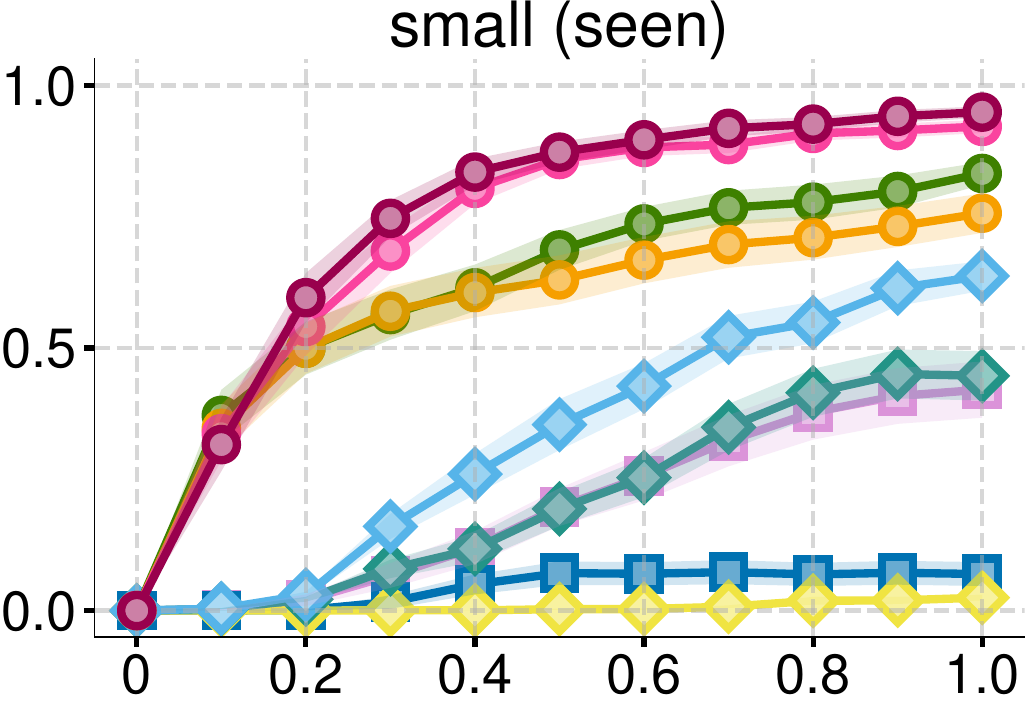}
    \includegraphics[width=0.27\linewidth]{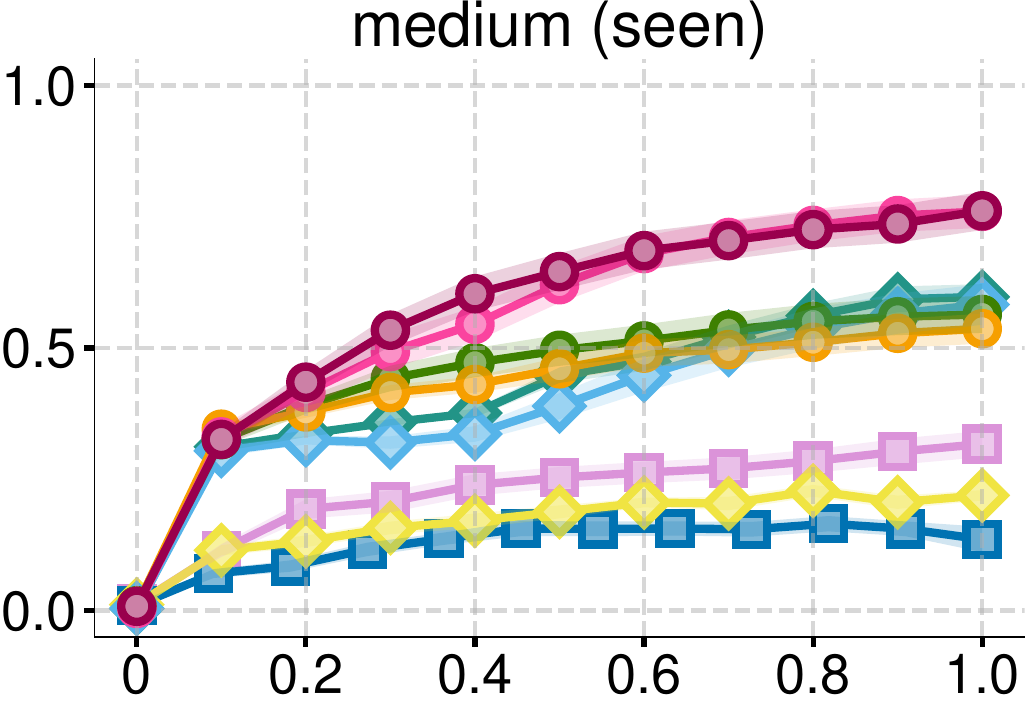}
    \includegraphics[width=0.27\linewidth]{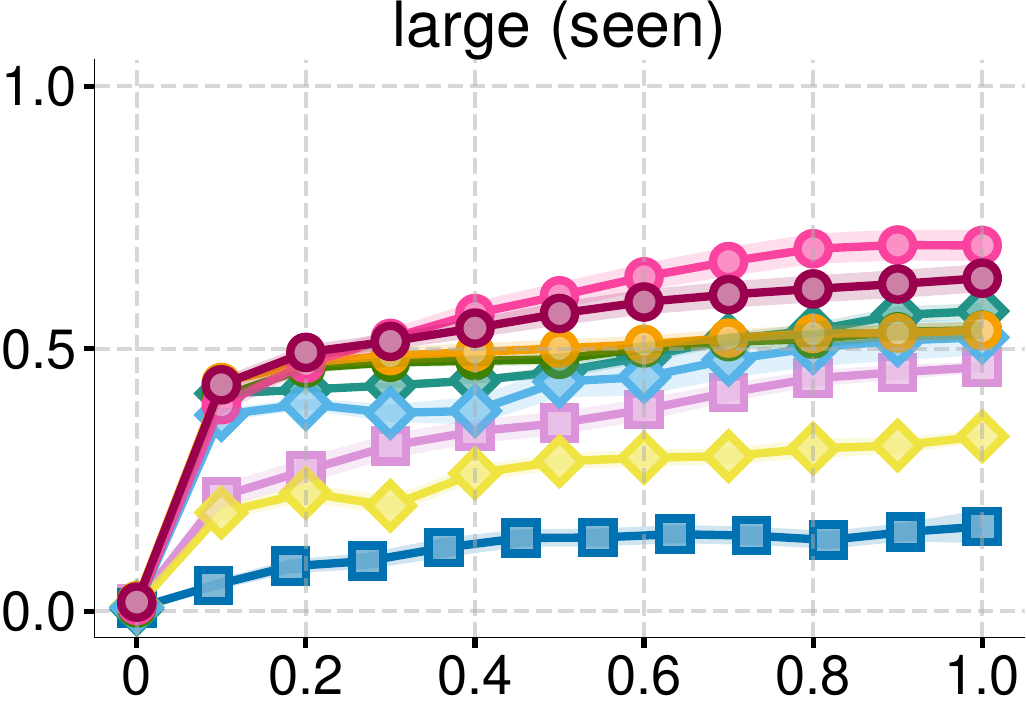}
    \includegraphics[width=0.53\linewidth]{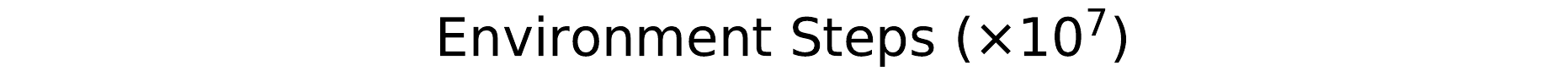}
    \vspace{0.05in}
    
    \raisebox{0.1in}{\includegraphics[width=0.02\linewidth]{pdf/ylabel.pdf}}
    \includegraphics[width=0.27\linewidth]{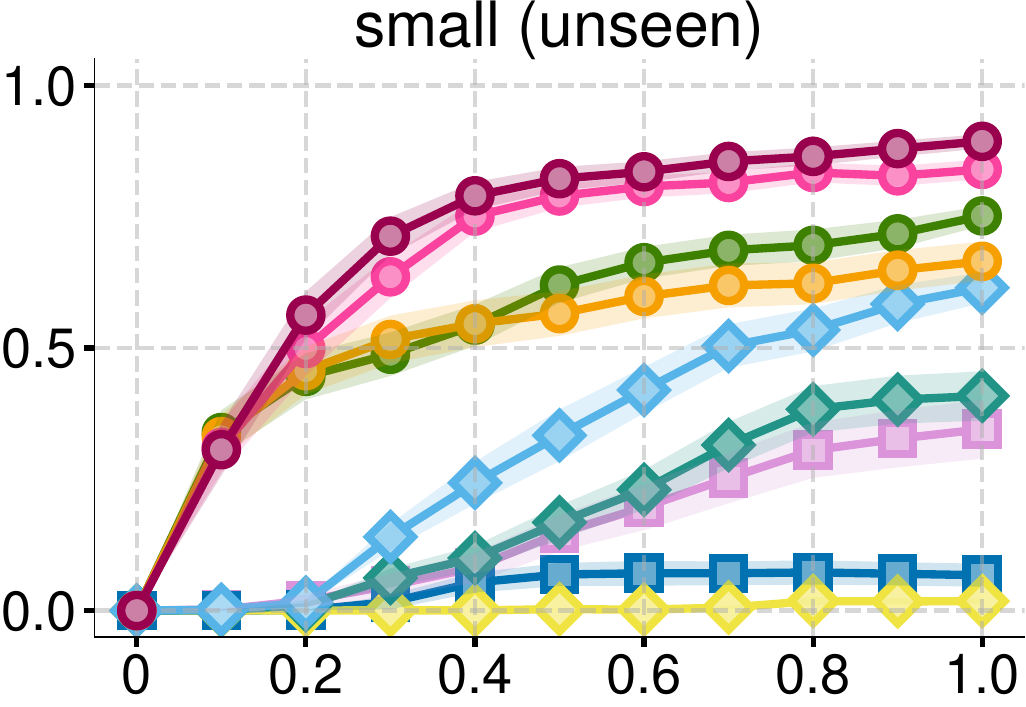}
    \includegraphics[width=0.27\linewidth]{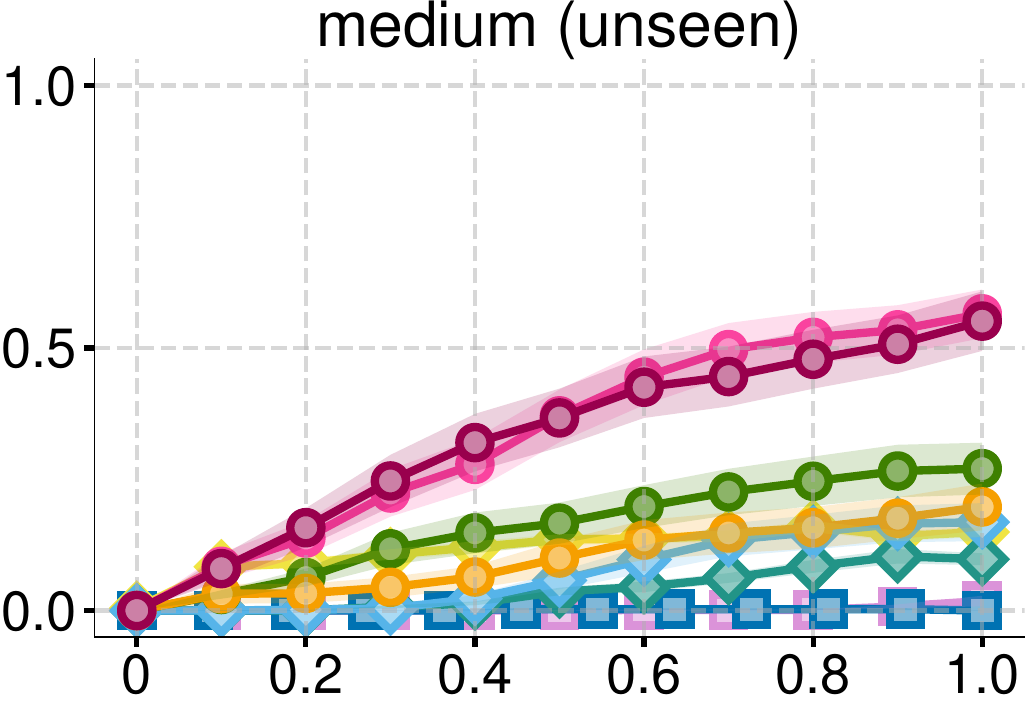}
    \includegraphics[width=0.27\linewidth]{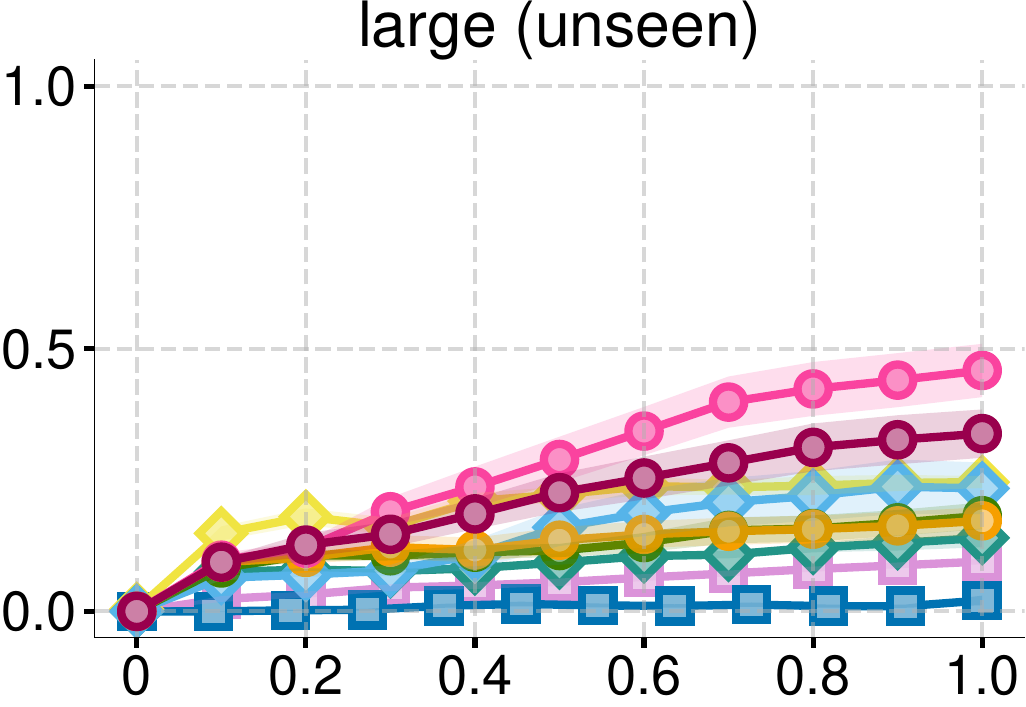}
    \includegraphics[width=0.53\linewidth]{pdf/xlabel_long.pdf}
    
    \caption{\textbf{Aggregated learning curves on SMAX.}
    Win rates are averaged across five tasks for each scale and reported on seen and unseen exogenous configurations.}
    \label{fig:main_results}
\end{figure}

Table~\ref{tab:main_results} summarizes performance across \cm{small}, \cm{medium}, and \cm{large} scales.
LEICA outperforms all non-LEICA baselines in both seen and unseen regimes at every scale, with statistically significant differences in all columns, improving the overall score over MAPPO from 61.0 to 78.2 in seen regimes and from 34.5 to 59.5 in unseen regimes. MAPPO+ctx improves over vanilla MAPPO in some settings, but remains far below LEICA.
This suggests that recognizing the current regime is not enough; the policy also needs a learning signal that identifies which directly induced state changes can affect the later opponent response.
Indeed, robust max-min training is insufficient in this setting, as M3DDPG performs poorly overall and M3PPO remains below MAPPO.

Notably, the default LEICA variant with automatic partitioning performs comparably to LEICA (manual) overall (no significant difference in the \cm{Overall} columns), indicating that LEICA can recover effective action-responsiveness partitions without prior domain knowledge of the state structure. The domain-informed partition remains significantly better at the \cm{large} scale, whereas the automatic one is better on \cm{small} unseen configurations.
Figure~\ref{fig:main_results} complements the final performance table by showing that LEICA improves both training dynamics and unseen-regime performance.

\begin{table}[t]
\centering
\small
\renewcommand{\arraystretch}{1.1}
\begin{minipage}[t]{0.56\linewidth}
\centering
\caption{\textbf{Ablations} on the \cm{medium} scale. Bold: highest mean.}
\label{tab:ablations}
\resizebox{\linewidth}{!}{
\begin{tabular}{llcc}
\toprule
\textbf{Category} & \textbf{Variant} & \textbf{Seen} & \textbf{Unseen} \\
\midrule
\multirow{3}{*}{\shortstack{Partition\\mode}}
& \textbf{LEICA}    & \textbf{76.1}\pmsem{3.7} & 55.1\pmsem{5.7} \\
& \textbf{LEICA (manual)}  & 76.0\pmsem{3.1} & \textbf{56.5}\pmsem{4.6} \\
& full-state reconstruction & 69.3\pmsem{3.5} & 43.6\pmsem{5.3} \\
\midrule
\multirow{3}{*}{\shortstack{Component\\evaluation}}
& w.o.\ intrinsic reward              & 59.4\pmsem{3.5} & 28.1\pmsem{4.8} \\
& w.o.\ counterfactual design         & 29.2\pmsem{3.4} & 17.0\pmsem{3.7} \\
& vanilla MAPPO                       & 53.7\pmsem{2.9} & 19.7\pmsem{4.5} \\
\midrule
\multirow{4}{*}{$\alpha$ scheduling}
& $\alpha {=} 1$ (influence-only)     & 67.6\pmsem{3.2} & 40.2\pmsem{5.1} \\
& $\alpha {=} 0$ (value-only)         & 65.7\pmsem{3.5} & 37.1\pmsem{5.6} \\
& $\alpha {=} 0.5$ (constant)         & 70.4\pmsem{3.3} & 46.6\pmsem{5.5} \\
& linear decaying $\alpha$            & 71.5\pmsem{3.5} & 49.6\pmsem{5.7} \\
\midrule
\multirow{3}{*}{\shortstack{Partition\\robustness}}
& swapping $10\%$                     & 65.4\pmsem{3.5} & 39.1\pmsem{5.6} \\
& swapping $30\%$                     & 67.2\pmsem{3.9} & 37.0\pmsem{5.8} \\
& swapping $50\%$                     & 55.6\pmsem{3.7} & 22.7\pmsem{5.2} \\
\bottomrule
\end{tabular}
}
\end{minipage}\hfill
\begin{minipage}[t]{0.415\linewidth}
\centering
\caption{\textbf{MABrax validation}: returns on seen / unseen regimes.}
\label{tab:mabrax}
\resizebox{\linewidth}{!}{
\begin{tabular}{lcc}
\toprule
\textbf{Environment} & MAPPO & \textbf{LEICA} \\
\midrule
\cm{Walker2d 6x1}     & 677\pmsem{71} / 544\pmsem{31}  & \textbf{987}\pmsem{62} / \textbf{706}\pmsem{18} \\
\cm{Ant 8x1}          & 785\pmsem{83} / 514\pmsem{24}  & \textbf{1084}\pmsem{52} / \textbf{677}\pmsem{26} \\
\cm{HalfCheetah 6x1}  & 1799\pmsem{31} / 665\pmsem{88} & \textbf{2280}\pmsem{131} / \textbf{1704}\pmsem{91} \\
\bottomrule
\end{tabular}
}

\vspace{0.4em}
\includegraphics[width=\linewidth]{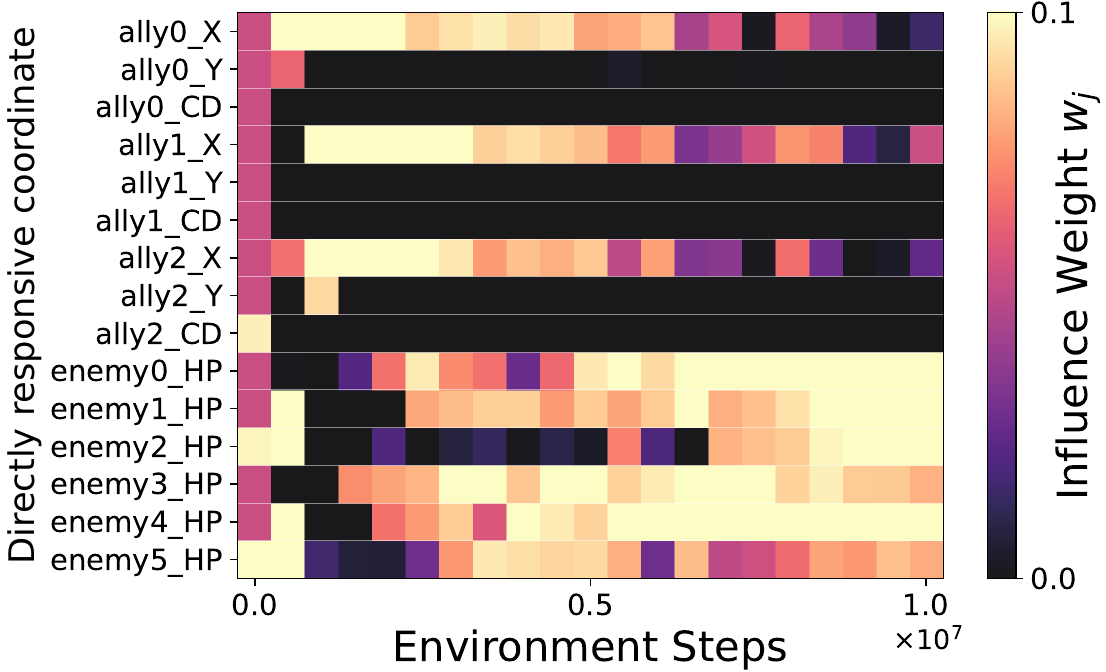}
\captionof{figure}{\textbf{Influence weights $w_j$ over training on \cm{3s\_vs\_6z}.}}
\label{fig:influence-3s6z}
\end{minipage}
\vspace{-0.15in}
\end{table}

\subsection{Learned Influence Weights}
\label{sec:influence_weight}

To illustrate what LEICA emphasizes during training, we visualize the coordinate weights $w$ on \cm{3s\_vs\_6z} in Figure~\ref{fig:influence-3s6z} (X/Y: spatial coordinates, CD: weapon cooldown, HP: health point).

Early in training, the weights are focused on directly action-responsive movement coordinates such as \cm{ally0\_X}, \cm{ally1\_X}, and \cm{ally2\_X}, which are predictive of subsequent opponent responses.
As training progresses, the emphasis shifts towards task-relevant outcome coordinates such as enemy HP, consistent with the decaying schedule on $\alpha$:
LEICA first prioritizes coordinates with high mediated influence and later shifts toward value-sensitive coordinates as the critic stabilizes.

\subsection{Component Evaluation and Ablation}
\label{sec:ablations-components}

We ablate LEICA to isolate which components drive generalization under exogenous variability. All results in Table~\ref{tab:ablations} are reported in the \cm{medium} scale setting, and the base algorithm is LEICA (manual).

\paragraph{Partition mode}
LEICA performs strongly without a hand-specified partition, while LEICA (manual) gives the better unseen generalization when a reliable domain partition is available. The full-state reconstruction variant also improves over MAPPO, indicating that influence-weighted shaping is useful even without an explicit split, but partitioned variants provide cleaner signals.

\paragraph{Component evaluation}
We test whether context inference alone is sufficient and whether counterfactual endogenous change is necessary. Removing intrinsic reward keeps the context modules but removes the shaping signal, yielding a context-only variant. Table 2 shows that this variant remains above vanilla MAPPO but is far below full LEICA. Removing the counterfactual design causes a sharper collapse, indicating that policy-relative endogenous change is crucial for stable shaping.

\paragraph{$\alpha$ scheduling}
We compare variants of the influence coefficient $\alpha$, which balances mediated influence and value sensitivity when forming $w$. Linear decay and a moderate constant trade-off perform well, while using only one signal consistently degrades performance. This supports our design: early training benefits from influence and later training benefits from value sensitive coordinates.

\paragraph{Partition robustness}
Randomly swapping coordinates between $S^{\mathrm{d}}$ and $S^{\mathrm{r}}$ degrades performance gradually, approaching MAPPO at 50\%.
This ablation measures robustness to misassigned coordinates, not whether a partition is causally correct: better alignment with one-step action dependence improves the influence signal, but a perfect partition is not required.

\subsection{Additional Validation Beyond SMAX}
\label{sec:additional_validation}

We evaluate LEICA on MABrax \citep{rutherford2024jaxmarl, brax2021github}, multi-agent continuous-control locomotion tasks where agents jointly control different parts of a robot. The details of regime design and implementation are provided in Appendix~\ref{app:mabrax}. As shown in Table~\ref{tab:mabrax}, LEICA improves over MAPPO on both seen and unseen regimes across all three tasks.
These results are complementary to and less extensive than the SMAX study. 
We also report a mixed-setting validation on MPE-Tag \citep{mordatch2017emergence} in Appendix~\ref{app:mpe-tag}.
While predators and prey are trained by concurrent self-play, cross-play shows that LEICA-trained predators and prey each improve their own returns against MAPPO, which we regard as preliminary evidence beyond fixed opponents~(Appendix).

\section{Conclusion and Limitations}
We addressed cooperative MARL under exogenous variability by introducing ED-POMDP, a Dec-
POMDP formalism that separates an episode-level exogenous regime from an approximate one-step partition of the state. Building on this view, we proposed \textbf{LEICA}, which shapes policy updates with agent-wise counterfactual changes in the directly action-responsive state weighted by their influence on the later response of the opponent or environment. Experiments on SMAX with opponent strategy variation show that LEICA improves performance on unseen regimes and reduces train-test gaps relative to strong baselines, with ablations confirming the contribution of each component.

\paragraph{Limitations}
LEICA relies on identifying action-dependent state structure from training rollouts. Its effectiveness may decrease with noisy transitions, hidden variables, changing dynamics, or limited data. It remains effective under mild and moderate action corruption, but its advantage disappears under severe corruption.
LEICA also does not guarantee disentanglement of entangled state inputs. Since Eq.~\ref{eq:jacobian} measures local influence, it may miss effects that emerge only after several steps without observable intermediate changes.
Our main experiments assume fixed but state-reactive opponents. Learning opponents and equilibrium adaptation remain outside our scope.
Finally, the automatic partition remains fixed after early estimation and may require updating when the environment changes. SMAX is also our only comprehensive benchmark.

\begin{ack}
  This work was supported in part by the Institute of Information \& Communications Technology Planning \& Evaluation (IITP) grant funded by the Korean government (MSIT) (No. RS-2024-00457882, National AI Research Lab Project), by the National Research Foundation of Korea (NRF) grant funded by the Korean government (MSIT) (No. RS-206-25510069), and by the “Advanced GPU Utilization Support Program” funded by the Government of the Republic of Korea (Ministry of Science and ICT) (No. 02-26-01-0110). The authors declare no competing interests.
\end{ack}

\bibliography{references}

\newpage
\appendix
\onecolumn
\section{Broader Impact}
\label{app:broader-impact}

This work is primarily methodological and aims to improve robustness in multi-agent reinforcement learning under changing external conditions. Potential applications include coordinated control, logistics, and resource allocation systems where agents must adapt to regime shifts. As with other reinforcement-learning methods, deployment in real-world settings should be accompanied by task-specific validation and monitoring.

\section{Experiment Details}
\label{app:environment}

We provide a complete specification of our experimental setup to support reproducibility and to clarify the evaluation protocol.
We first describe the platform, scenarios, and the SMAX interfaces in Sec.~\ref{app:exp:env}.
We then describe two partitioning choices: the automatic action-sensitivity partition used by default in LEICA, and the domain-informed SMAX partition used for LEICA (manual), in Secs.~\ref{app:auto-partition}--\ref{app:exp:partition}.
We finally define the scripted opponent configuration space and our train-test split construction in Sec.~\ref{app:exp:strategies}.

\subsection{Environment and Evaluation Protocol}
\label{app:exp:env}

We run all experiments in SMAX using \textsc{JaxMARL}~\citep{rutherford2024jaxmarl}, which provides a fully vectorized JAX implementation of SMAC-style micromanagement tasks.
Table~\ref{tab:smax_scenarios} lists the five scenarios used in the main paper together with their episode horizons.

\begin{table}[h]
    \centering
    \small
    \caption{SMAX scenarios used in the experiments.}
    \label{tab:smax_scenarios}
    \begin{tabular}{lccc}
        \toprule
        \textbf{Name}       & \textbf{Ally Team}    & \textbf{Enemy Team}   & \textbf{Max Steps} \\
        \midrule
        \cm{5m\_vs\_6m}     & 5 Marines             & 6 Marines             & 70                 \\
        \cm{10m\_vs\_11m}   & 10 Marines            & 11 Marines            & 150                \\
        \cm{3s5z\_vs\_3s6z} & 3 Stalkers, 5 Zealots & 3 Stalkers, 6 Zealots & 170                \\
        \cm{6h\_vs\_9z}     & 6 Hydralisks          & 9 Zealots             & 150                \\
        \cm{3s\_vs\_6z}     & 3 Stalkers            & 6 Zealots             & 250                \\
        \bottomrule
    \end{tabular}
\end{table}

Each agent receives a local observation defined by SMAX.
The observation contains:
(i) relative position, (ii) current health, (iii) \cm{last\_movement\_x}, \cm{last\_movement\_y}, (iv) \cm{last\_targeted}, and (v) weapon cooldown, for observable entities within sight range.
We use the default SMAX visibility and featurization provided by \textsc{JaxMARL}.

The discrete action set consists of 4 move actions (\cm{up, down, left, right}), \cm{stop}, and \cm{attack} actions for each attackable enemy unit.
As in SMAC/SMAX, invalid actions are masked.
The environment reward matches the SMAX default:
(i) \emph{damage shaping} proportional to total enemy damage dealt over the episode, and (ii) a \emph{win bonus} upon victory.

\subsection{Automatic Action-Sensitivity Partition}
\label{app:auto-partition}

The default LEICA does not use manually assigned state partition. Instead, it constructs the partition $S=(S^{\mathrm{d}},S^{\mathrm{r}})$ from training rollouts using action sensitivity. The principle is that coordinates whose next-step values change substantially under local counterfactual action perturbations are treated as directly action-responsive, while coordinates with weak action sensitivity are treated as not directly action-responsive at the current step.

\paragraph{Terminology}
A \emph{state coordinate} $s_j$ is one scalar feature of the centralized training state $S_t\in\mathbb{R}^d$.
A \emph{counterfactual action perturbation} replaces one agent's action with another valid action while holding the state and the other agents' actions fixed.
The \emph{action-sensitivity score} $\chi_j$ below is the normalized predicted next-step change of coordinate $j$ under such perturbations.
The \emph{value-sensitivity score} $c^v_j$ in Eq.~(\ref{eq:cv}) is the absolute gradient of the centralized critic with respect to coordinate $j$ of $S^{\mathrm{d}}$.
We say that a partition is \emph{aligned with one-step action dependence} when the coordinates assigned to $S^{\mathrm{d}}$ are those whose next-step values depend on the current team action given the state and regime.

Concretely, we first fit a lightweight one-step transition predictor $\hat f_\omega(S_t,A_t)$ on training rollouts to predict the next global state. For each state coordinate $j$, we compute an action-sensitivity score
\begin{equation}
    \chi_j
    =
    \mathbb{E}_{(S_t,A_t)\sim\mathcal{D}}
    \left[
    \mathbb{E}_{\tilde A_t\sim \mathcal{P}(A_t)}
    \frac{
        \left|
        \hat f_{\omega,j}(S_t,\tilde A_t)
        -
        \hat f_{\omega,j}(S_t,A_t)
        \right|
    }{\hat\sigma_j+\epsilon}
    \right],
\end{equation}
where $\mathcal{P}(A_t)$ denotes valid counterfactual action perturbations and $\hat\sigma_j$ is the empirical standard deviation of coordinate $j$ in the training rollouts.

For discrete SMAX actions, $\tilde A_t$ is formed by replacing one agent's action with another valid action, and averaging over sampled agents and valid alternatives.
In our implementation, we normalize the scores $\{\chi_j\}_{j=1}^d$ and apply a one-dimensional two-cluster split on a logarithmic scale. Coordinates in the higher-sensitivity cluster are assigned to $S^{\mathrm{d}}$, while the remaining coordinates are assigned to $S^{\mathrm{r}}$.
The partition is estimated using rollouts collected during the first 10\% of training and is then kept fixed for the remaining policy optimization. Until the partition is fixed, both predictors are trained on all coordinates and no intrinsic reward is applied.

\paragraph{Stationarity assumption}
This procedure does not assume that the full dynamics are stationary, since the regime and the policies change during training. It does assume that the action-responsiveness structure, i.e., which coordinates respond directly to the current team action, remains sufficiently stable after the estimation window. Settings in which this structure changes substantially during training would require online repartitioning, which is outside the scope of this work.

\subsection{Domain-Informed Partition and SMAX Example}
\label{app:exp:partition}
\label{app:exp:di_partition}
For LEICA (manual), we use a domain-informed partition.

\paragraph{Procedure}
We provide (i) the list of centralized-state coordinates and (ii) domain knowledge about which current team actions can directly affect each coordinate's next-step update. Each coordinate is then assigned by the following rule:
\begin{itemize}[leftmargin=*,itemsep=0pt,topsep=2pt]
    \item if the team action directly updates the coordinate within one decision step, it is assigned to $S^{\mathrm{d}}$;
    \item if the coordinate is primarily updated by the opponent or the environment, it is assigned to $S^{\mathrm{r}}$;
    \item if both mechanisms act on the coordinate, it is assigned according to the dominant one-step mechanism, so that the resulting partition is approximate.
\end{itemize}
The resulting split corresponds to the approximate one-step action-responsiveness condition in Definition~\ref{def:team_exo} with small but nonzero $\varepsilon_{\mathrm{r}},\varepsilon_{\mathrm{d}}$, rather than to an exact factorization of the simulator.

In SMAX, ally moves immediately change ally positions, and ally attacks (when valid) immediately update ally weapon cooldown and reduce enemy health.
Accordingly, we define $S^{\mathrm{d}}$ as the union of:
(i) ally positions $(x^{\text{ally}},y^{\text{ally}})$,
(ii) ally weapon cooldowns, and
(iii) enemy health values.

\paragraph{Not directly action-responsive coordinates $S^{\mathrm{r}}$}
Enemy positions and enemy cooldowns evolve according to the opponent micro-policy, and ally health changes as a consequence of enemy attacks.
Thus, we define $S^{\mathrm{r}}$ as:
(i) enemy positions $(x^{\text{enemy}},y^{\text{enemy}})$,
(ii) enemy weapon cooldowns, and
(iii) ally health values.

This partition captures the dominant one-step update mechanism. Ally actions can influence the response coordinates through later interaction, for example by changing ally positions and thereby changing subsequent opponent behavior. It does not assert exact absence of within-step coupling.
This matches the ED-POMDP interpretation used in the main paper. Because contacts between units and sub-step interactions within a decision step can couple the two groups, this SMAX split is an $\varepsilon$-approximate controllability partition that captures the dominant one-step mechanism rather than an exact simulator factorization.



\subsection{Opponent Strategy Space and Train-Test Regimes}
\label{app:exp:strategies}

\paragraph{Scripted Opponent Strategy Axes.}
To evaluate generalization under exogenous regime shifts, we vary the built-in scripted opponent configuration along the following discrete axes:
\[
    \text{\cm{enemy\_shoots}}\in\text{\cm{\{T,F\}}},
    \quad
    \text{\cm{attack\_mode}}\in\text{\cm{\{\text{closest},\text{weakest},\text{random}\}}},
\]
\[
    \text{\cm{persistency}}\in\text{\cm{\{T,F\}}},
    \quad
    \text{\cm{move\_mode}}\in\text{\cm{\{\text{center},\text{ally},\text{random}\}}},
\]
\[
    \text{\cm{target\_mode}}\in\text{\cm{\{\text{closest},\text{weakest},\text{random}\}}},
    \quad
    \text{\cm{distance}}\in\text{\cm{\{4,6,8\}}}.
\]
When \cm{enemy\_shoots}~=~\cm{F}, attack-specific axes are inactive; the remaining behavior is governed by movement/targeting logic.
Counting unique opponent scripts yields 63 configurations over the non-distance axes; crossing with \cm{distance} produces $63\times 3=189$ total configurations.

\paragraph{Spawn proximity (distance)}
We parameterize the initial ally-enemy separation by $\rho\in\{0.4,0.6,0.8\}$ (abbreviated as \cm{4,6,8}).
With map width $W=32$, the nominal team-center separation is $d^\star(\rho)=W\cdot\rho+2$, where $+2$ enforces a minimum separation after per-unit jitter.
All ally (enemy) units are placed by adding i.i.d.\ noise of $\pm 1$ around their team center, so realized separations concentrate around $d^\star(\rho)$.
A configuration is sampled at episode start and fixed for the entire episode.

\paragraph{Strategy encoding}
We encode each configuration with a 6-character token \cm{[S][A][P][M][T][D]}:
\cm{S} $\in\{$\cm{T}, \cm{F}$\}$ for \cm{enemy\_shoots};
\cm{A} $\in\{$\cm{C}, \cm{W}, \cm{R}$\}$ for \cm{attack\_mode}~=~\{\cm{closest}, \cm{weakest}, \cm{random}\};
\cm{P} $\in\{$\cm{T}, \cm{F}$\}$ for \cm{persistency};
\cm{M} $\in\{$\cm{C}, \cm{A}, \cm{R}$\}$ for \cm{move\_mode}~=~\{\cm{center}, \cm{ally}, \cm{random}\};
\cm{T} $\in\{$\cm{C}, \cm{W}, \cm{R}$\}$ for \cm{target\_mode}~=~\{\cm{closest}, \cm{weakest}, \cm{random}\};
\cm{D} $\in\{4, 6, 8\}$ for \cm{distance}.
When an axis is inactive (e.g., some attack-related axes under \cm{enemy\_shoots}~=~\cm{F}), we use a placeholder symbol (e.g., \cm{X}) for that position in the token.

\paragraph{Train-test splits}
\label{app:exp:splits}
We construct train-test regimes by selecting disjoint sets of strategy tokens.
For each scale, training and test sets contain disjoint configuration tokens. A held-out token can differ from a training token in only one axis, so this protocol tests generalization to unseen configurations within the defined configuration space rather than enforcing a minimum separation beyond token disjointness.

\paragraph{Small (3 train / 3 test)}
Train=\{\cm{TCTCC6}, \cm{TWTCW6}, \cm{TRFAR6}\};
Test=\{\cm{TCTCR6}, \cm{TWFAC6}, \cm{TRTCW6}\}.

\paragraph{Medium (10 train / 5 test)}
Train=\{\cm{TCTCR6}, \cm{TCFAR8}, \cm{TCFRW4}, \cm{TWTRC6}, \cm{TWTRR4}, \cm{TWFAC8}, \cm{TRTCC8}, \cm{TRTCW6}, \cm{TRFAW4}, \cm{FXXAR6}\};
Test=\{\cm{TCTCC4}, \cm{TCTAW6}, \cm{TWTCW8}, \cm{TWTAC4}, \cm{TRFCC4}\}.

\paragraph{Large (30 train / 10 test)}
Train=\{\cm{TCTCW4}, \cm{TCTCR6}, \cm{TCTAC6}, \cm{TCTRC6}, \cm{TCTRR4}, \cm{TCFCW6}, \cm{TCFAC8}, \cm{TCFAR8}, \cm{TCFRW4}, \cm{TWTCR4}, \cm{TWTRC6}, \cm{TWTRR4}, \cm{TWTRR6}, \cm{TWFCC8}, \cm{TWFAC8}, \cm{TWFAR6}, \cm{TWFRC8}, \cm{TRTCC8}, \cm{TRTCW6}, \cm{TRTAW8}, \cm{TRTRW6}, \cm{TRFCW8}, \cm{TRFAW4}, \cm{TRFRC8}, \cm{TRFRR6}, \cm{FXXAW4}, \cm{FXXCR4}, \cm{FXXCC8}, \cm{FXXRW8}, \cm{FXXAR6}\};
Test=\{\cm{TCTCC4}, \cm{TCTAW4}, \cm{TCTAW6}, \cm{TWTCW4}, \cm{TWTCW8}, \cm{TWTAC4}, \cm{TWTAC6}, \cm{TWTRW6}, \cm{TRTCR4}, \cm{TRFCC4}\}.

\paragraph{Standard SMAX (single-opponent)}
For completeness, we also report results on the default SMAX setting without opponent variation (Appendix~\ref{app:experiments}), using the built-in heuristic configuration \cm{TCTCR6}.

\subsection{Relation to Other Benchmarks}
\label{app:exp:related_benchmarks}

Our evaluation protocol is complementary to recent SMAC-derived benchmarks that were introduced to stress-test generalization beyond the default single-opponent setting.

SMAC-HARD~\cite{deng2024smac} is closely aligned with our motivation. It explicitly targets opponent diversity by enabling customizable opponent strategy scripts, randomization over adversarial behaviors, and an explicit black-box testing protocol.
Our SMAX-based protocol shares the same motivation, but differs in emphasis. We construct multi-axis exogenous regime shifts by composing multiple explicit opponent-behavior knobs, rather than focusing on edited scripts and self-play interfaces.

SMACv2~\cite{ellis2023smacv2} addresses a different but related generalization gap by procedurally generating scenarios and requiring agents to generalize to previously unseen settings sampled from the same distribution at evaluation time, with additional variants designed to enforce meaningful partial observability.
In contrast, our protocol keeps the underlying scenario fixed and focuses on exogenous regime shifts driven by opponent micro-policy variations, isolating robustness to opponent/environment changes from robustness to scenario generation.
SMACv2 primarily tests distributional variation in initial conditions and scenario structure, while our benchmark tests behavioral variation of the opponent script under a fixed task.

\section{Method Details}
\label{app:method}

\subsection{Theoretical Results}

\paragraph{Proof of Lemma~\ref{thm:factorization}}
Throughout, let $\varepsilon_{\mathrm{r}}=\varepsilon_{\mathrm{d}}=0$ and fix $t$. We use the fact that, for random variables $U,V,W$, $I(U;V\mid W)=0$ holds if and only if $U\perp V\mid W$, i.e., $P(U\mid V,W)=P(U\mid W)$ for every $(V,W)$ in the support.

\emph{Part (a): marginal invariance.}
The condition $\varepsilon_{\mathrm{r}}=0$ in Definition~\ref{def:team_exo} gives $I(S^{\mathrm{r}}_{t+1};A_t\mid S_t,Y_t^y)=0$, i.e., $S^{\mathrm{r}}_{t+1}\perp A_t\mid(S_t,Y_t^y)$. Hence
\[
    P(S^{\mathrm{r}}_{t+1}\mid S_t,A_t,Y_t^y)
    =
    P(S^{\mathrm{r}}_{t+1}\mid S_t,Y_t^y)
    \equiv P_{\mathrm{r}}(S^{\mathrm{r}}_{t+1}\mid S_t,Y_t^y).
\]
Similarly, $\varepsilon_{\mathrm{d}}=0$ gives $I(S^{\mathrm{d}}_{t+1};Y_t^y\mid S_t,A_t)=0$, i.e., $S^{\mathrm{d}}_{t+1}\perp Y_t^y\mid(S_t,A_t)$, and therefore
\[
    P(S^{\mathrm{d}}_{t+1}\mid S_t,A_t,Y_t^y)
    =
    P(S^{\mathrm{d}}_{t+1}\mid S_t,A_t)
    \equiv P_{\mathrm{d}}(S^{\mathrm{d}}_{t+1}\mid S_t,A_t).
\]
These two identities concern only the one-step marginals of $S^{\mathrm{d}}_{t+1}$ and $S^{\mathrm{r}}_{t+1}$; they do not constrain the dependence between the two components.

\emph{Part (b): factorization.}
Since $S_{t+1}=(S^{\mathrm{d}}_{t+1},S^{\mathrm{r}}_{t+1})$, the chain rule gives
\[
    P(S_{t+1}\mid S_t,A_t,Y_t^y)
    =
    P(S^{\mathrm{d}}_{t+1}\mid S_t,A_t,Y_t^y)\,
    P(S^{\mathrm{r}}_{t+1}\mid S^{\mathrm{d}}_{t+1},S_t,A_t,Y_t^y).
\]
Without further assumptions, the second factor may depend on $S^{\mathrm{d}}_{t+1}$ and, through it, on $A_t$; Definition~\ref{def:team_exo} alone therefore does not imply a product form.
Under the additional assumption $S^{\mathrm{d}}_{t+1}\perp S^{\mathrm{r}}_{t+1}\mid(S_t,A_t,Y_t^y)$,
\[
    P(S^{\mathrm{r}}_{t+1}\mid S^{\mathrm{d}}_{t+1},S_t,A_t,Y_t^y)
    =
    P(S^{\mathrm{r}}_{t+1}\mid S_t,A_t,Y_t^y).
\]
Substituting this identity and the two marginal identities from part (a) into the chain-rule expansion yields
\[
    P(S_{t+1}\mid S_t,A_t,Y_t^y)
    =
    P_{\mathrm{d}}(S^{\mathrm{d}}_{t+1}\mid S_t,A_t)\,
    P_{\mathrm{r}}(S^{\mathrm{r}}_{t+1}\mid S_t,Y_t^y).
\]
\qed

The exact factorization in part (b) should be understood as an idealized case that requires the additional conditional-independence assumption. In practical environments, the partition is approximate, and weak coupling between the two components may remain. LEICA neither enforces nor requires this factorization. The predictors and influence signal are learned directly from rollouts and rely only on approximate one-step action responsiveness.


\subsection{Context Inference and Predictor Training}
\label{app:method:inference}
For each agent $i$ and time $t$, we infer two stochastic contexts with reparameterization,
\[
    q_{\phi_z^i}\!\big(z_t^i \mid \tau_{\le t}^i\big)=\mathcal{N}\!\big(\mu_{z,t}^i,\operatorname{diag}(\sigma_{z,t}^{i\,2})\big),\qquad
    q_{\phi_y^i}\!\big(y_t^i \mid \tau_{\le t}^i\big)=\mathcal{N}\!\big(\mu_{y,t}^i,\operatorname{diag}(\sigma_{y,t}^{i\,2})\big),
\]
where the history encoder for contexts is a shared GRU \citep{chung2014empirical} over the agent’s trajectory $\tau_{\le t}^i$, and $z_t^i,y_t^i$ are sampled via the standard reparameterization trick. We use filtered priors in the style of variBAD~\citep{zintgraf2019varibad}:
\begin{equation}
    \label{eq:priors}
    p(z_t^i)=\begin{cases}
        \mathcal{N}(0,I),                              & t=0, \\
        q_{\phi_z^i}(z_{t-1}^i \mid \tau_{\le t-1}^i), & t>0,
    \end{cases}
    \qquad
    p(y_t^i)=\begin{cases}
        \mathcal{N}(0,I),                              & t=0, \\
        q_{\phi_y^i}(y_{t-1}^i \mid \tau_{\le t-1}^i), & t>0.
    \end{cases}
\end{equation}

GRU is not shared with the actor or critic encoders. The two contexts share the same GRU and branch into separate heads for each $\mu$ and $\sigma$.
With $(z_t^i,y_t^i)$, we train two predictors on multiple indices $k$ sampled from the same episode:
\[
    \hat S^{\mathrm{d},i}_{k+1\mid t} = p_{\psi_z^i}\!\big(S_k, a_k^i, z_t^i\big),\qquad
    \hat S^{\mathrm{r},i}_{k+1\mid t} = p_{\psi_y^i}\!\big(S_k, y_t^i\big),\qquad k\in\mathcal{K}_t\subseteq\{0,\dots T\}.
\]
Here $S_k=(S_k^{\mathrm{d}},S_k^{\mathrm{r}})$ includes both components for inputs to both predictors. Each agent-specific predictor receives the centralized state together with the context inferred from agent $i$'s local history. To respect the condition in Definition~\ref{def:team_exo}, the response predictor does not take actions as input. The direct predictor conditions on the individual action $a_k^i$. 

We sample a fixed number $|\mathcal{K}_t|=K$ of valid indices. For a rollout window of length $T$, predictor evaluation therefore costs $\mathcal{O}(KT)$ per batch element. The current same-episode index sampler additionally constructs $\mathcal{O}(T^2)$ time-pair masks, so the linear claim applies to predictor evaluation rather than the entire sampling pipeline. Decoder training uses dropout with $p=0.5$. We use $K\in\{4,8\}$ in the reported configurations.

\paragraph{Predictive Objective and KL Regularization}
For each anchor $(i,t)$ we optimize two separate ELBOs. One for the direct context $z_t^i$ and one for the regime context $y_t^i$.

Direct (for $z_t^i$ and $p_{\psi_z}$):
\[
    \mathcal{L}^{\,i,t}_{\mathrm{ELBO},z}
    =\;\mathbb{E}_{q_{\phi_z^i}(z_t^i)}
    \!\left[
        \sum_{k\in\mathcal{K}_t}
        \log p_{\psi_z}\!\big(S^{\mathrm{d}}_{k+1}\mid S_k,a_k^i,z_t^i\big)
        \right]
    -\beta_z\,\mathrm{KL}\!\big(q_{\phi_z^i}(z_t^i)\,\|\,p(z_t^i)\big),
\]
Response (for $y_t^i$ and $p_{\psi_y}$):
\[
    \mathcal{L}^{\,i,t}_{\mathrm{ELBO},y}
    =\;\mathbb{E}_{q_{\phi_y^i}(y_t^i)}
    \!\left[
        \sum_{k\in\mathcal{K}_t}
        \log p_{\psi_y}\!\big(S^{\mathrm{r}}_{k+1}\mid S_k,y_t^i\big)
        \right]
    -\beta_y\,\mathrm{KL}\!\big(q_{\phi_y^i}(y_t^i)\,\|\,p(y_t^i)\big),
\]
with filtered priors from Eq.~\ref{eq:priors}.

In practice, we replace the Gaussian log-likelihood terms with reconstruction MSE losses scaled by $\lambda_{\mathrm{recon}}$, and we use a single reparameterized sample from each posterior to estimate the expectations. The total objective maximized is
\[
    \mathcal{L}_{\mathrm{ELBO}}
    =\lambda_{\mathrm{ELBO}} \sum_{t=0}^{T} \sum_{i=1}^{N} (\mathcal{L}^{\,i,t}_{\mathrm{ELBO},z}+\mathcal{L}^{\,i,t}_{\mathrm{ELBO},y}).
\]

Context encoders and predictors are updated only by $\mathcal{L}_{\mathrm{ELBO}}$ above. When intrinsic rewards are formed, gradients are stopped through the influence weights $w$ and through the counterfactual direct change $\Delta_t^i$. Thus, RL losses do not update the representation modules through the reward path.

\subsection{Details of Weight Construction}
\label{app:exp:weight_impl}

This subsection complements Sec.~\ref{sec:shaping} by specifying the practical details.

Let $\mathcal{B}$ denote the set of decision-time indices $(t{+}1)$ in a PPO minibatch.
From each transition in $\mathcal{B}$, we compute (i) influence scores $c^{\mathrm{r}}_t\in\mathbb{R}^{d_{\mathrm{d}}}$ from Eq.~\ref{eq:cx}, and (ii) value-sensitivity scores $c^v_t\in\mathbb{R}^{d_{\mathrm{d}}}$ from Eq.~\ref{eq:cv}.
We aggregate them across the minibatch by averaging:
\begin{equation}
    c^{\mathrm{r}} ~=~ \frac{1}{|\mathcal{B}|}\sum_{(t)\in\mathcal{B}} c^{\mathrm{r}}_t,
    \qquad
    c^v ~=~ \frac{1}{|\mathcal{B}|}\sum_{(t)\in\mathcal{B}} c^v_t.
\end{equation}
We standardize each vector across coordinates:
\begin{equation}
    \bar c^{\mathrm{r}} ~=~ \frac{c^{\mathrm{r}} - \mu(c^{\mathrm{r}})\mathbf{1}}{\sigma(c^{\mathrm{r}})+\epsilon},
    \qquad
    \bar c^v ~=~ \frac{c^v - \mu(c^v)\mathbf{1}}{\sigma(c^v)+\epsilon},
\end{equation}
where $\mu(\cdot)$ and $\sigma(\cdot)$ denote the mean and standard deviation over $j\in\{1,\ldots,d_{\mathrm{d}}\}$, $\mathbf{1}\in\mathbb{R}^{d_{\mathrm{d}}}$ is the ones vector, and $\epsilon$ is a small constant for numerical stability.

Given the combined score vector $c$ from Eq.~\ref{eq:combine}, we map it to a positive weight vector via a temperature-scaled softmax:
\begin{equation}
    \tilde w ~=~ \softmax\!\left(\frac{c}{\tau}\right),
    \qquad \tilde w\in\mathbb{R}^{d_{\mathrm{d}}}_{>0},~ \sum_{j=1}^{d_{\mathrm{d}}}\tilde w_j = 1,
\end{equation}
where $\tau>0$ controls the sharpness of the distribution.

To reduce noise from Jacobians and value gradients, we apply an exponential moving average (EMA) to the weights:
\begin{equation}
    w \leftarrow (1-\beta)\,w + \beta\,\tilde w,
    \qquad \beta\in(0,1].
\end{equation}
Here $w$ is maintained across policy updates and initialized as the uniform vector $w_j = 1/d_{\mathrm{d}}$ at the start of training.
This smoothing yields a temporally more consistent weighting signal and prevents abrupt changes of the intrinsic reward when the critic or predictors are still adapting.

\paragraph{Schedule of $\alpha$}
We implement the exponentially decaying trade-off $\alpha$ using the update index $x$ (number of PPO updates):
\begin{equation}
    \alpha(x) ~=~ \exp(-kx),
\end{equation}
with $k>0$ chosen so that $\alpha$ remains large in early training and decreases gradually thereafter. We use $k=10/X$, where $X$ is the total number of PPO updates.

\subsection{Details regarding the influence-based reward}
\label{app:influence}

\paragraph{Counterfactual Baseline}
Counterfactual baselines are widely used to reduce variance and to provide local credit assignment in policy gradient methods. In the single-agent case, adding a state-dependent baseline $b(S_t)$ leaves the gradient unbiased since
\[
    \mathbb{E}_{\pi}\!\left[\nabla_{\theta}\log\pi_{\theta}(A_t\!\mid\!O_t)\, b(S_t)\right]
    = \mathbb{E}_{\pi}\!\left[b(S_t)\,\nabla_{\theta}\!\sum_{a}\pi_{\theta}(a\!\mid\!O_t)\right]
    = 0,
\]
so one may replace returns with advantages $G_t - b(S_t)$ without bias \citep{williams1992simple}.

In cooperative multi-agent settings under CTDE, this counterfactual idea is used to isolate the marginal effect of a single agent while marginalizing over its own action. Practically, this choice provides variance reduction and agent-level credit signals, and has been widely adopted in related multi-agent policy gradient methods \citep{foerster2018counterfactual,shao2023counterfactual,kim2026gpae}. The canonical example is COMA \citep{foerster2018counterfactual}, which defines the agent-wise counterfactual advantage
\[
    A^i_{\mathrm{COMA}}(S_t,A_t)
    = Q(S_t,A_t)
    - \sum_{a'^i}\pi^i(a'^i\!\mid\!o_t^i)\,Q\big(S_t,(a'^i,A_t^{-i})\big),
\]
holding other agents fixed at $A_t^{-i}$.

In LEICA, the same principle is applied not to $Q$ values but to predicted transitions of the directly action-responsive state, inspired by recent study \citep{park2025center}. The baseline for agent $i$ is defined as a policy expectation at time $t$,
\[
    b^{i}_t
    =\mathbb{E}_{\bar a_t^i\sim\pi^i(\cdot\mid\tau_{\le t}^i)}
    \Big[\big|p_{\psi_z^i}(S_t,\bar a_t^i,z_t^i)-S^{\mathrm{d}}_t\big|\Big],
\]
the expected magnitude of the direct change under the agent's own policy, computed exactly over the valid actions with the policy renormalized over them. Contrasting this baseline with the realized action gives $\Delta_t^i$ in Eq.~\ref{eq:cf}. The comparison keeps the recorded state and agent-specific context fixed. Unlike COMA, the predictor does not condition on the other agents' realized actions and the teammate behavior is marginalized through the on-policy training distribution. This substitution preserves the spirit of counterfactual baselines, marginalizing the agent’s own action under its current policy, while tailoring the signal to LEICA’s objective of weighting direct changes by their mediated influence on the later response and by value sensitivity.

We summarize the full training pipeline of LEICA with Algorithm~\ref{alg:leica}.

\begin{algorithm}[h]
\caption{\textbf{LEICA}}
\label{alg:leica}
\begin{algorithmic}[1]
    \STATE Initialize parameters $\Theta$, partition predictor
    $\hat f_\omega$, and normalization statistics.
    \STATE $w\leftarrow\mathbf{0}$ (auto) or
    $\mathbf{1}/d_{\mathrm{d}}$ (manual).

    \FOR{$x=0,\ldots,X-1$}
        \STATE \textbf{(A) Rollout and targets}
        \STATE $\pi_{\mathrm{old}}\leftarrow\pi_\theta$;
        $\mathcal{D}\leftarrow\text{Rollout}(\pi_{\mathrm{old}})$.
        \STATE $\Delta_t^i\leftarrow
        \text{CounterfactualChange}(S_t,a_t^i,z_t^i;
        \pi_{\mathrm{old}})$ \quad (Eq.~\ref{eq:cf})
        \STATE $r_t^{\mathrm{int},i}\leftarrow
        \langle w,\Delta_t^i\rangle$
        \STATE $\tilde r_t^i\leftarrow
        r_t^{\mathrm{ext}}+\lambda r_t^{\mathrm{int},i}$;
        $\tilde r_t\leftarrow N^{-1}\sum_i\tilde r_t^i$.
        \STATE $\hat A^i\leftarrow
        \text{GAE}(\tilde r^i,V_\phi)$;
        $\hat R\leftarrow V_\phi+
        \text{GAE}(\tilde r,V_\phi)$.

        \STATE \textbf{(B) Joint optimization}
        \FOR{each PPO epoch}
            \STATE $z_t^i\sim q_{\phi_z^i}$,
            $y_t^i\sim q_{\phi_y^i}$;
            $\mathcal{K}_t\leftarrow
            \text{SampleSameEpisode}(\mathcal{D})$ ~~ (Appendix~\ref{app:method:inference}).
            \STATE $\mathcal{L}_{\mathrm{total}}\leftarrow
            \mathcal{L}_{\mathrm{actor}}
            +\mathcal{L}_{\mathrm{critic}}
            +\mathcal{L}^{i,t}_{\mathrm{ELBO},z}+\mathcal{L}^{i,t}_{\mathrm{ELBO},y}$.
            \STATE Update $\Theta$ with $\mathcal{L}_{\mathrm{total}}$.
        \ENDFOR

        \STATE \textbf{(C) Weights for the next rollout}
        \IF{automatic partition is unresolved and warm-up has ended}
            \STATE $\chi\leftarrow
            \text{ActionSensitivity}(\hat f_\omega,\mathcal{D})$.
            \STATE $(S^{\mathrm{d}},S^{\mathrm{r}})
            \leftarrow\text{TwoMeans}
            \bigl(\log(\chi/(\max_j\chi_j+10^{-8})+0.01)\bigr)$.
            \STATE Freeze the partition;
            $w\leftarrow\mathbf{1}/d_{\mathrm{d}}$.
        \ENDIF
        \IF{the partition is available}
            \STATE Compute $c^{\mathrm{r}},c^v$
            by Eqs.~\ref{eq:cx} and~\ref{eq:cv}.
            \STATE $\bar c^{\mathrm{r}}\leftarrow
            \text{Std}(c^{\mathrm{r}})$;
            $\bar c^v\leftarrow\text{Std}(c^v)$.
            \STATE $\alpha\leftarrow e^{-kx}$;
            $c\leftarrow\alpha\bar c^{\mathrm{r}}
            +(1-\alpha)\bar c^v$.
            \STATE $w\leftarrow\text{sg}
            \bigl[(1-\beta)w+
            \beta\,\softmax(c/\tau)\bigr]$.
        \ENDIF
    \ENDFOR
\end{algorithmic}
\end{algorithm}

\section{Implementation Details}
\label{app:implementation}
All experiments use JAX/Flax with JaxMARL and vectorized environments, collecting fixed-length rollouts from multiple parallel environments and training on minibatches on GPU \citep{rutherford2024jaxmarl}. We apply parameter sharing for homogeneous agents and action masking at the policy head. For MAPPO \citep{yu2022surprising}, QMIX \citep{rashid2018qmix}, and SHAQ \citep{wang2022shaq}, we follow the JaxMARL implementation. For LAIES \citep{liu2023lazy}, as there was no code implemented in a version suitable for SMAX, we re-implement on top of QMIX by adding the original intrinsic influence term from the authors’ code and paper,\footnote{\url{https://github.com/liuboyin/LAIES}} keeping the QMIX mixing and target update structure \citep{liu2023lazy}. We implement M3DDPG \citep{li2019robust} by following EPyMARL \citep{papoudakis2020benchmarking} and the official M3DDPG codebase \footnote{\url{https://github.com/dadadidodi/m3ddpg}}, reusing only the buffer and sampling structure of QMIX implementation while re-implementing the algorithmic components. For M3PPO, we modify MAPPO to learn a state-action value instead of a state value, and apply the same max-min update scheme used in M3DDPG.

For M3DDPG and M3PPO, the adversarial component is trained only on the training opponent-regime set and never observes held-out test regimes. The adversarial rate controls the strength of the max-min update and is selected from the search space in Tables~\ref{tab:mappo-space} and~\ref{tab:qmix-space}. In M3PPO, we use the same PPO rollout data as MAPPO but replace the state-value critic with a state-action critic so that the inner minimization can evaluate adversarial action perturbations. This makes M3PPO the closest PPO-style analogue of M3DDPG under our CTDE implementation.

\subsection{Training and Evaluation Protocol}
\label{app:exp:protocol}

At evaluation time, each episode samples a configuration uniformly from the union of the training and test sets and holds it fixed for the full episode. Win rates are reported separately for the seen and unseen configurations.
Each evaluation runs 256 parallel environments for 256 steps with greedy actions, and all reported results are averaged over 8 random seeds.
Our primary metric is average win rate.

\paragraph{Computational cost}
Experiments were run on systems equipped with NVIDIA RTX 3090 (24GB) or RTX 4090 (24GB) GPUs. We report both parameter overhead and runtime overhead relative to MAPPO in Table~\ref{tab:computational-cost}. LEICA introduces additional train-time computation from context inference, transition predictors, and influence-weight updates. These components are used for training-time shaping only. At execution time, the policy does not use the predictors or Jacobian scores.

\begin{table}[h]
    \centering
    \small
    \caption{Computational overhead relative to MAPPO. Auxiliary predictors and Jacobian computations are training-time components.}
    \label{tab:computational-cost}
    \begin{tabular}{lccc}
        \toprule
        Method & Train-time parameters & Execution-time parameters & Runtime overhead                             \\
        \midrule
        MAPPO  & $1.00\times$          & $1.00\times$              & $1.00\times$                                 \\
        LEICA  & $1.45\times$          & $1.06\times$              & $1.13\times$ train / $<1.05\times$ execution \\
        \bottomrule
    \end{tabular}
\end{table}

\subsection{Hyperparameters}
\label{app:hyperparameter}
Hyperparameters were tuned with Bayesian optimization over a fixed budget~\citep{snoek2012practical}, and search spaces are detailed below.

\subsubsection{Default parameters}

\begin{table}[h]
    \centering
    \small
    \setlength{\tabcolsep}{8pt}
    \caption{Default hyperparameters for MAPPO, MAPPO+ctx, M3PPO, and LEICA.}
    \label{tab:mappo-defaults}
    \begin{tabular}{ll}
        \toprule
        \textbf{Hyperparameter}  & \textbf{Value}                  \\
        \midrule
        Discount factor $\gamma$ & 0.99                            \\
        GAE parameter $\lambda$  & 0.95                            \\
        Value loss coefficient   & 0.5                             \\
        Number of minibatches    & 1                               \\
        Critic Network           & [128, 128]                      \\
        Actor Network            & [128, 128]                      \\
        Recurrent Network        & GRU                             \\
        Recurrent Hidden         & 128                             \\
        Activation               & tanh                            \\
        Learning-rate schedule   & linear decay                    \\
        Value-loss clipping      & PPO clip $\epsilon$             \\
        \textbf{LEICA specific}  &                                 \\
        Predictor network        & [64]                            \\
        $\alpha$                 & exponentially decaying $\alpha$ \\
        Reward weight $\lambda$  & 0.03                            \\
        EMA ratio $\beta$        & 0.2                             \\
        Softmax $\tau$           & 0.5                             \\
        \bottomrule
    \end{tabular}

\end{table}

\begin{table}[h]
    \centering
    \small
    \setlength{\tabcolsep}{8pt}
    \caption{Default hyperparameters for QMIX, LAIES, SHAQ, and M3DDPG.}
    \begin{tabular}{ll}
        \toprule
        \textbf{Hyperparameter}             & \textbf{Value} \\
        \midrule
        Discount factor $\gamma$            & 0.99           \\
        Buffer batch size                   & 64             \\
        Network size                        & [128, 128]     \\
        Mixer embedding size                & 32             \\
        Mixer hypernet size                 & 64             \\
        Recurrent Network                   & GRU            \\
        Recurrent Hidden                    & 128            \\
        Activation                          & ReLU           \\
        Maximum gradient norm               & 10             \\
        Minimum exploration coef $\epsilon$ & 0.05           \\
        Epochs                              & 4              \\
        \bottomrule
    \end{tabular}

    \label{tab:qmix-defaults}
\end{table}

\newpage
\subsubsection{Hyperparameter search spaces}

\begin{table}[h]
    \centering
    \small
    \setlength{\tabcolsep}{8pt}
    \caption{Hyperparameter search spaces for MAPPO, MAPPO+ctx, M3PPO, and LEICA.}
    \label{tab:mappo-space}
    \begin{tabular}{ll}
        \toprule
        \textbf{Hyperparameter}                 & \textbf{Value}             \\
        \midrule
        Entropy Coefficient                     & \{0, 0.01\}                \\
        Learning Rate                           & \{2e-4, 5e-4, 1e-3, 2e-3\} \\
        Maximum gradient norm                   & \{1, 5, 10\}               \\
        PPO clipping $\epsilon$                 & \{0.05, 0.1, 0.2\}         \\
        PPO epochs                              & \{4 ,8, 16\}               \\
        \textbf{M3PPO specific}                 &                            \\
        Adversarial rate                        & \{1e-1, 1e-2, 1e-3\}       \\
        \textbf{MAPPO+ctx specific}             &                            \\
        latent size                             & \{8, 16, 32\}              \\
        Reconstruction weight $\lambda_{recon}$ & \{1e2, 1e3, 1e4\}          \\
        KL regularization coef                  & \{0.5, 1\}                 \\
        \textbf{LEICA specific}                 &                            \\
        latent $z, y$ size                      & \{8, 16, 32\}              \\
        Sampling number $K$                     & \{4, 8, 16\}               \\
        Reconstruction weight $\lambda_{recon}$ & \{1e2, 1e3, 1e4\}          \\
        KL regularization $\beta_z, \beta_y$    & \{0.5, 1\}                 \\
        \bottomrule
    \end{tabular}
\end{table}

\begin{table}[h]
    \centering
    \small
    \setlength{\tabcolsep}{8pt}
    \caption{Hyperparameter search spaces for QMIX, LAIES, SHAQ, and M3DDPG.}
    \label{tab:qmix-space}
    \begin{tabular}{ll}
        \toprule
        \textbf{Hyperparameter}  & \textbf{Value}             \\
        \midrule
        Learning Rate            & \{2e-4, 5e-4, 1e-3, 2e-3\} \\
        Rollout length           & \{8, 16, 32\}              \\
        Replay buffer size       & \{1000, 2000, 4000\}       \\
        Target update interval   & \{100,200,400,800\}        \\
        \textbf{SHAQ specific}   &                            \\
        Sample size              & \{1, 5, 10\}               \\
        \textbf{M3DDPG specific} &                            \\
        Adversarial rate         & \{1e-1, 1e-2, 1e-3\}       \\
        \bottomrule
    \end{tabular}
\end{table}

\subsubsection{Guidelines for Hyperparameter Selection}

\label{app:hyperparam_guidelines}

The softmax temperature $\tau$ controls how selectively the weights focus on influential directly action-responsive coordinates. We find empirically that values in the range $0.1$-$1.0$ work well.
The EMA coefficient $\beta$ smooths weight updates and improves stability. Its effective behavior is correlated with the update interval. When updates occur frequently with small batches, a lower $\beta$ is preferable, whereas larger batches or slower update intervals allow for higher $\beta$.
The intrinsic reward should guide adaptation without overwhelming the task reward. $\lambda \in [0.01, 0.1]$ work reliably in SMAX, but this coefficient is generally a sensitive parameter and should be tuned per environment.

\section{LLM Usage}
We used LLMs for polishing text. All algorithmic design, mathematics and experiments were written by the authors.

\section{Full Experimental Results}
\label{app:experiments}

\subsection{Component Evaluation and Ablation}

This appendix complements Table~\ref{tab:ablations} by visualizing the full learning dynamics of LEICA ablations on the \cm{medium}-scale benchmark.
Each figure reports win-rate learning curves evaluated separately on \cm{seen} and \cm{unseen} configurations.
Curves are averaged across the five SMAX scenarios and eight random seeds, and shaded regions indicate standard error of the mean.

\begin{figure}[h]
    \centering
    \includegraphics[width=0.45\linewidth]{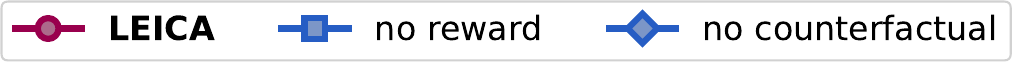}
    \vspace{0.05in}

    \raisebox{0.1in}{\includegraphics[width=0.0222\linewidth]{pdf/ylabel.pdf}}
    \includegraphics[width=0.3\linewidth]{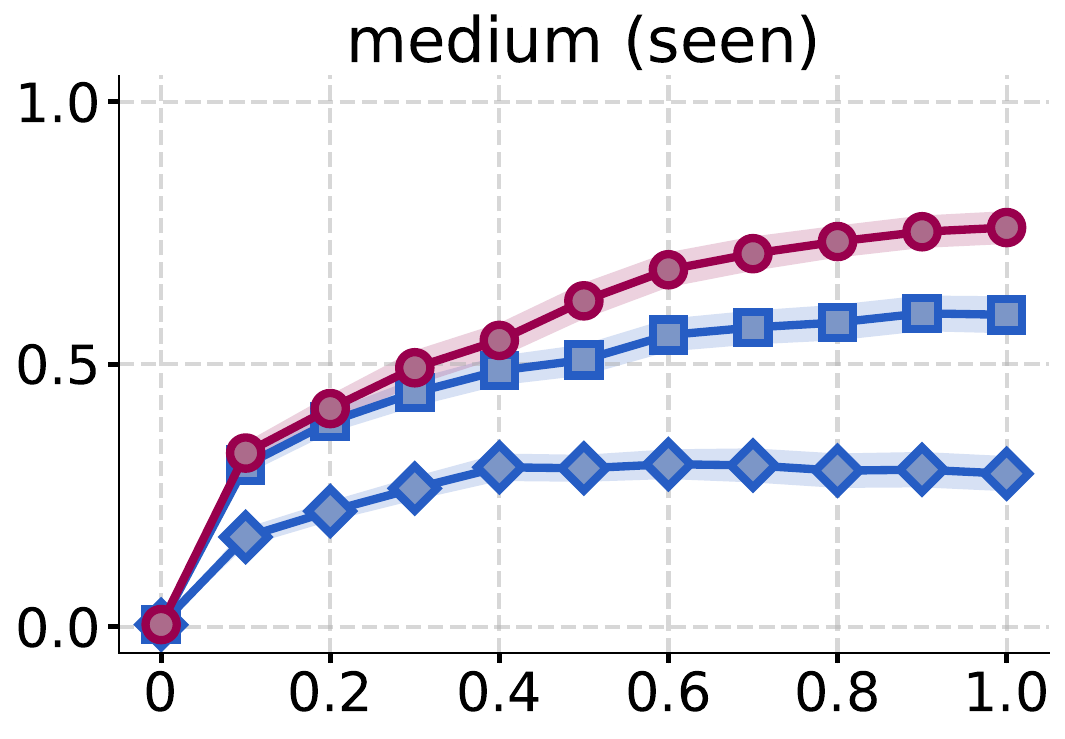}
    \includegraphics[width=0.3\linewidth]{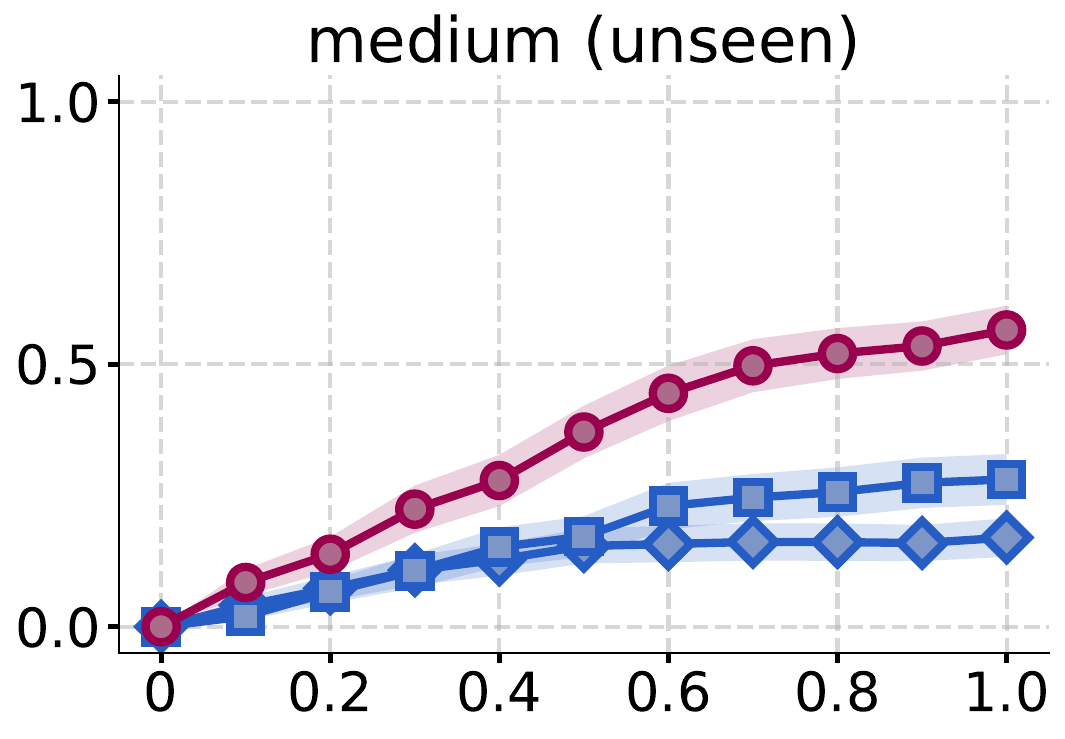}
    \vspace{-0.05in}

    \includegraphics[width=0.5\linewidth]{pdf/xlabel_long.pdf}
    \vspace{0.05in}

    \caption{\textbf{Ablation on reward design.}
        Win-rate learning curves on the medium-scale benchmark.
        We compare LEICA (default) against two variants: no intrinsic reward (contexts retained but intrinsic reward removed) and no counterfactual design (intrinsic reward retained but the counterfactual baseline in Eq.~(\ref{eq:cf}) removed).}
    \label{fig:ablation_reward}
\end{figure}
\begin{figure}[h]
    \centering
    \includegraphics[width=0.8\linewidth]{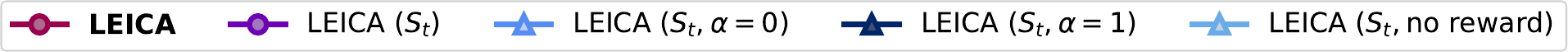}
    \vspace{0.05in}

    \raisebox{0.1in}{\includegraphics[width=0.0222\linewidth]{pdf/ylabel.pdf}}
    \includegraphics[width=0.3\linewidth]{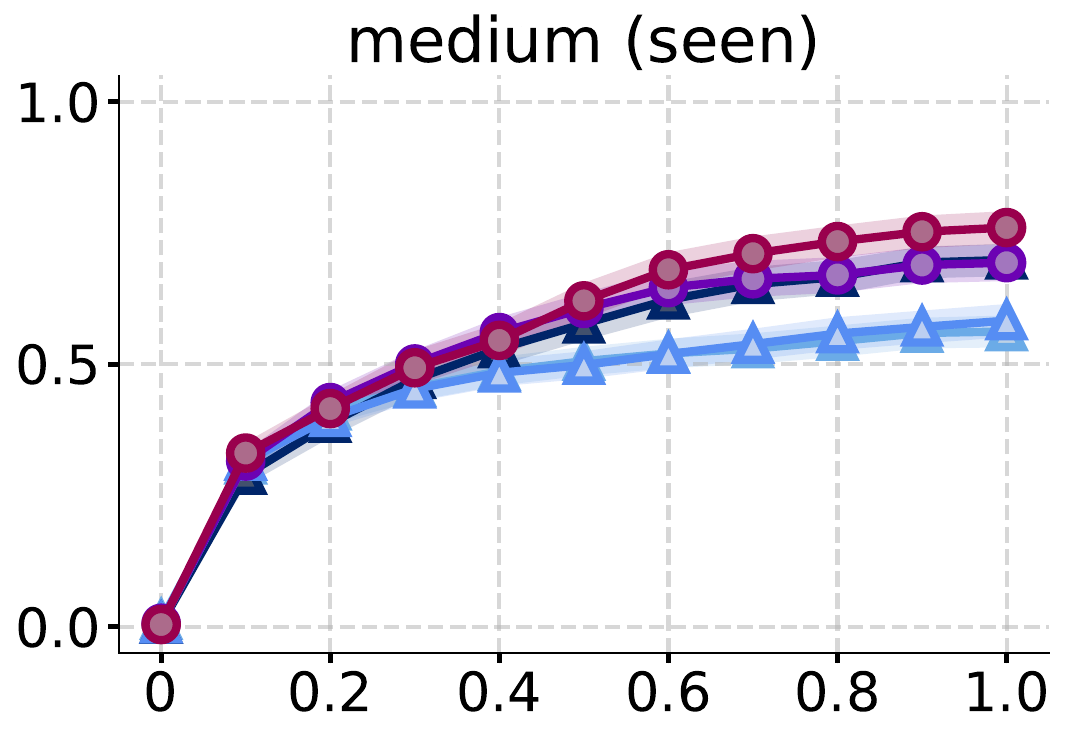}
    \includegraphics[width=0.3\linewidth]{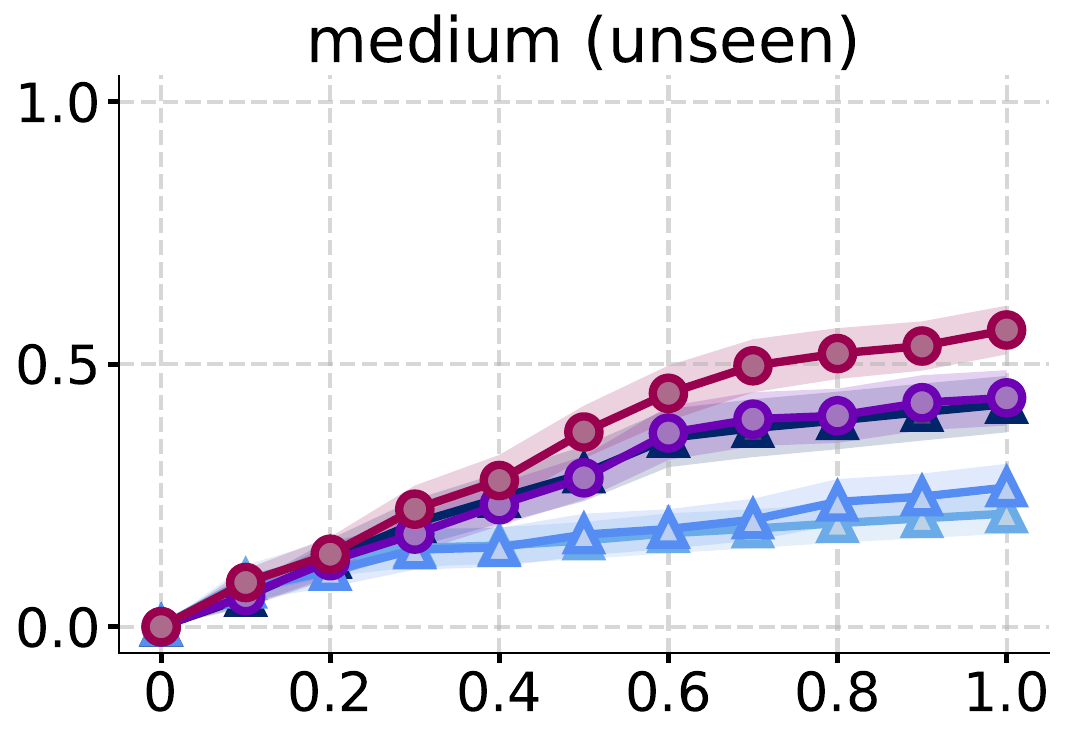}
    \vspace{-0.05in}

    \includegraphics[width=0.5\linewidth]{pdf/xlabel_long.pdf}
    \vspace{0.05in}

    \caption{\textbf{Ablation on full-state reconstruction.}
        LEICA($S_t$) removes the state partition and trains predictors by reconstructing the full state.
        We additionally report LEICA($S_t, \alpha=0$) (value-only weighting), LEICA($S_t, \alpha=1$) (influence-only weighting), and LEICA($S_t$, no reward).}
    \label{fig:ablation_fullstate}
\end{figure}
\begin{figure}[h]
    \centering
    \includegraphics[width=0.6\linewidth]{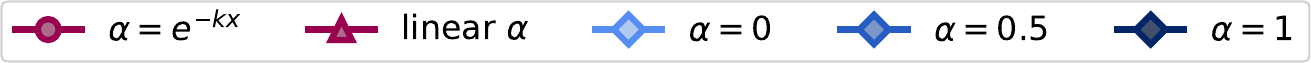}
    \vspace{0.05in}

    \raisebox{0.1in}{\includegraphics[width=0.0222\linewidth]{pdf/ylabel.pdf}}
    \includegraphics[width=0.3\linewidth]{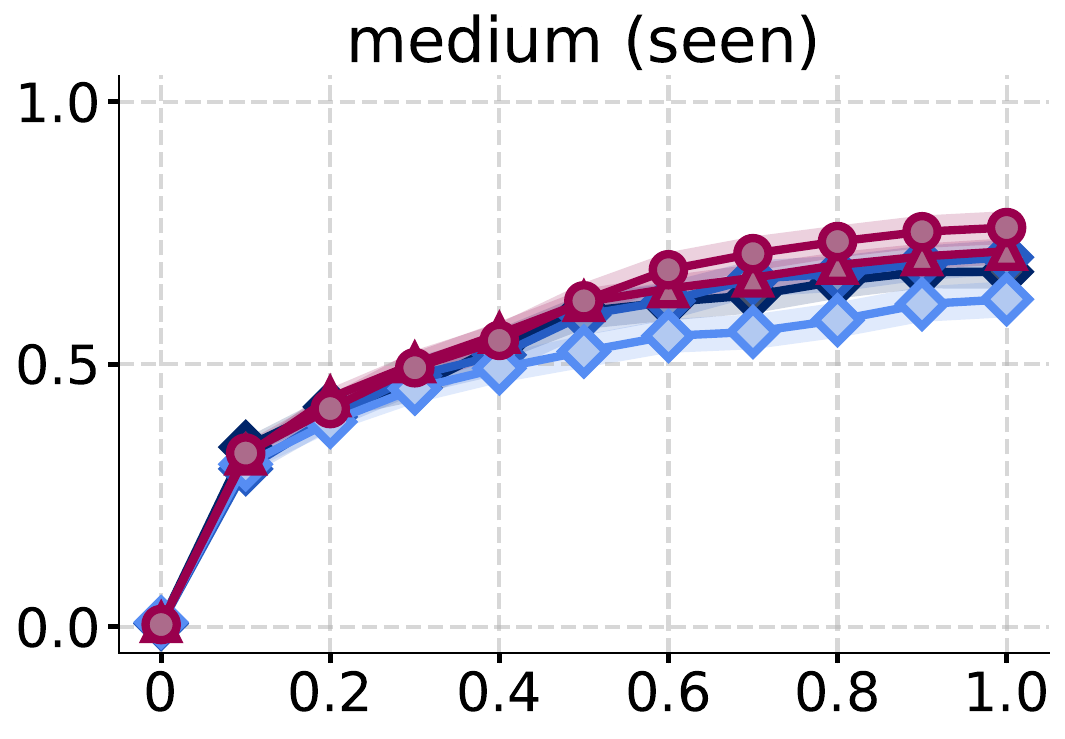}
    \includegraphics[width=0.3\linewidth]{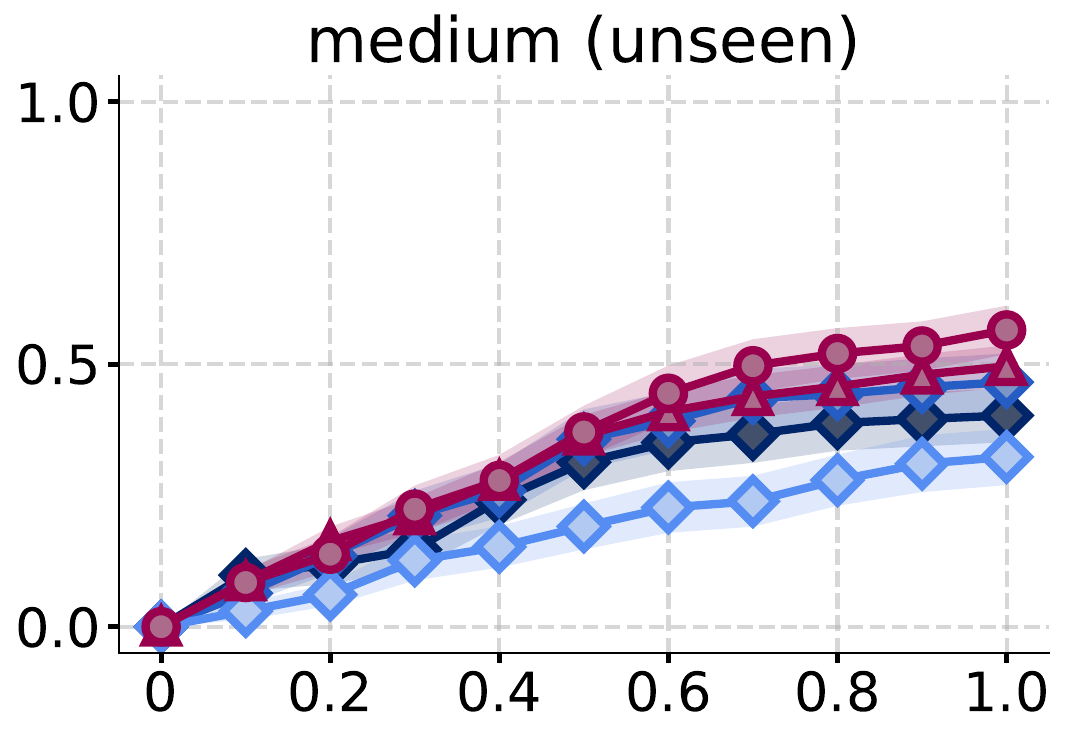}
    \vspace{-0.05in}

    \includegraphics[width=0.5\linewidth]{pdf/xlabel_long.pdf}
    \vspace{0.05in}

    \caption{\textbf{Ablation on $\alpha$ scheduling.}
        We compare the default exponentially decaying schedule $\alpha(x)=\exp(-kx)$ (training update index $x$) against linear decay, and constant choices $\alpha\in\{0, 0.5, 1\}$.}
    \label{fig:ablation_alpha}
\end{figure}

\begin{figure}[h]
    \centering
    \includegraphics[width=0.8\linewidth]{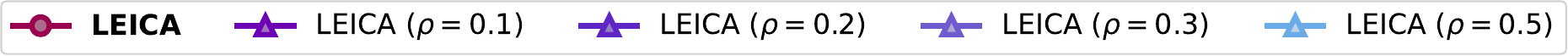}
    \vspace{0.05in}

    \raisebox{0.1in}{\includegraphics[width=0.0222\linewidth]{pdf/ylabel.pdf}}
    \includegraphics[width=0.3\linewidth]{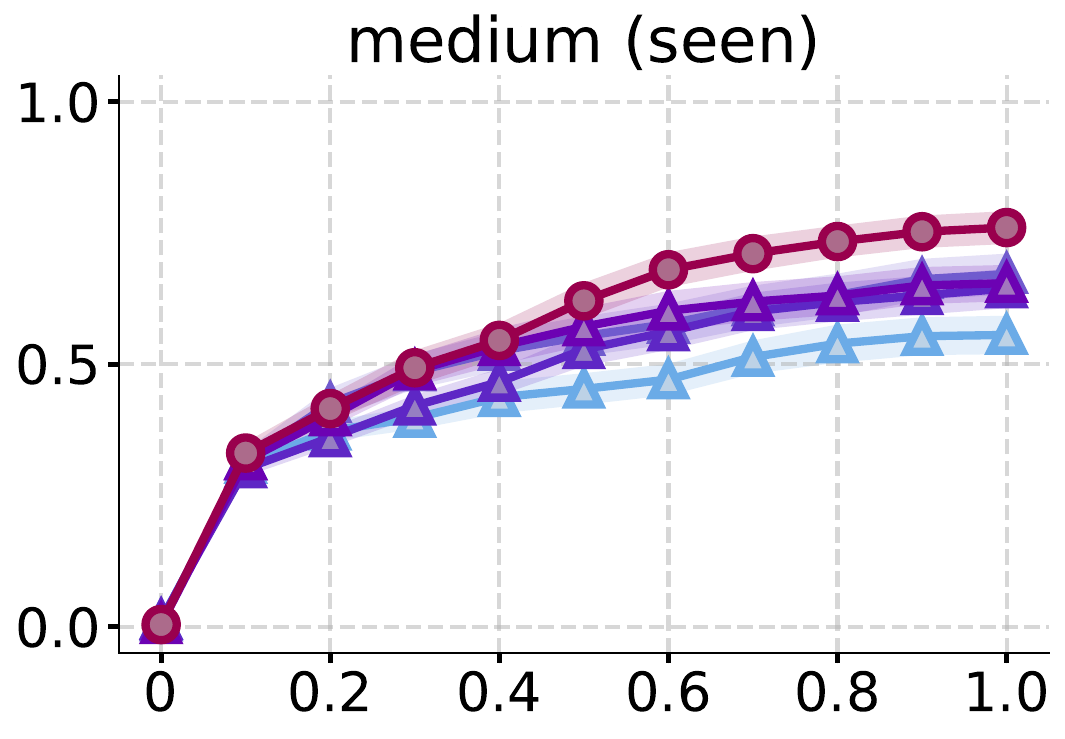}
    \includegraphics[width=0.3\linewidth]{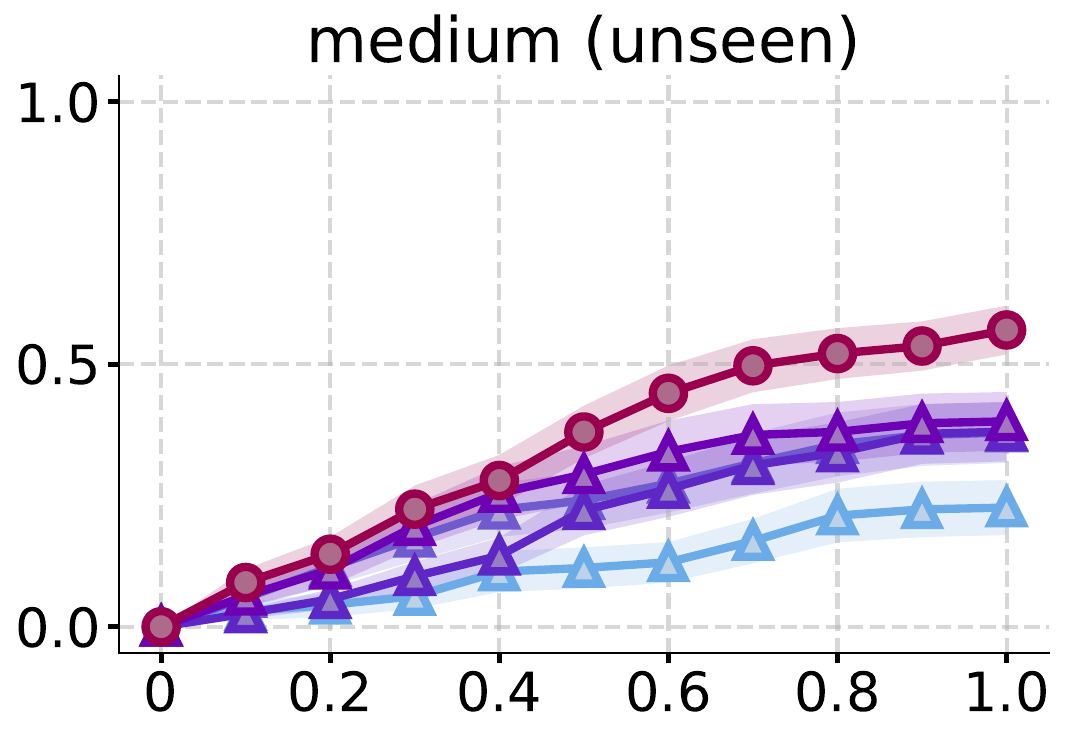}
    \vspace{-0.05in}

    \includegraphics[width=0.5\linewidth]{pdf/xlabel_long.pdf}
    \vspace{0.05in}

    \caption{\textbf{Robustness to imperfect partitions.}
        We randomly swap a fraction $\rho$ of coordinates between $S^{\mathrm{d}}$ and $S^{\mathrm{r}}$ (swapping rate $\rho\in\{0.1,0.2,0.3,0.5\}$) and compare against the default LEICA ($\rho=0$).}
    \label{fig:ablation_partition}
\end{figure}

\FloatBarrier

\subsection{Statistical Analysis of the Main Results}
\label{app:stats}

\paragraph{Evaluation units and procedure}
Each independently trained run is scored by its final win rate on seen or unseen configurations, with runs grouped by the five SMAX tasks. Since shared seeds do not ensure paired runs across methods, we use unpaired comparisons. Differences $\Delta$ are block-balanced mean differences (percentage points). We apply within-task stratified permutation tests, a two-stage cluster bootstrap (tasks, then runs) for 95\% confidence intervals, and Holm-Bonferroni correction across baselines within each scale and metric. The probability of improvement (PoI) is the block-balanced fraction of run pairs favoring LEICA, with ties counted as $1/2$. \cm{Overall} pools all $15$ task-scale blocks.

Table~\ref{tab:stats} shows that LEICA significantly outperforms all non-LEICA baselines in every column. Compared with LEICA (manual), it performs significantly better on \cm{small} unseen configurations but worse at \cm{large} scale, with no significant difference in \cm{Overall}.

\begin{table}[h]
    \centering
    \scriptsize
    \setlength{\tabcolsep}{3pt}
    \caption{Unpaired statistical comparisons of LEICA against each method in Table~\ref{tab:main_results}: block-balanced difference $\Delta$ (percentage points) with 95\% cluster-bootstrap CI, PoI with 95\% CI, and Holm-adjusted stratified permutation $p$-value.}
    \label{tab:stats}
    \resizebox{\linewidth}{!}{%
        \begin{tabular}{llcccccc}
            \toprule
                  &                & \multicolumn{3}{c}{\cm{Seen}} & \multicolumn{3}{c}{\cm{Unseen}}                                                                             \\
            \cmidrule(lr){3-5}\cmidrule(lr){6-8}
            Scale & LEICA vs.      & $\Delta$ [95\% CI]            & PoI [95\% CI]                   & $p$      & $\Delta$ [95\% CI]           & PoI [95\% CI]        & $p$      \\
            \midrule
            \multirow{8}{*}{\cm{Small}}
                  & LEICA (manual) & 2.72 [$-$3.45, 8.56]          & 0.614 [0.345, 0.866]            & 0.053    & 5.35 [0.54, 10.35]           & 0.681 [0.463, 0.891] & 0.001    \\
                  & MAPPO          & 19.25 [6.22, 34.40]           & 0.853 [0.681, 0.984]            & $<$0.001 & 22.85 [10.20, 36.21]         & 0.994 [0.969, 1.000] & $<$0.001 \\
                  & MAPPO+ctx      & 11.69 [4.43, 18.82]           & 0.847 [0.675, 0.981]            & $<$0.001 & 14.21 [8.39, 19.59]          & 0.903 [0.772, 0.991] & $<$0.001 \\
                  & QMIX           & 31.21 [21.99, 40.44]          & 0.997 [0.981, 1.000]            & $<$0.001 & 27.89 [20.54, 35.37]         & 1.000 [1.000, 1.000] & $<$0.001 \\
                  & LAIES          & 50.23 [35.13, 68.62]          & 0.997 [0.981, 1.000]            & $<$0.001 & 48.50 [32.95, 67.77]         & 0.970 [0.894, 1.000] & $<$0.001 \\
                  & SHAQ           & 92.47 [87.90, 96.80]          & 1.000 [1.000, 1.000]            & $<$0.001 & 87.58 [82.51, 94.03]         & 1.000 [1.000, 1.000] & $<$0.001 \\
                  & M3DDPG         & 87.99 [74.93, 96.75]          & 1.000 [1.000, 1.000]            & $<$0.001 & 82.64 [69.01, 93.87]         & 1.000 [1.000, 1.000] & $<$0.001 \\
                  & M3PPO          & 52.85 [35.27, 70.01]          & 0.972 [0.900, 1.000]            & $<$0.001 & 54.84 [35.74, 72.73]         & 0.986 [0.938, 1.000] & $<$0.001 \\
            \midrule
            \multirow{8}{*}{\cm{Medium}}
                  & LEICA (manual) & 0.11 [$-$7.29, 6.90]          & 0.562 [0.362, 0.753]            & 0.971    & $-$1.39 [$-$12.08, 7.16]     & 0.541 [0.330, 0.725] & 0.709    \\
                  & MAPPO          & 22.45 [11.93, 32.45]          & 0.806 [0.631, 0.947]            & $<$0.001 & 35.39 [18.49, 52.83]         & 0.891 [0.794, 0.972] & $<$0.001 \\
                  & MAPPO+ctx      & 19.74 [6.38, 32.41]           & 0.791 [0.600, 0.944]            & $<$0.001 & 28.09 [6.84, 47.63]          & 0.853 [0.698, 0.963] & $<$0.001 \\
                  & QMIX           & 17.85 [$-$0.31, 36.64]        & 0.825 [0.550, 1.000]            & $<$0.001 & 38.29 [18.07, 58.02]         & 0.819 [0.550, 1.000] & $<$0.001 \\
                  & LAIES          & 16.37 [$-$5.71, 38.75]        & 0.637 [0.306, 0.950]            & $<$0.001 & 45.35 [15.72, 74.43]         & 0.802 [0.520, 1.000] & $<$0.001 \\
                  & SHAQ           & 54.18 [39.33, 68.19]          & 0.975 [0.900, 1.000]            & $<$0.001 & 40.16 [13.74, 67.26]         & 0.753 [0.459, 0.975] & $<$0.001 \\
                  & M3DDPG         & 60.56 [43.61, 77.72]          & 0.988 [0.941, 1.000]            & $<$0.001 & 54.86 [27.18, 81.56]         & 0.981 [0.938, 1.000] & $<$0.001 \\
                  & M3PPO          & 44.42 [28.69, 60.90]          & 0.959 [0.850, 1.000]            & $<$0.001 & 53.31 [26.94, 78.42]         & 0.963 [0.867, 1.000] & $<$0.001 \\
            \midrule
            \multirow{8}{*}{\cm{Large}}
                  & LEICA (manual) & $-$6.22 [$-$12.84, $-$0.39]   & 0.378 [0.188, 0.569]            & 0.025    & $-$12.03 [$-$22.98, $-$2.63] & 0.339 [0.203, 0.484] & 0.015    \\
                  & MAPPO          & 9.97 [1.61, 20.97]            & 0.706 [0.525, 0.866]            & 0.002    & 16.61 [2.79, 35.69]          & 0.775 [0.597, 0.927] & 0.002    \\
                  & MAPPO+ctx      & 9.90 [$-$0.80, 25.80]         & 0.675 [0.472, 0.875]            & 0.002    & 15.64 [$-$1.41, 41.59]       & 0.697 [0.491, 0.894] & 0.007    \\
                  & QMIX           & 11.27 [$-$1.31, 23.90]        & 0.781 [0.547, 0.981]            & $<$0.001 & 10.43 [$-$4.61, 24.70]       & 0.759 [0.463, 0.991] & 0.015    \\
                  & LAIES          & 6.28 [$-$7.56, 22.07]         & 0.569 [0.269, 0.869]            & 0.034    & 19.87 [$-$0.23, 44.88]       & 0.700 [0.425, 0.944] & $<$0.001 \\
                  & SHAQ           & 30.14 [18.87, 41.45]          & 1.000 [1.000, 1.000]            & $<$0.001 & 9.23 [$-$10.43, 31.93]       & 0.512 [0.169, 0.853] & 0.029    \\
                  & M3DDPG         & 48.34 [40.19, 59.44]          & 1.000 [1.000, 1.000]            & $<$0.001 & 32.88 [16.75, 54.29]         & 1.000 [1.000, 1.000] & $<$0.001 \\
                  & M3PPO          & 16.94 [6.83, 28.48]           & 0.869 [0.731, 0.975]            & $<$0.001 & 24.37 [7.85, 45.68]          & 0.887 [0.694, 1.000] & $<$0.001 \\
            \midrule
            \multirow{8}{*}{\cm{Overall}}
                  & LEICA (manual) & $-$1.13 [$-$5.43, 2.99]       & 0.518 [0.381, 0.656]            & 0.404    & $-$2.69 [$-$9.21, 3.09]      & 0.520 [0.392, 0.653] & 0.171    \\
                  & MAPPO          & 17.22 [10.25, 24.87]          & 0.789 [0.685, 0.883]            & $<$0.001 & 24.95 [15.32, 35.36]         & 0.886 [0.801, 0.956] & $<$0.001 \\
                  & MAPPO+ctx      & 13.78 [6.89, 21.29]           & 0.771 [0.656, 0.874]            & $<$0.001 & 19.31 [9.22, 30.54]          & 0.818 [0.713, 0.909] & $<$0.001 \\
                  & QMIX           & 20.11 [10.95, 29.13]          & 0.868 [0.739, 0.973]            & $<$0.001 & 25.54 [15.15, 36.11]         & 0.859 [0.715, 0.974] & $<$0.001 \\
                  & LAIES          & 24.29 [10.28, 38.60]          & 0.734 [0.549, 0.900]            & $<$0.001 & 37.91 [22.81, 53.16]         & 0.824 [0.681, 0.946] & $<$0.001 \\
                  & SHAQ           & 58.93 [44.47, 73.19]          & 0.992 [0.967, 1.000]            & $<$0.001 & 45.66 [25.38, 65.44]         & 0.755 [0.565, 0.919] & $<$0.001 \\
                  & M3DDPG         & 65.63 [54.33, 77.03]          & 0.996 [0.979, 1.000]            & $<$0.001 & 56.79 [40.62, 72.38]         & 0.994 [0.978, 1.000] & $<$0.001 \\
                  & M3PPO          & 38.07 [26.79, 49.67]          & 0.933 [0.867, 0.983]            & $<$0.001 & 44.17 [29.80, 58.64]         & 0.945 [0.867, 0.996] & $<$0.001 \\
            \bottomrule
        \end{tabular}}
\end{table}

\subsection{Sensitivity to Stochasticity}
\label{app:stochasticity}

To examine how LEICA behaves when its relative predictive signal is perturbed by stochastic dynamics, we conduct two sensitivity experiments on the five \cm{medium}-scale tasks with eight seeds per task, using the same statistical procedure as in Appendix~\ref{app:stats}.
In the first, each opponent action is replaced by a randomly sampled valid action with probability $p\in\{0.1,0.2,0.4\}$. In the second, we use an action-attack protocol common in robust MARL: each cooperative agent's executed action is replaced by a randomly sampled valid action with probability $p\in\{0.1,0.2,0.4\}$.
Table~\ref{tab:stochasticity} reports the final win rates (as fractions).

\begin{table}[h]
    \centering
    \scriptsize
    \setlength{\tabcolsep}{4pt}
    \caption{Sensitivity to stochasticity on the \cm{medium} scale: block-balanced win rates, difference with 95\% cluster-bootstrap CI, PoI with 95\% CI, and Holm-adjusted stratified permutation $p$-value.}
    \label{tab:stochasticity}
    \begin{tabular}{llcccccc}
        \toprule
        Perturbation & Regime      & $p$ & LEICA & MAPPO & Difference [95\% CI]    & PoI [95\% CI]        & $p$-value \\
        \midrule
        \multirow{6}{*}{\shortstack[l]{Opponent                                                                       \\random action}}
                     & \cm{Seen}   & 0.1 & 0.713 & 0.278 & 0.435 [0.306, 0.563]    & 0.795 [0.719, 0.867] & $<$0.001  \\
                     & \cm{Seen}   & 0.2 & 0.837 & 0.640 & 0.196 [0.124, 0.274]    & 0.709 [0.644, 0.783] & $<$0.001  \\
                     & \cm{Seen}   & 0.4 & 0.964 & 0.956 & 0.008 [-0.004, 0.016] & 0.486 [0.378, 0.606] & 0.270     \\
                     & \cm{Unseen} & 0.1 & 0.681 & 0.241 & 0.441 [0.318, 0.605]    & 0.777 [0.697, 0.861] & $<$0.001  \\
                     & \cm{Unseen} & 0.2 & 0.819 & 0.554 & 0.265 [0.175, 0.357]    & 0.784 [0.716, 0.853] & $<$0.001  \\
                     & \cm{Unseen} & 0.4 & 0.962 & 0.931 & 0.031 [0.017, 0.040]    & 0.720 [0.633, 0.809] & 0.0014    \\
        \midrule
        \multirow{6}{*}{\shortstack[l]{Ally action                                                                    \\attack}}
                     & \cm{Seen}   & 0.1 & 0.604 & 0.522 & 0.082 [0.044, 0.121]    & 0.691 [0.578, 0.800] & 0.0049    \\
                     & \cm{Seen}   & 0.2 & 0.500 & 0.396 & 0.104 [0.083, 0.123]    & 0.747 [0.647, 0.856] & 0.0003    \\
                     & \cm{Seen}   & 0.4 & 0.301 & 0.294 & 0.007 [-0.007, 0.021] & 0.545 [0.416, 0.694] & 0.371     \\
                     & \cm{Unseen} & 0.1 & 0.320 & 0.184 & 0.136 [0.078, 0.194]    & 0.645 [0.534, 0.756] & 0.0047    \\
                     & \cm{Unseen} & 0.2 & 0.120 & 0.027 & 0.094 [0.062, 0.120]    & 0.694 [0.634, 0.750] & 0.0070    \\
                     & \cm{Unseen} & 0.4 & 0.000 & 0.000 & 0.000 [0.000, 0.000]    & 0.487 [0.463, 0.500] & 1.000     \\
        \bottomrule
    \end{tabular}
\end{table}

\paragraph{Opponent stochasticity}
LEICA remains significantly better than MAPPO at all tested levels except the seen regime at $p=0.4$, where both methods approach saturated performance. Random opponent actions also weaken the opponent, as reflected by the increasing win rates of both methods. These results show maintained task performance under the tested opponent-action stochasticity.

\paragraph{Ally action attacks}
Absolute performance decreases as the attack becomes stronger, but LEICA maintains a statistically supported advantage over MAPPO under mild and moderate attacks ($p\in\{0.1,0.2\}$) in both seen and unseen regimes. Under the strongest attack ($p=0.4$), task performance is severely degraded and LEICA no longer differs significantly from MAPPO. This experiment characterizes performance under corruption of the cooperative agents' execution channel; attributing the degradation specifically to partition recovery requires a separate quantitative analysis.

\subsection{MABrax Continuous-Control Validation}
\label{app:mabrax}
We additionally evaluate a latent predictive variant of LEICA on MABrax, a multi-agent continuous-control benchmark. Below are the details.

\paragraph{Exogenous context construction}
Each episode is associated with an exogenous context
\begin{equation}
    c = (s_a, v^\star, \alpha_w, g_w),
\end{equation}
where $s_a$ is an actuator-scale multiplier, $v^\star$ is the target forward velocity, $\alpha_w$ is the momentum coefficient of a reactive wind perturbation, and $g_w$ controls the strength with which forward velocity drives the wind. At timestep $t$, the action sent to the underlying MABrax environment is $\tilde{a}_t = s_a a_t + w_t$,
where $w_t$ is a scalar wind state shared across action dimensions. The wind evolves according to $w_{t+1} = \mathrm{clip}\left(\alpha_w w_t + (1-\alpha_w)g_w v_t, -1, 1\right)$,
where $v_t$ denotes the current forward velocity. The wind update uses the pre-step velocity $v_t$. Therefore, the current action cannot directly determine $w_{t+1}$ at the same decision step. Instead, it can affect the next robot velocity and thereby influence $w_{t+2}$ and later wind states. The task reward is further shaped by the context-specific target velocity, $r^{\mathrm{MABrax}}_t = r^{\mathrm{base}}_t - \eta (v_{t+1}-v^\star)^2$,
with $\eta=0.05$ in our experiments.

\paragraph{Train/test context split}
During training, each episode samples one context uniformly from the four training contexts in Table~\ref{tab:mabrax-contexts}. Evaluation is performed on both the seen training contexts and three held-out contexts. The held-out contexts simultaneously change actuation strength, target velocity, and wind dynamics, providing continuous physical exogenous shifts rather than opponent-script shifts.

\begin{table}[h]
    \centering
    \small
    \caption{MABrax exogenous contexts. Each context is $(s_a, v^\star, \alpha_w, g_w)$, corresponding to actuator scale, target velocity, wind momentum, and wind gain.}
    \label{tab:mabrax-contexts}
    \begin{tabular}{lcccc}
        \toprule
        Split   & $s_a$ & $v^\star$ & $\alpha_w$ & $g_w$ \\
        \midrule
        Train 1 & 0.90  & 0.80      & 0.85       & 0.05  \\
        Train 2 & 0.95  & 1.00      & 0.80       & 0.10  \\
        Train 3 & 1.05  & 1.20      & 0.75       & 0.08  \\
        Train 4 & 1.10  & 1.00      & 0.90       & 0.12  \\
        \midrule
        Test 1  & 0.85  & 0.60      & 0.70       & 0.15  \\
        Test 2  & 1.00  & 1.40      & 0.95       & 0.18  \\
        Test 3  & 1.15  & 1.10      & 0.65       & 0.06  \\
        \bottomrule
    \end{tabular}
\end{table}

\paragraph{Latent-space LEICA implementation}
Let $h_t^i$ denote the recurrent latent state produced by agent $i$'s encoder from its local observation history. The MABrax implementation uses two stochastic latent heads, $z_{\mathrm{int},t}^i$ and $z_{\mathrm{ext},t}^i$. The direct decoder predicts the next latent state using the current latent state and the action,
\begin{equation}
    \hat{h}^{\mathrm{endo},i}_{t+1} = f_{\mathrm{endo}}(h_t^i, a_t^i, z_{\mathrm{int},t}^i),
\end{equation}
whereas the response decoder predicts the next latent state without the action,
\begin{equation}
    \hat{h}^{\mathrm{exo},i}_{t+1} = f_{\mathrm{exo}}(h_t^i, z_{\mathrm{ext},t}^i).
\end{equation}
This mirrors the SMAX design in latent space: the direct branch captures action-sensitive transitions, while the response branch captures action-free latent transition structure. The intrinsic reward compares the predicted latent change under the sampled action with the predicted change under the policy mean action,
\begin{equation}
    r^{\mathrm{int},i}_t
    = \sum_j w_j \left(
    \left|\hat{h}^{\mathrm{endo},i}_{t+1,j}(a_t^i)-h^i_{t,j}\right|
    -
    \left|\hat{h}^{\mathrm{endo},i}_{t+1,j}(\overline{a_t^i})-h^i_{t,j}\right|
    \right),
\end{equation}
where $\overline{a_t^i}$ denotes the policy mean action. The coordinate weights $w$ are updated from the Jacobian norm of the response decoder with respect to the latent state with EMA smoothing. As in the main experiments, the auxiliary decoders and Jacobian computations are used during training only.

\subsection{MPE-Tag Mixed-Setting Validation}
\label{app:mpe-tag}
We also evaluate a mixed cooperative-competitive extension on MPE-Tag \citep{mordatch2017emergence}. MPE-Tag contains asymmetric predator and prey roles with opposing objectives. We train role-specific policies with either MAPPO or LEICA and evaluate whether replacing one role's MAPPO policy with a LEICA policy improves that role's own return while holding the counterpart fixed. Table~\ref{tab:mpe-tag} reports role-wise comparisons.

\begin{table}[h]
    \centering
    \caption{MPE-Tag role-wise cross-play validation.}
    \label{tab:mpe-tag}
    \begin{tabular}{llccc}
        \toprule
        Evaluated role & Fixed counterpart & MAPPO role policy & LEICA role policy & $\Delta$ \\
        \midrule
        Predator       & MAPPO prey        & 162.7             & \textbf{168.6}    & $+5.9$   \\
        Predator       & LEICA prey        & 159.2             & \textbf{165.2}    & $+6.0$   \\
        Prey           & MAPPO predator    & -224.3            & \textbf{-215.6}   & $+8.7$   \\
        Prey           & LEICA predator    & -232.1            & \textbf{-220.2}   & $+11.9$  \\
        \bottomrule
    \end{tabular}
\end{table}

The cross-play results isolate the role-wise effect of LEICA.  Positive $\Delta$ means that replacing MAPPO with LEICA improves the return of the replaced role. In all cases, replacing MAPPO policy with a LEICA policy improves the performance.
These results are intended as an additional role-wise validation rather than a full mixed-game benchmark. They suggest that the mediated-influence mechanism is not restricted to purely cooperative SMAX settings, while a broader study of mixed and competitive games, including self-play dynamics and opponent-learning baselines, remains future work.

\subsection{Single-Task Verification on Standard SMAX}
\begin{figure}[h]
    \centering
    \includegraphics[width=0.75\linewidth]{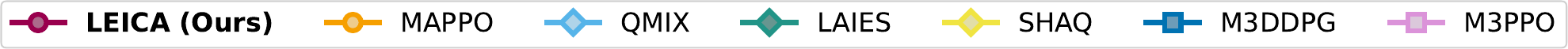}
    \vspace{0.05in}

    \raisebox{0.05in}{\includegraphics[width=0.015\linewidth]{pdf/ylabel.pdf}}
    \includegraphics[width=0.19\linewidth]{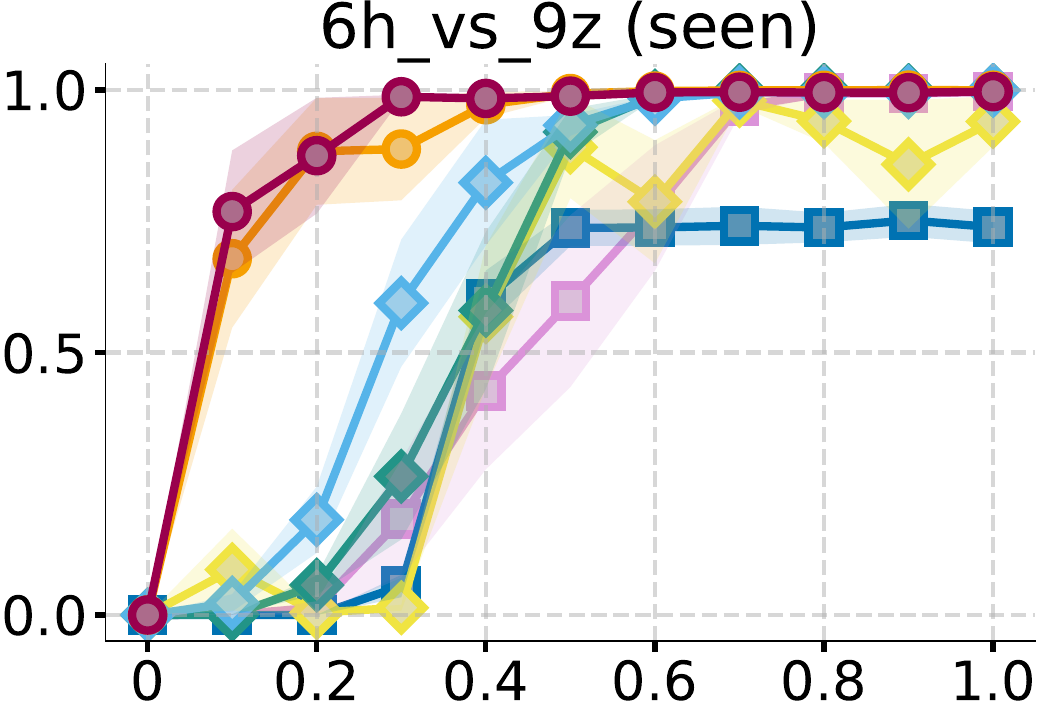}
    \includegraphics[width=0.19\linewidth]{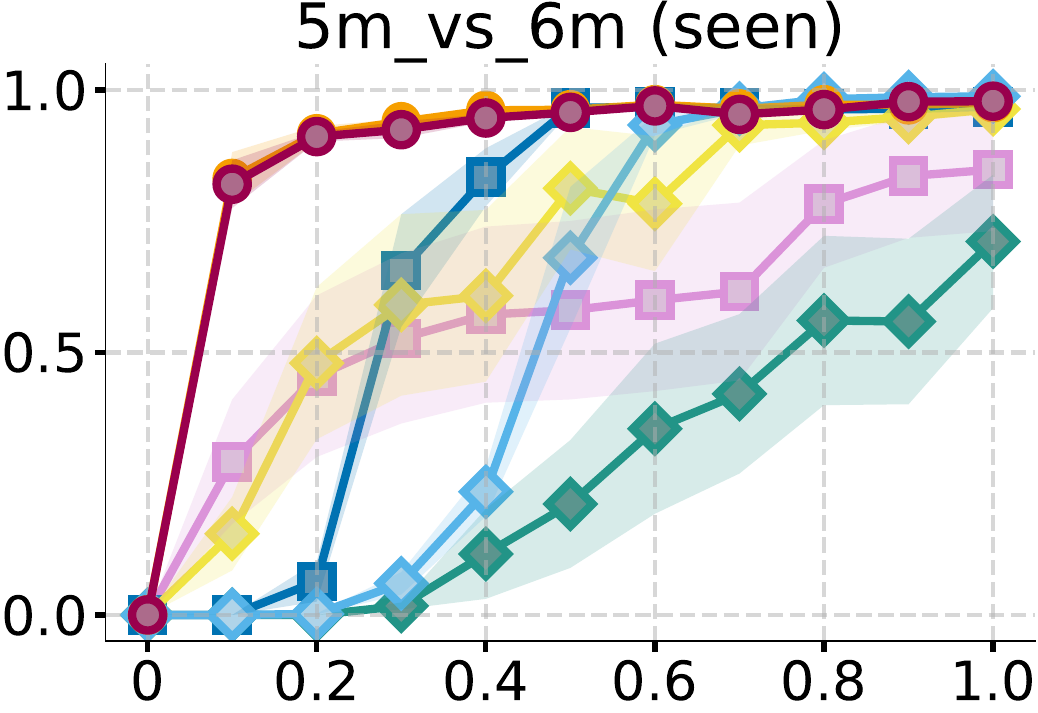}
    \includegraphics[width=0.19\linewidth]{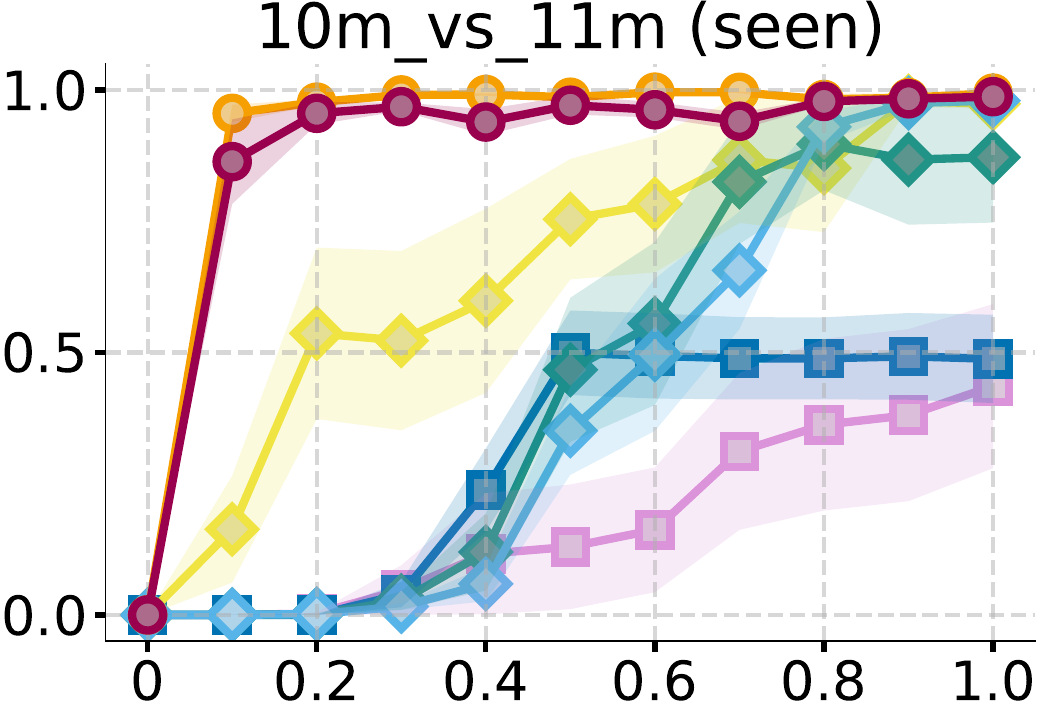}
    \includegraphics[width=0.19\linewidth]{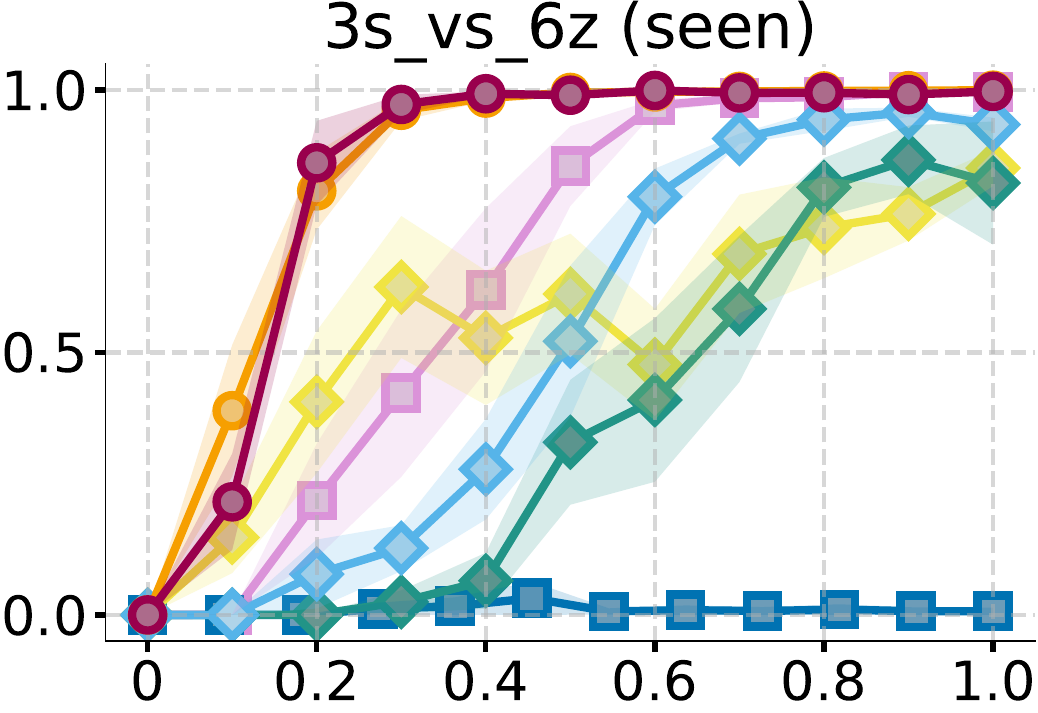}
    \includegraphics[width=0.19\linewidth]{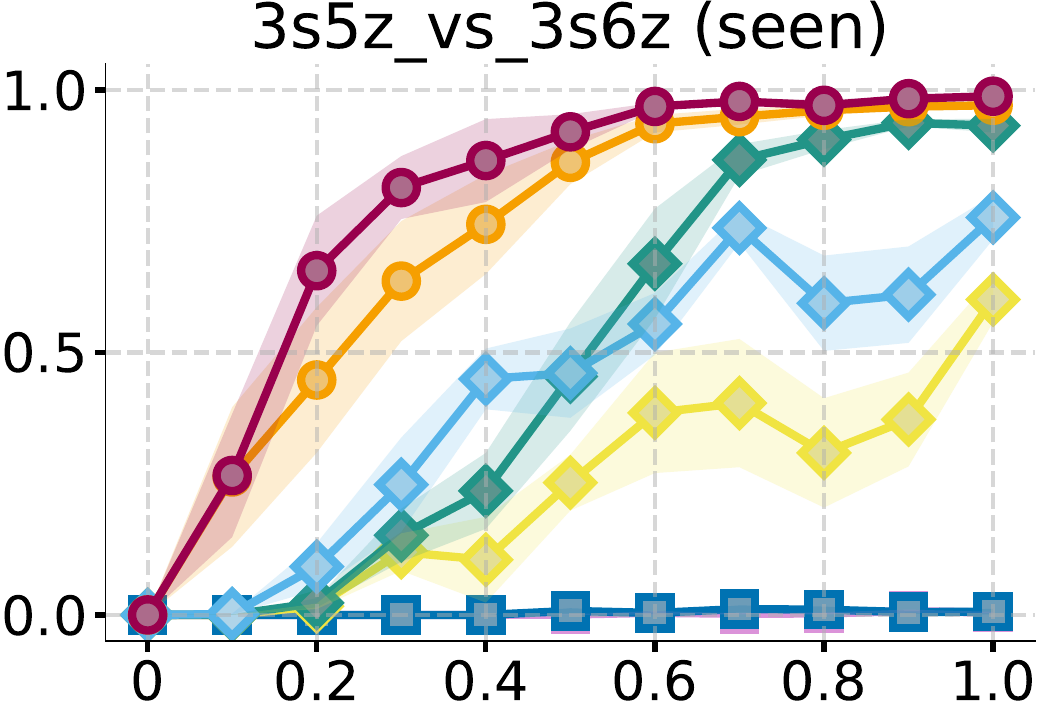}
    \vspace{-0.05in}

    \includegraphics[width=0.4\linewidth]{pdf/xlabel_long.pdf}

    \caption{\textbf{Single-task SMAX.} Learning curves on standard SMAX \textbf{without} exogenous variation. LEICA matches or exceeds baselines, indicating that the proposed method does not degrade conventional performance.}
    \label{fig:single_task}
\end{figure}

On the original single-strategy SMAX benchmarks, LEICA maintains competitive or superior performance relative to all baselines, as shown in Figure~\ref{fig:single_task}. This confirms that the proposed design is not specialized solely for distribution shifts but also remains effective in fixed-opponent settings.

\newpage

\subsection{Additional Results per All Task-Regime Pairs}
\begin{figure}[h]
    \vspace{-0.05in}
    \centering
    \includegraphics[width=0.99\linewidth]{pdf/legend_long.pdf}
    \vspace{0.02in}

    \raisebox{0.05in}{\includegraphics[width=0.015\linewidth]{pdf/ylabel.pdf}}
    \includegraphics[width=0.19\linewidth]{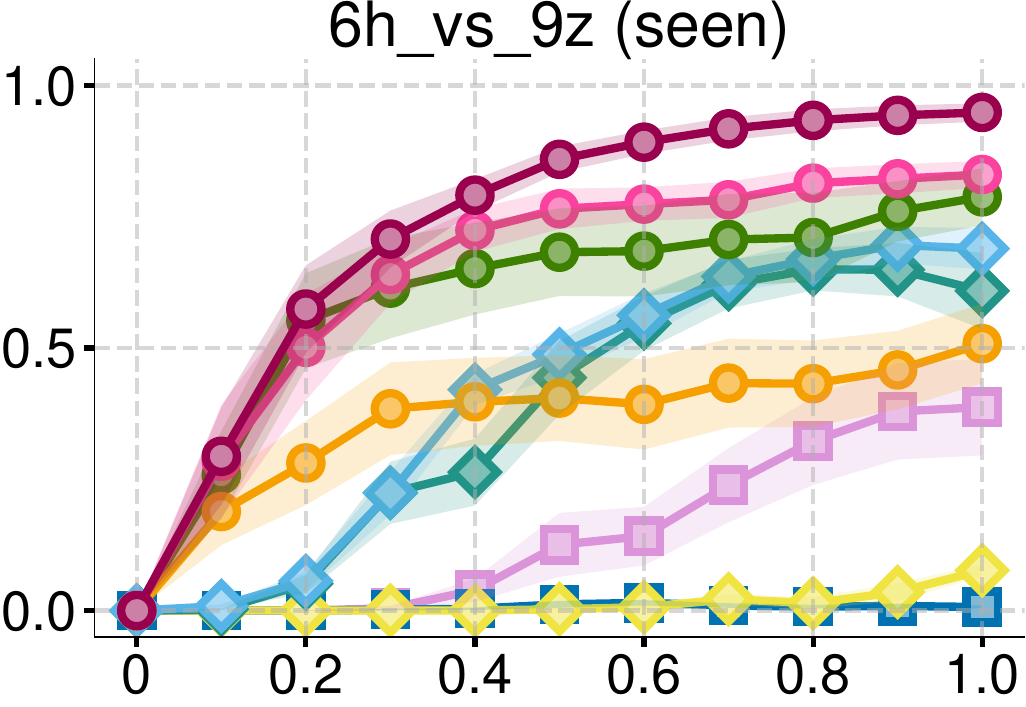}%
    \includegraphics[width=0.19\linewidth]{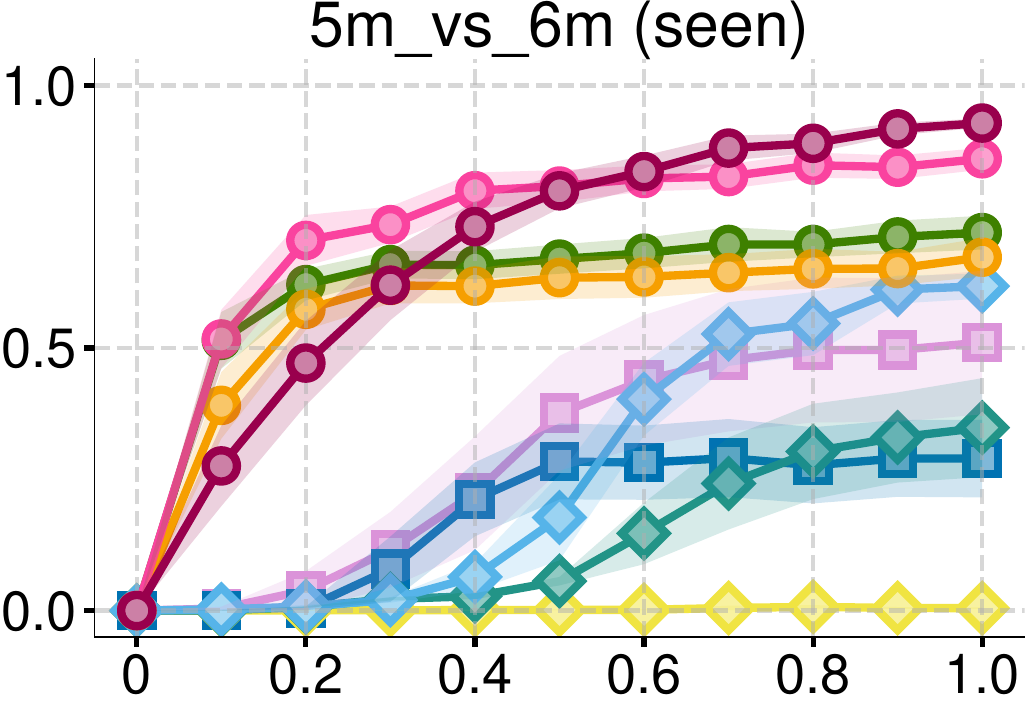}%
    \includegraphics[width=0.19\linewidth]{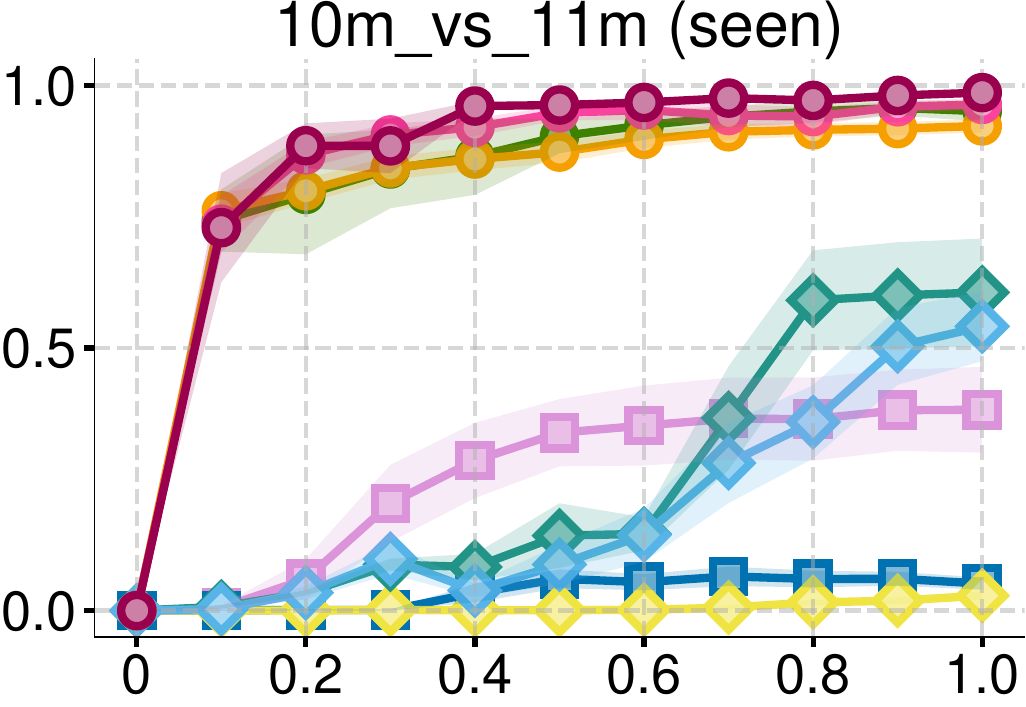}%
    \includegraphics[width=0.19\linewidth]{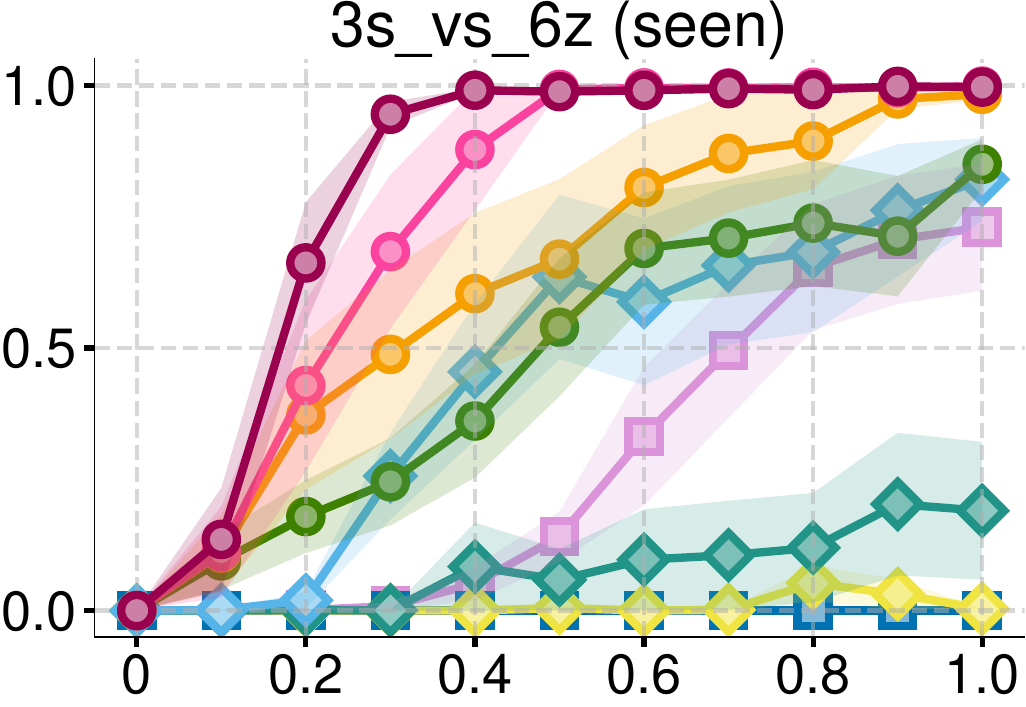}%
    \includegraphics[width=0.19\linewidth]{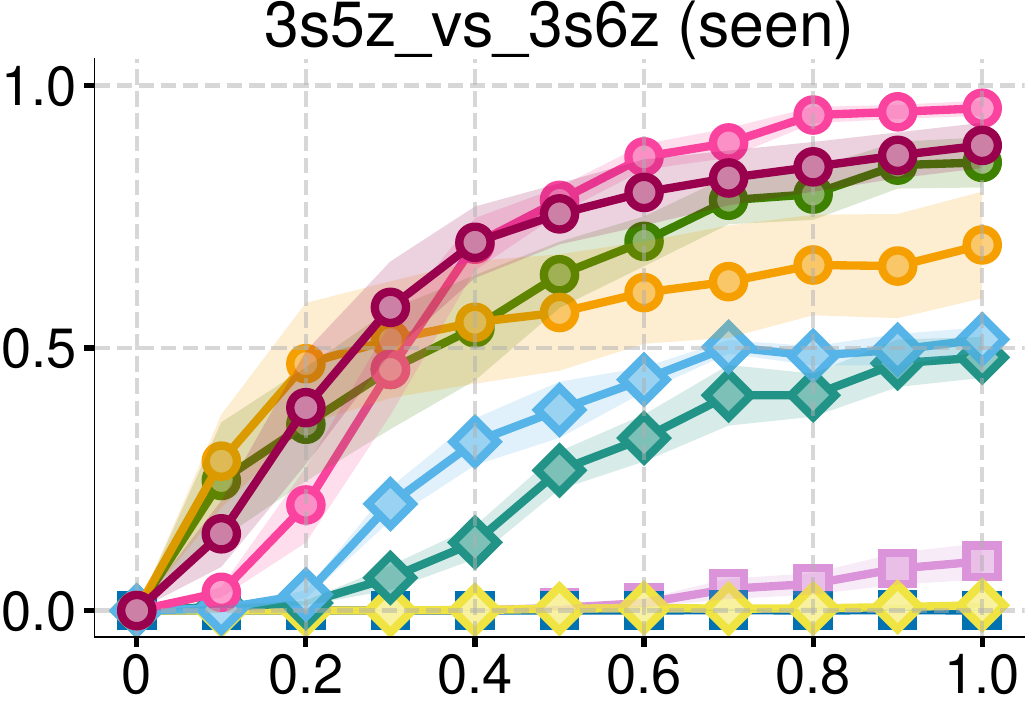}
    \vspace{-0.05in}

    \includegraphics[width=0.4\linewidth]{pdf/xlabel_long.pdf}
    \vspace{0.05in}

    \raisebox{0.05in}{\includegraphics[width=0.015\linewidth]{pdf/ylabel.pdf}}
    \includegraphics[width=0.19\linewidth]{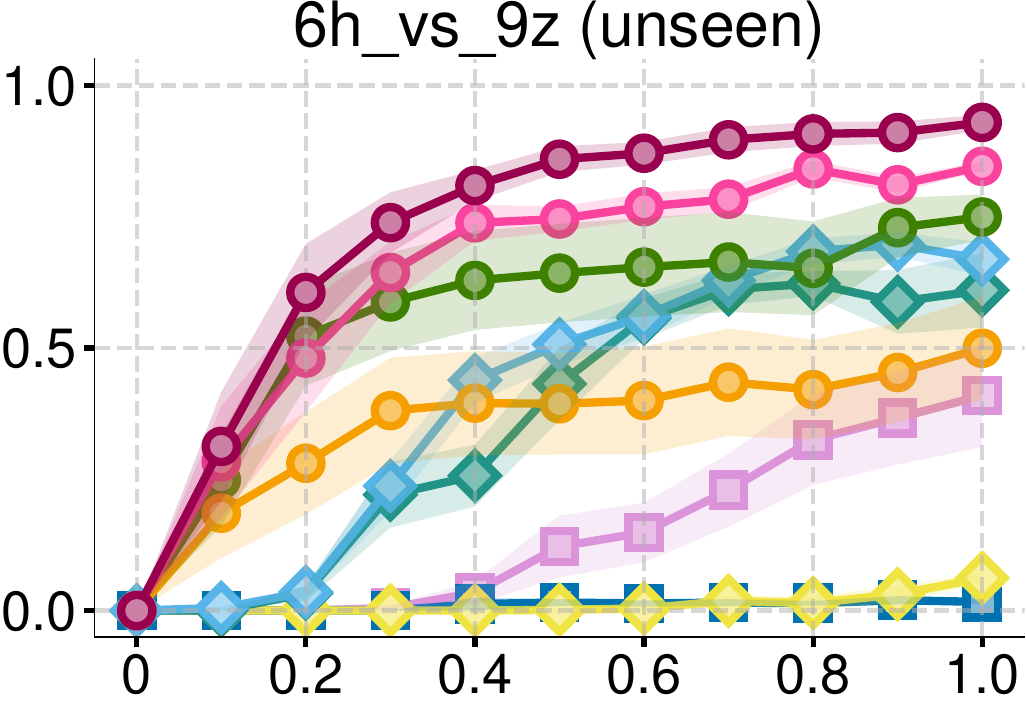}%
    \includegraphics[width=0.19\linewidth]{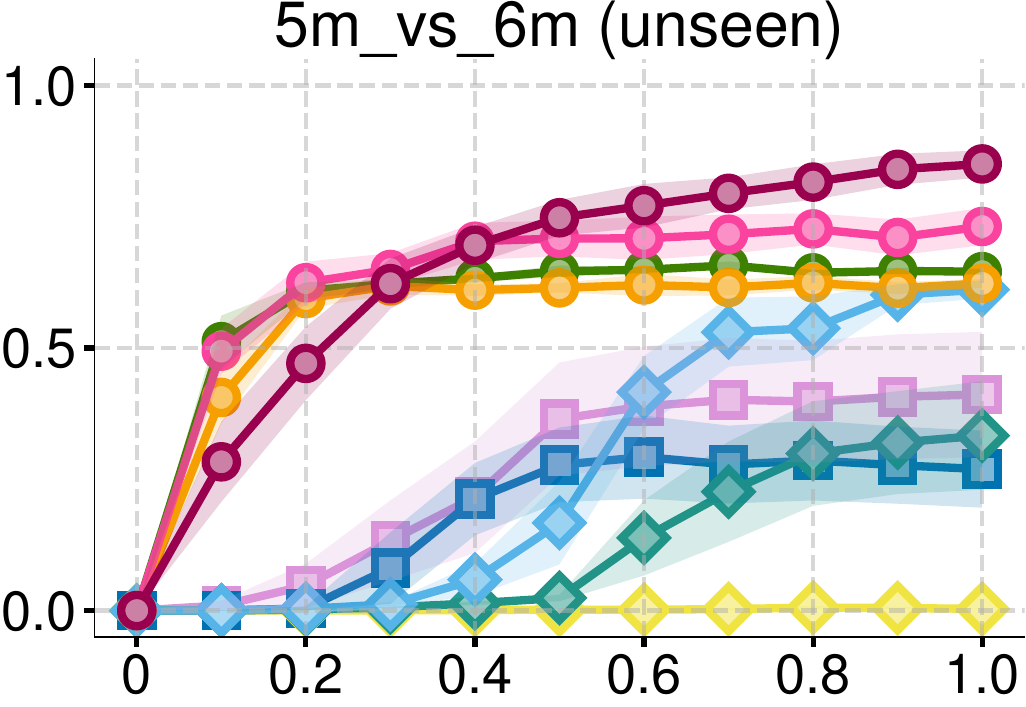}%
    \includegraphics[width=0.19\linewidth]{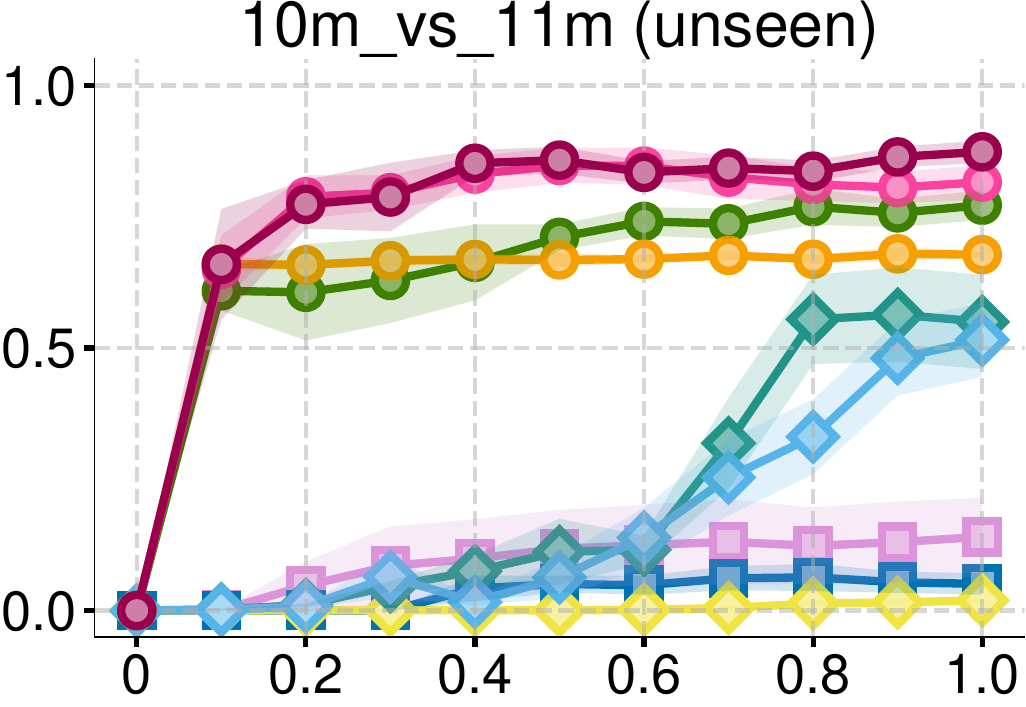}%
    \includegraphics[width=0.19\linewidth]{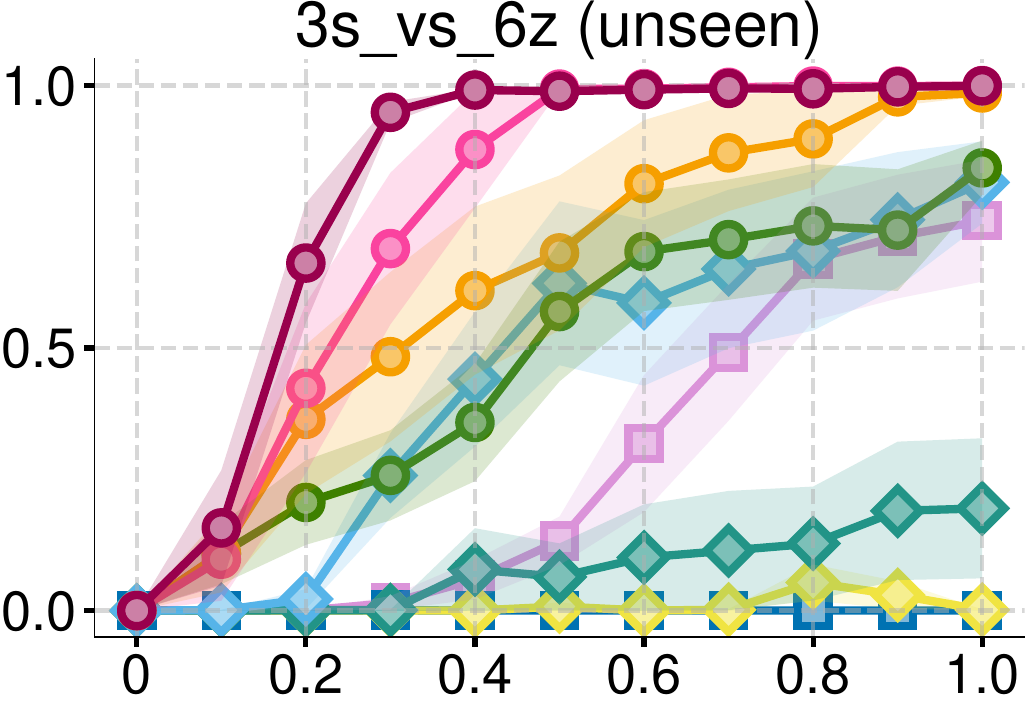}%
    \includegraphics[width=0.19\linewidth]{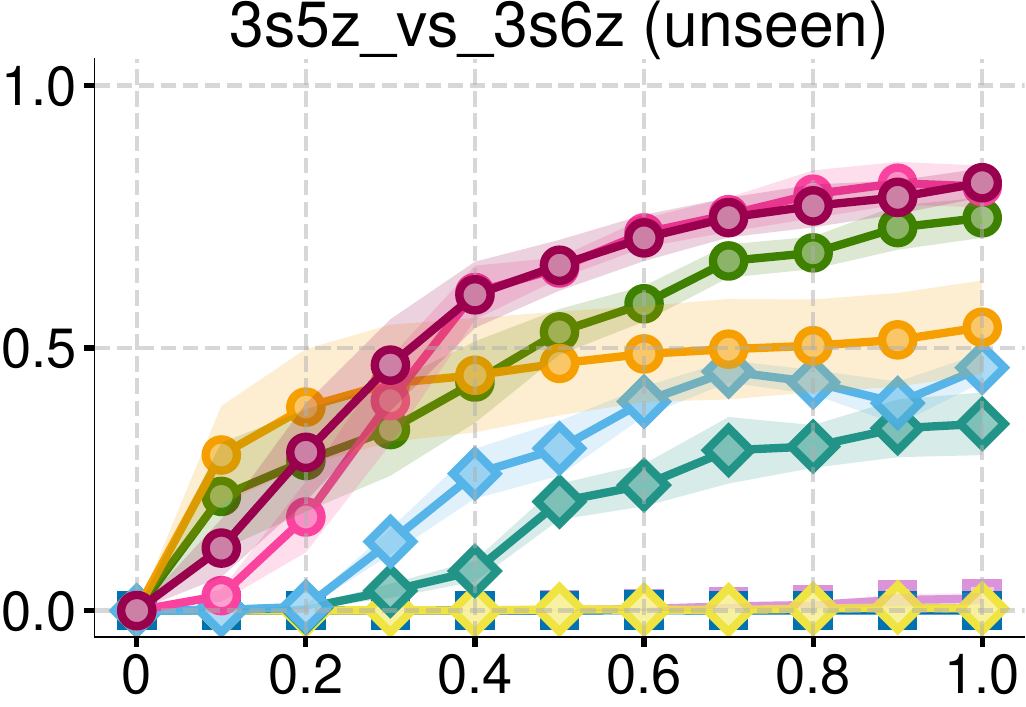}
    \vspace{-0.05in}

    \includegraphics[width=0.4\linewidth]{pdf/xlabel_long.pdf}
    \vspace{-0.05in}

    \caption{Average win rate for all task-regime pairs in the small-scale benchmark.}
    \vspace{0.05in}
    \label{fig:small_all}

    \raisebox{0.05in}{\includegraphics[width=0.015\linewidth]{pdf/ylabel.pdf}}
    \includegraphics[width=0.19\linewidth]{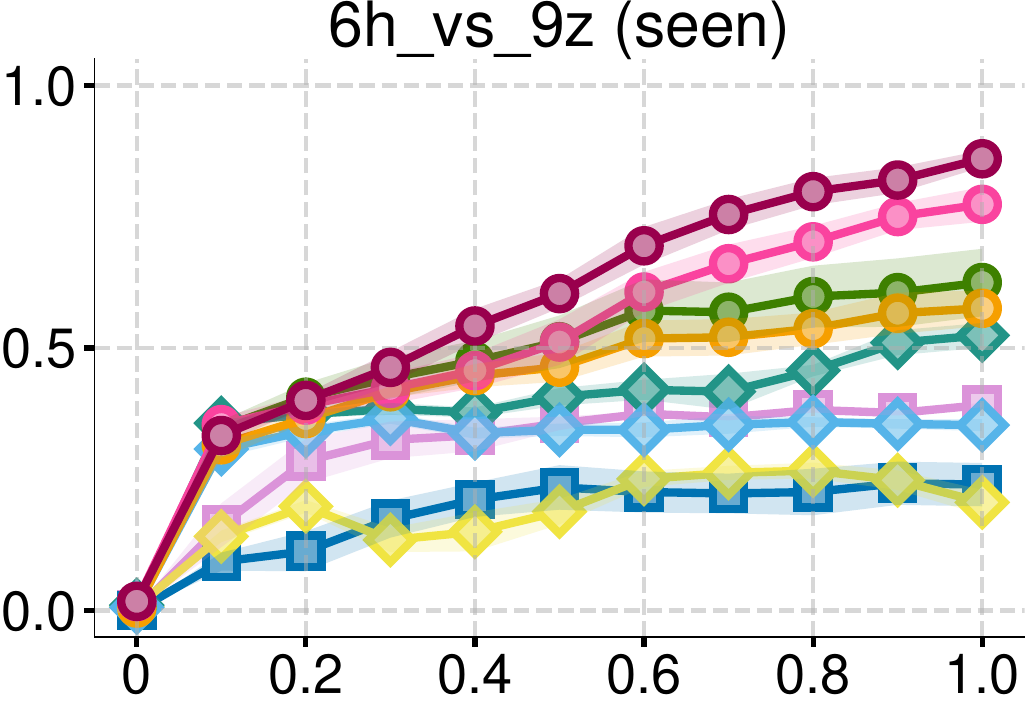}%
    \includegraphics[width=0.19\linewidth]{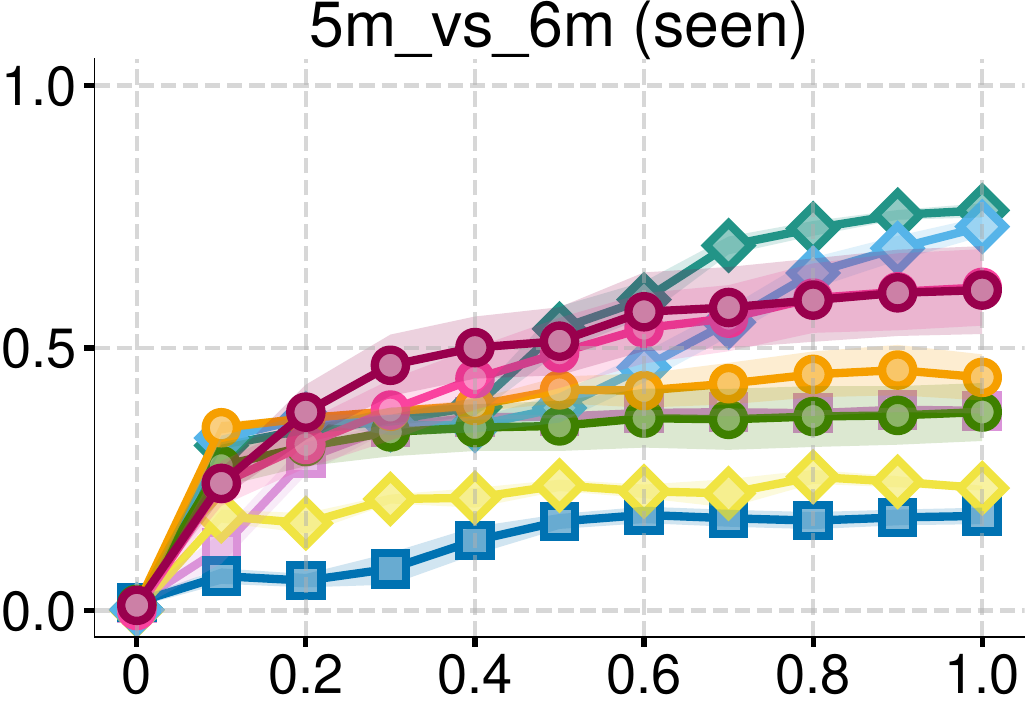}%
    \includegraphics[width=0.19\linewidth]{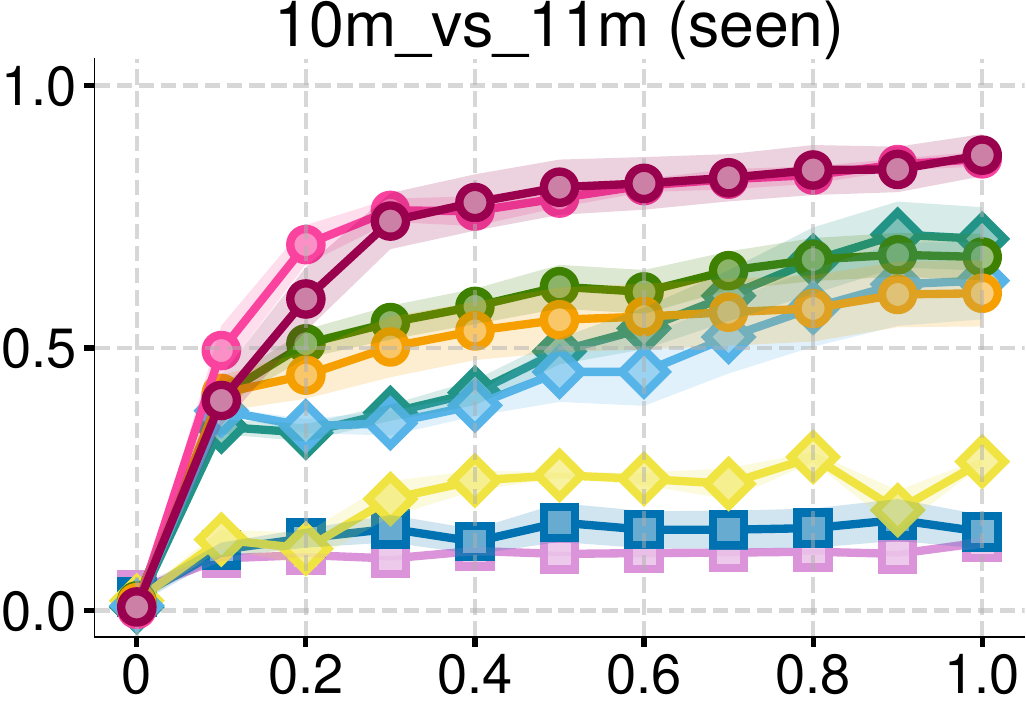}%
    \includegraphics[width=0.19\linewidth]{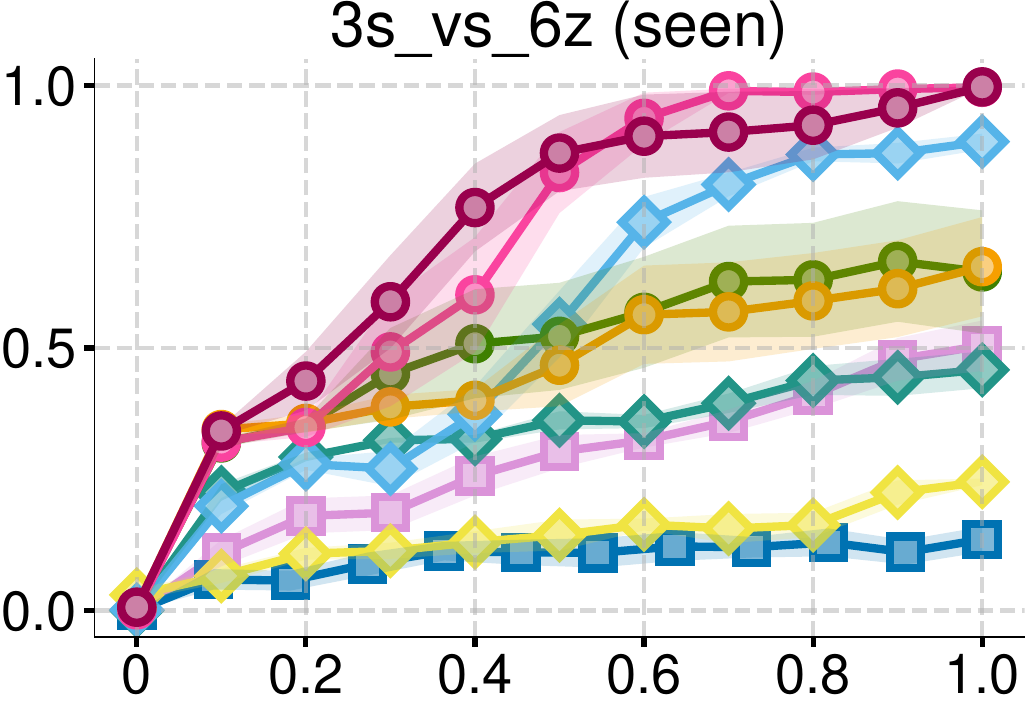}%
    \includegraphics[width=0.19\linewidth]{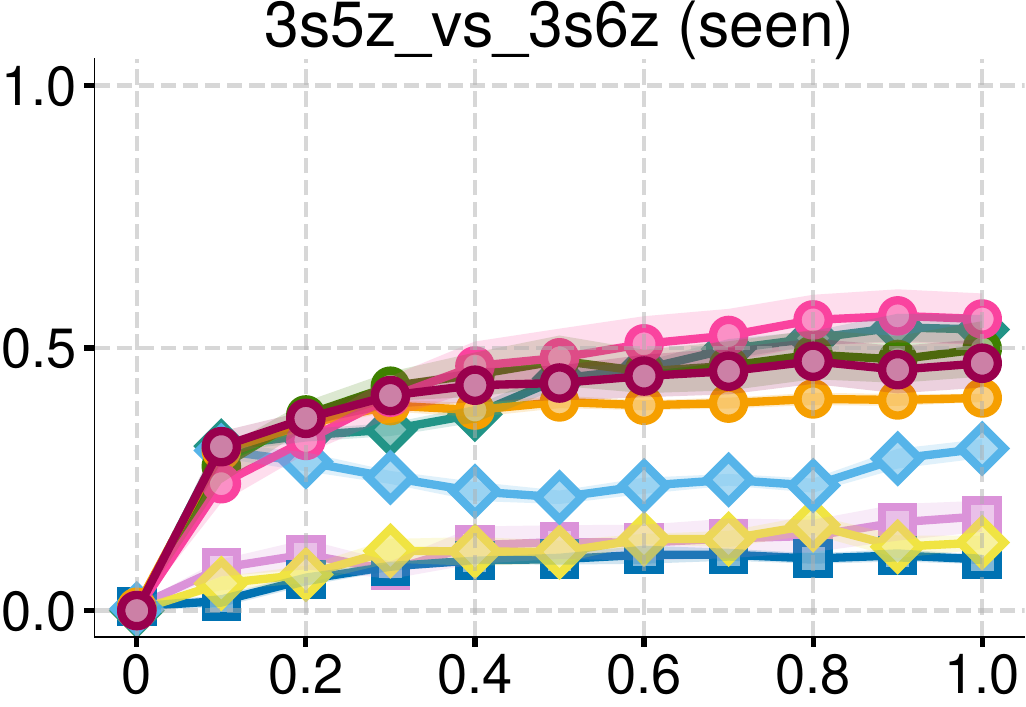}
    \vspace{-0.05in}

    \includegraphics[width=0.4\linewidth]{pdf/xlabel_long.pdf}
    \vspace{0.05in}

    \raisebox{0.05in}{\includegraphics[width=0.015\linewidth]{pdf/ylabel.pdf}}
    \includegraphics[width=0.19\linewidth]{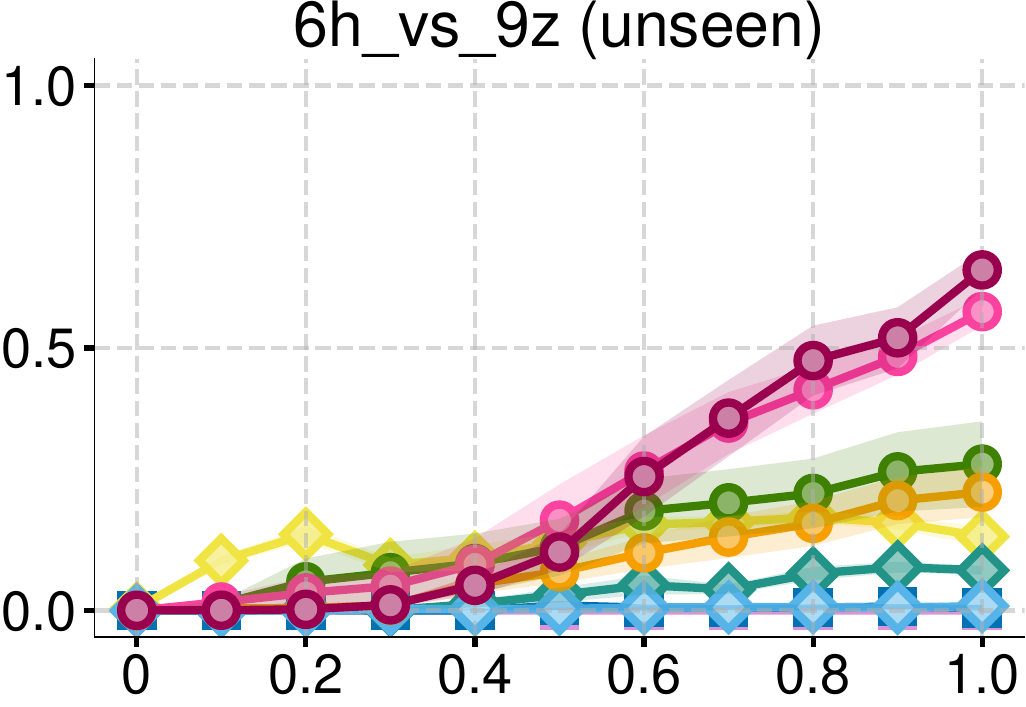}%
    \includegraphics[width=0.19\linewidth]{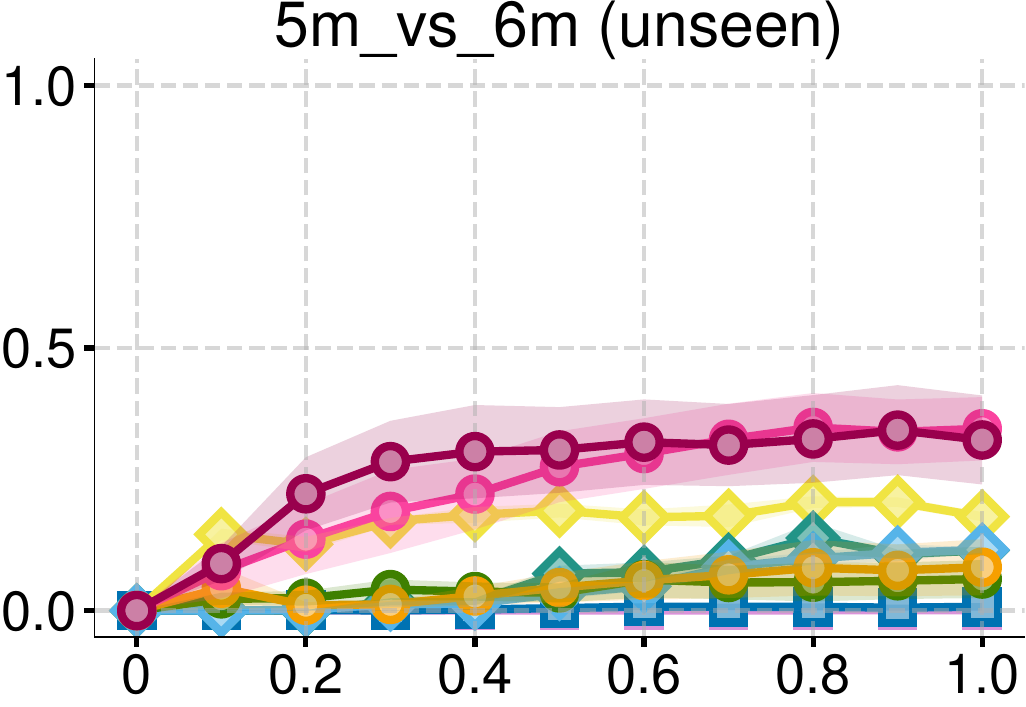}%
    \includegraphics[width=0.19\linewidth]{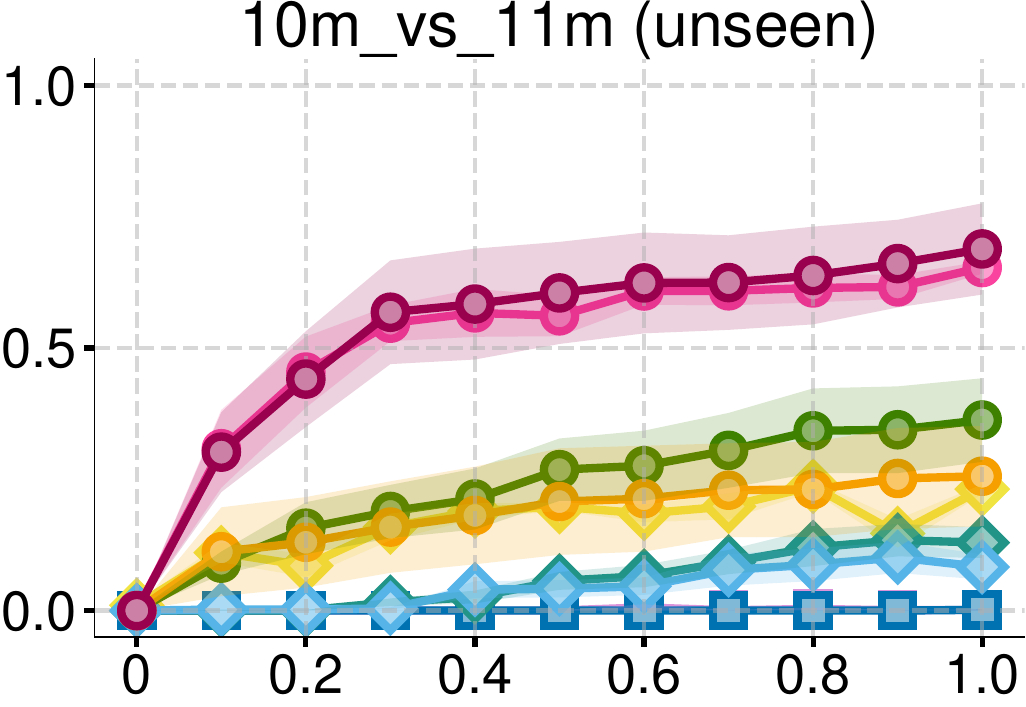}%
    \includegraphics[width=0.19\linewidth]{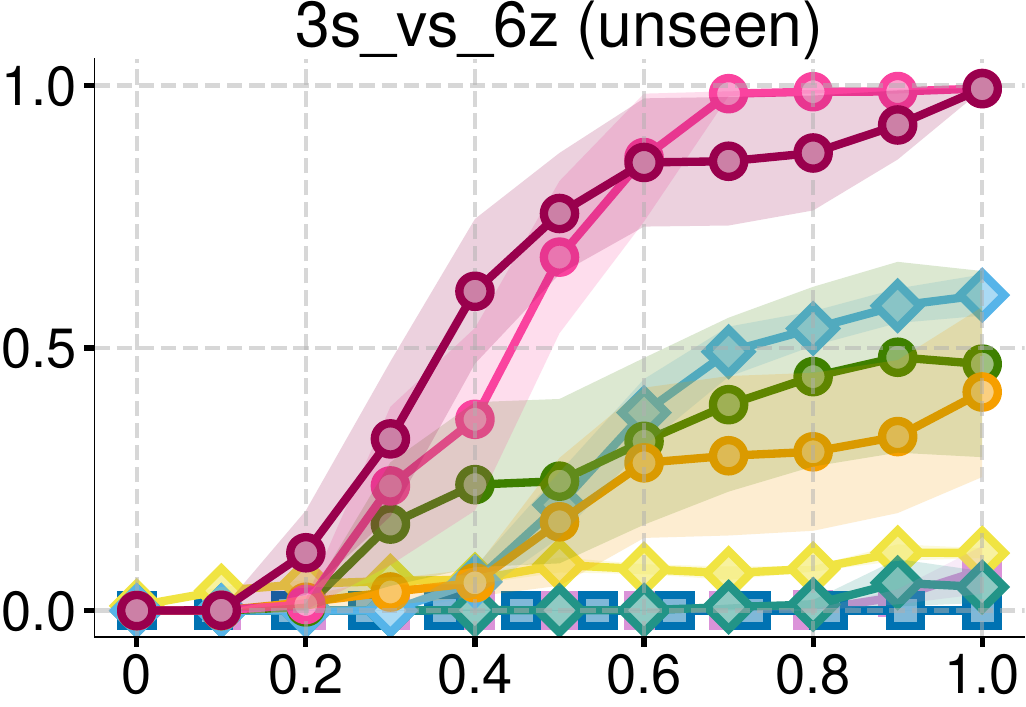}%
    \includegraphics[width=0.19\linewidth]{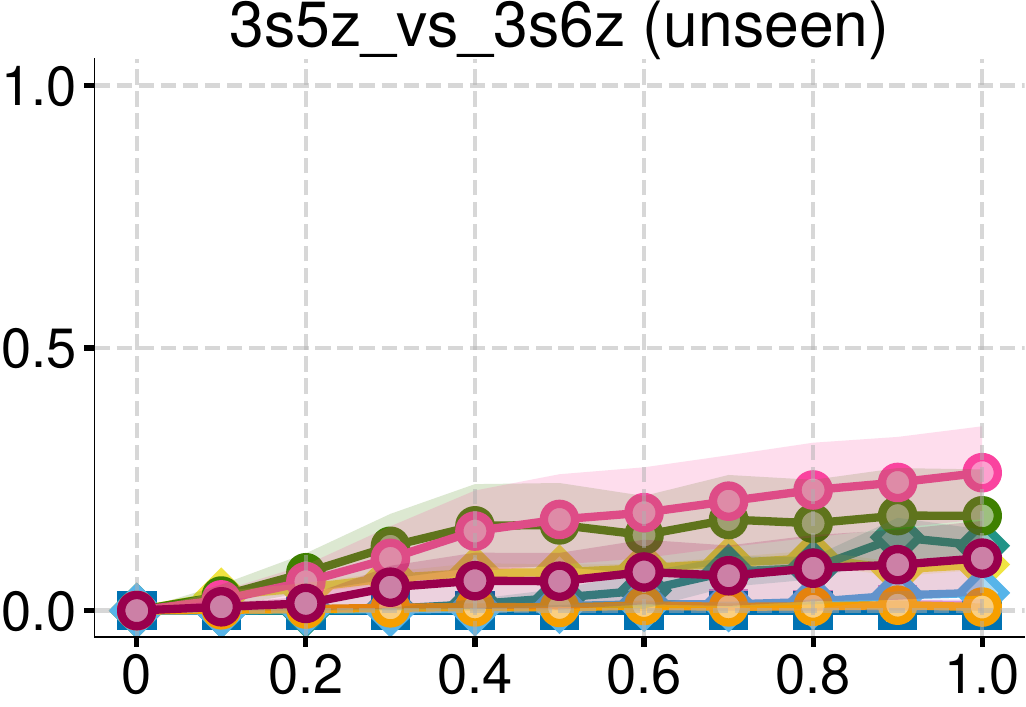}
    \vspace{-0.05in}

    \includegraphics[width=0.4\linewidth]{pdf/xlabel_long.pdf}
    \vspace{-0.05in}

    \caption{Average win rate for all task-regime pairs in the medium-scale benchmark.}
    \vspace{0.05in}
    \label{fig:medium_all}

    \raisebox{0.05in}{\includegraphics[width=0.015\linewidth]{pdf/ylabel.pdf}}
    \includegraphics[width=0.19\linewidth]{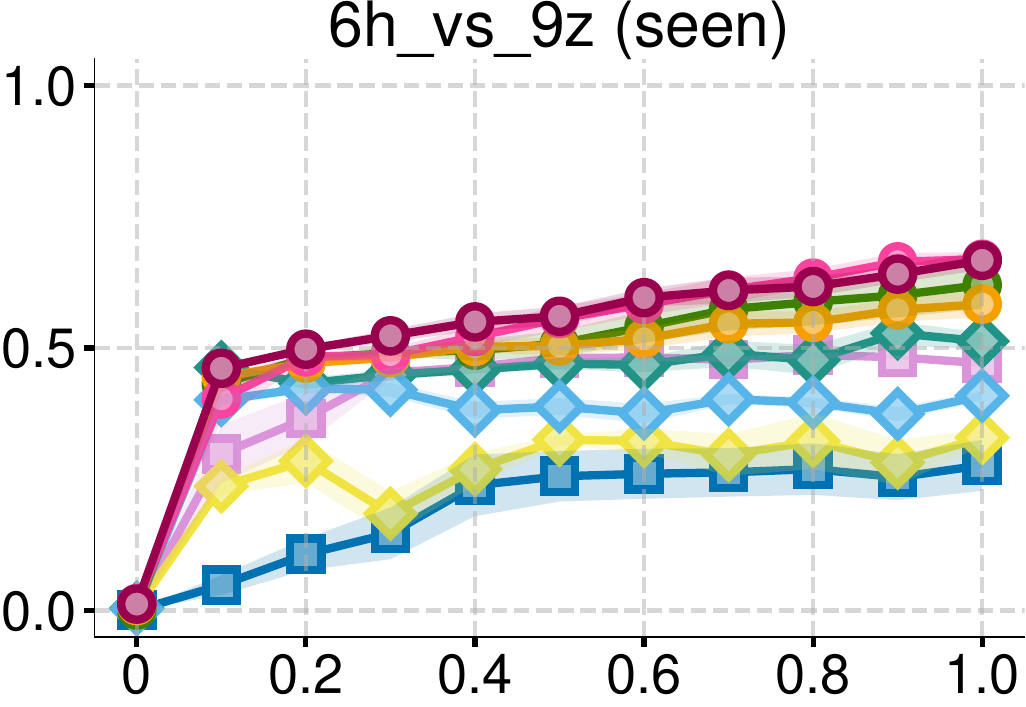}%
    \includegraphics[width=0.19\linewidth]{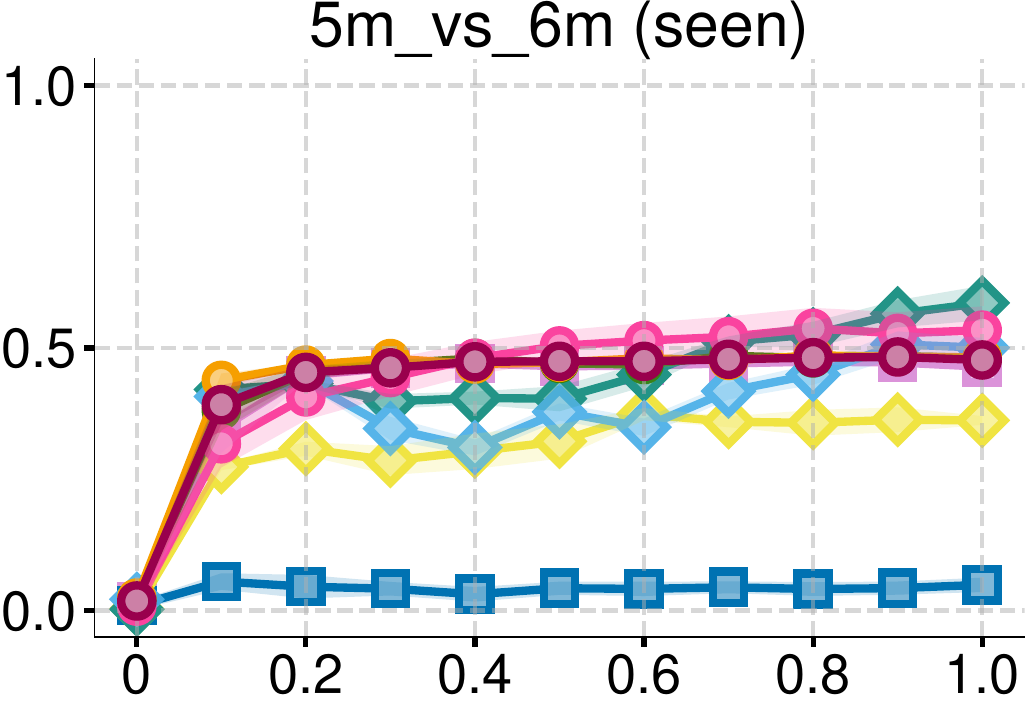}%
    \includegraphics[width=0.19\linewidth]{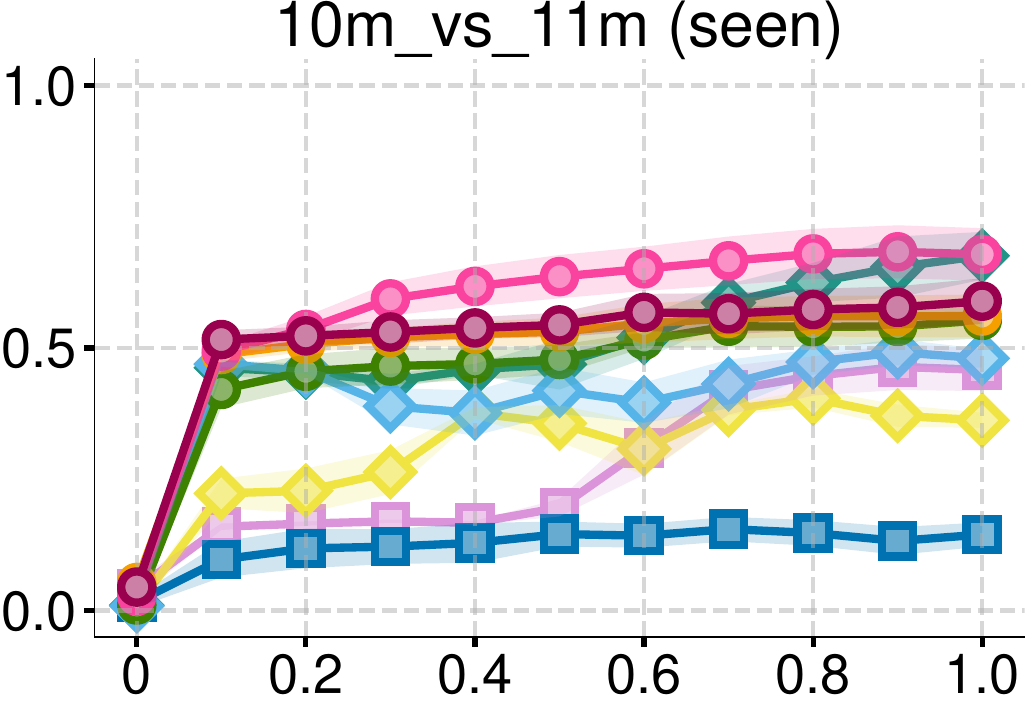}%
    \includegraphics[width=0.19\linewidth]{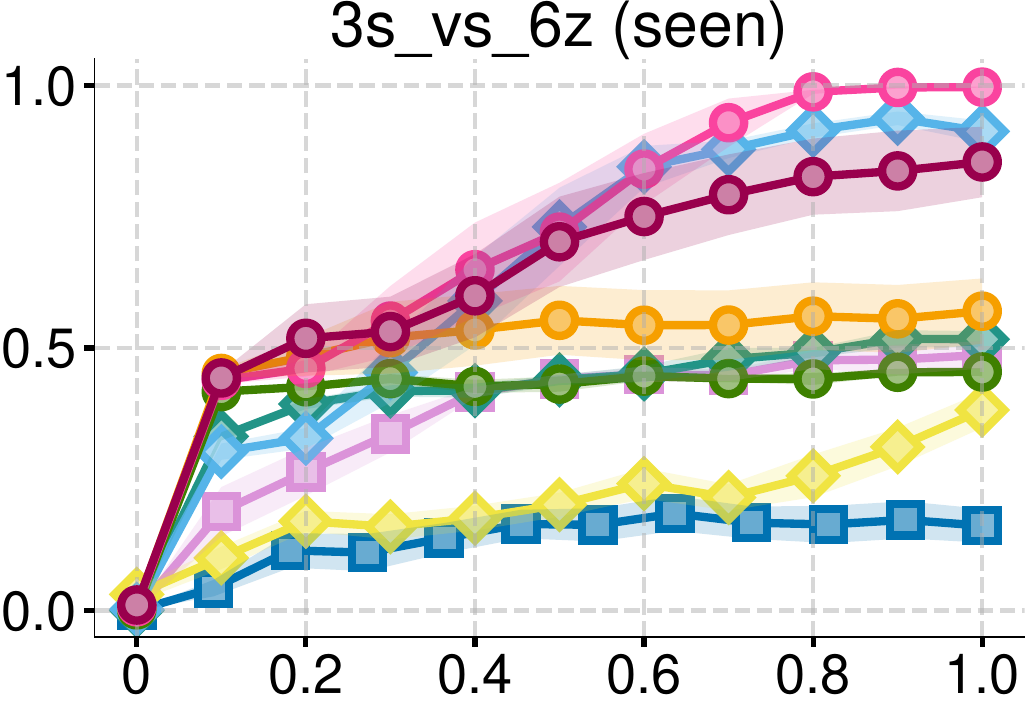}%
    \includegraphics[width=0.19\linewidth]{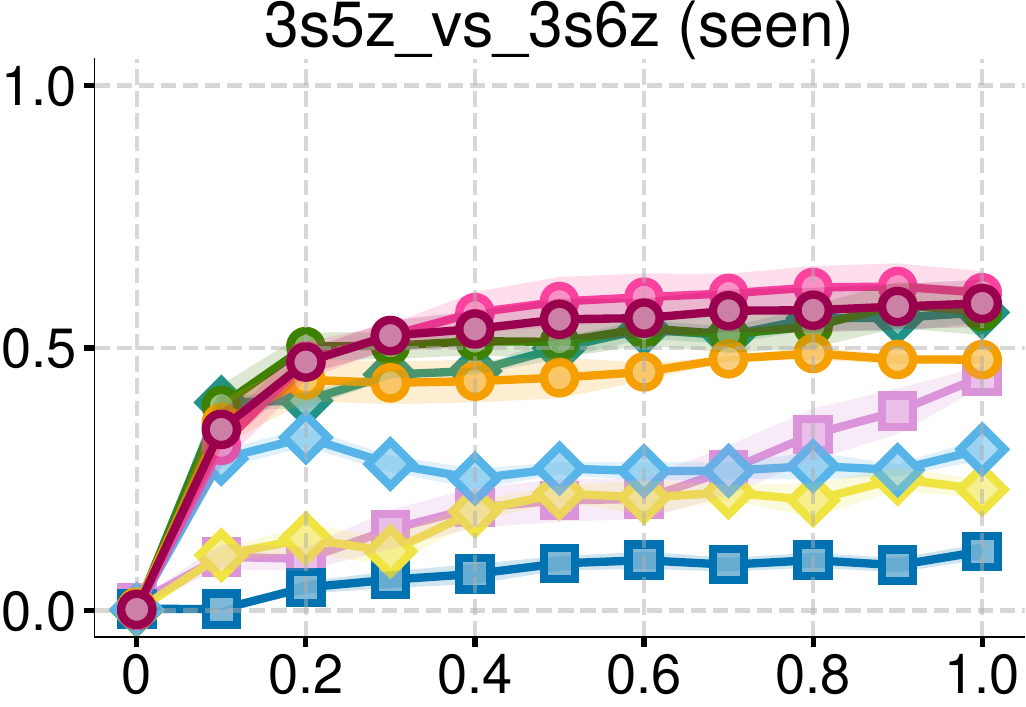}
    \vspace{-0.05in}

    \includegraphics[width=0.4\linewidth]{pdf/xlabel_long.pdf}
    \vspace{0.05in}

    \raisebox{0.05in}{\includegraphics[width=0.015\linewidth]{pdf/ylabel.pdf}}
    \includegraphics[width=0.19\linewidth]{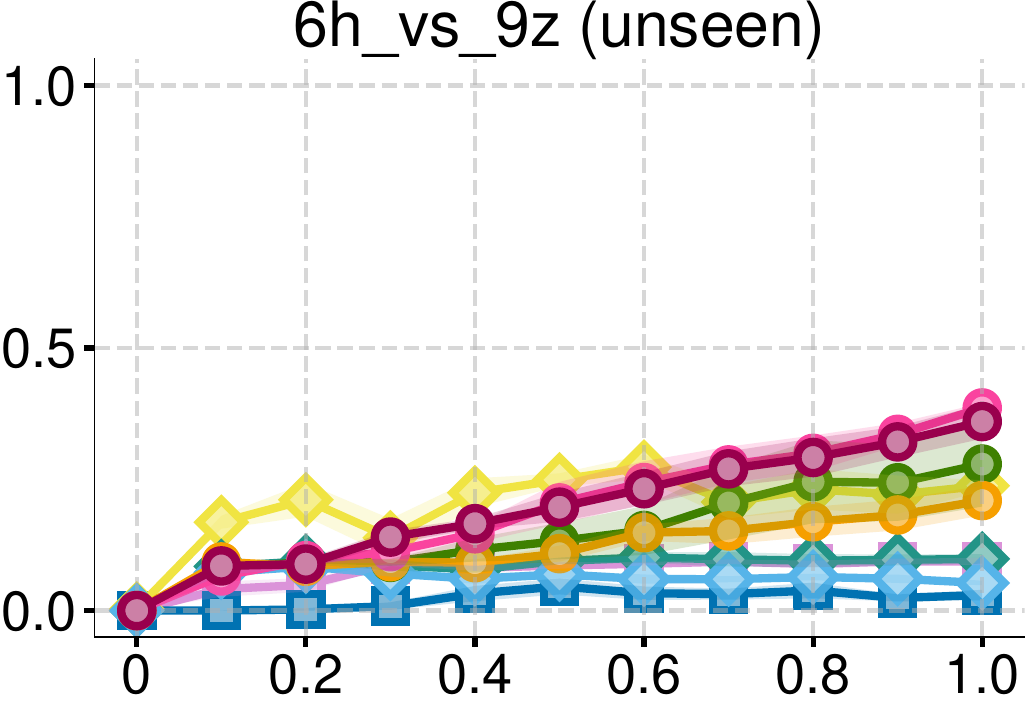}%
    \includegraphics[width=0.19\linewidth]{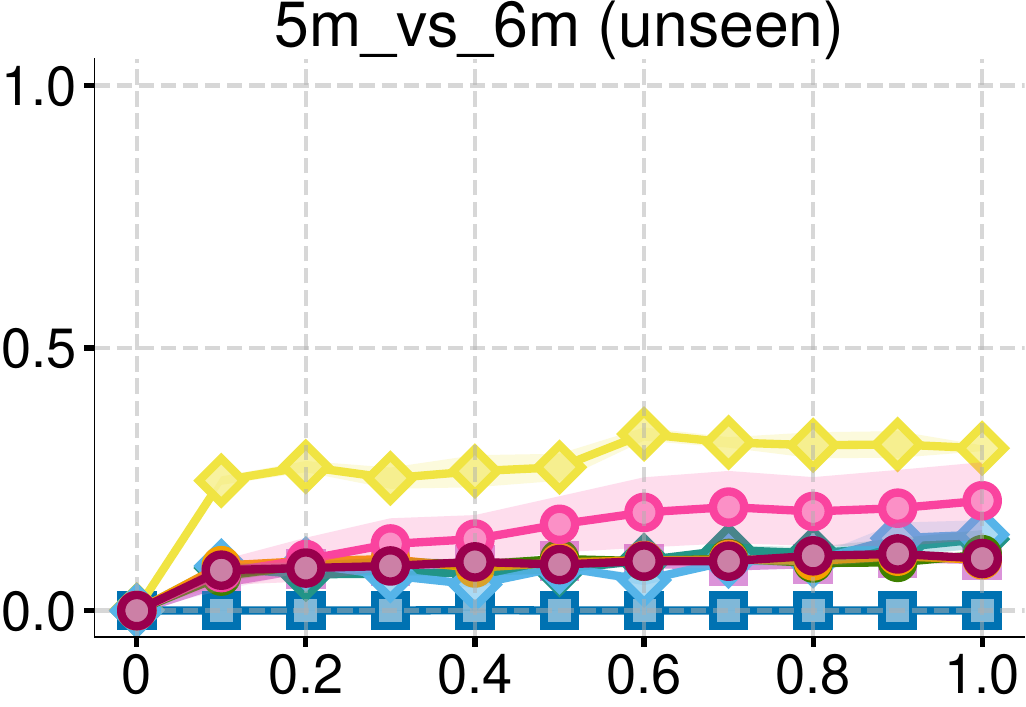}%
    \includegraphics[width=0.19\linewidth]{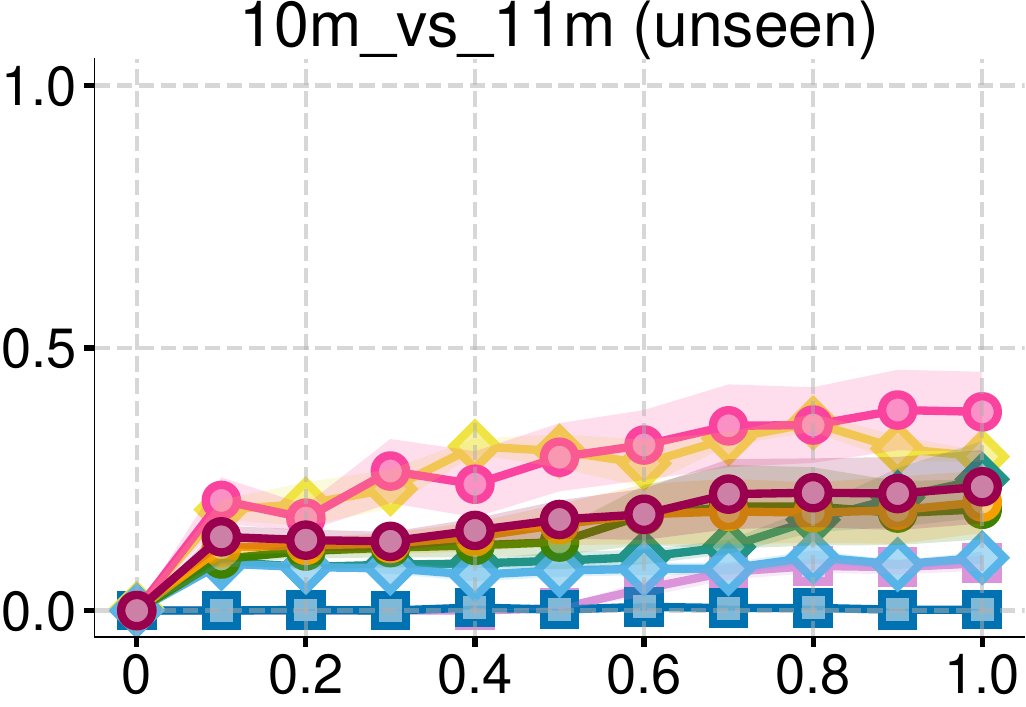}%
    \includegraphics[width=0.19\linewidth]{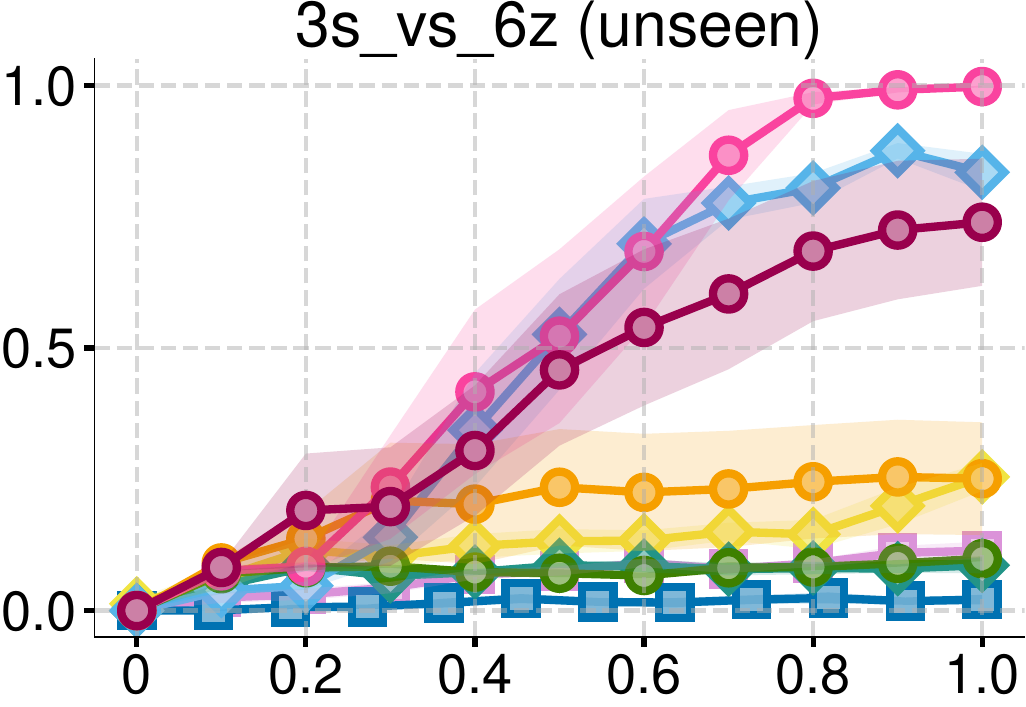}%
    \includegraphics[width=0.19\linewidth]{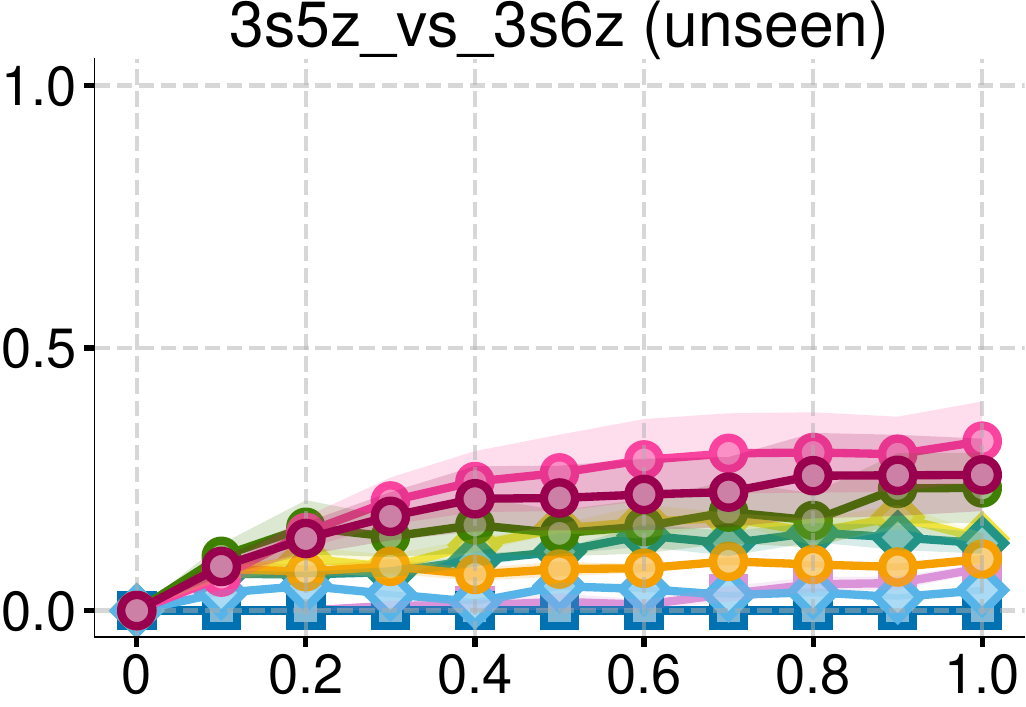}
    \vspace{-0.05in}

    \includegraphics[width=0.4\linewidth]{pdf/xlabel_long.pdf}
    \vspace{-0.05in}

    \caption{Average win rate for all task-regime pairs in the large-scale benchmark.}
    \label{fig:large_all}
\end{figure}

We provide all task results below. Figures~\ref{fig:small_all}-~\ref{fig:large_all} present the results for Small, Medium, Large scale respectively.

\end{document}